\documentclass[pra,preprint,showpacs]{revtex4}
\usepackage{amsmath,amssymb}
\usepackage[pdftex]{graphicx}
\usepackage{subfigure}
\usepackage{booktabs}
\newcommand \beq{\begin{eqnarray}}
\newcommand \eeq{\end{eqnarray}}
\newcommand \ah{\hat{a}}
\newcommand \ahd{\hat{a}^{\dagger}}

\newcommand \bhd{\hat{b}^{\dagger}}
\newcommand \bh{\hat{b}}
\newcommand \sk{\sum_{\mathbf{k}}}
\newcommand \ek{\epsilon_{\mathbf{k}}}
\newcommand \mk{\mathbf{k}}
\newcommand \mmk{\mathbf{-k}}

\begin{document}
\author{Shun Uchino$^{1}$}
\author{Michikazu Kobayashi$^{1}$}
\author{Masahito Ueda$^{1,2}$}
\affiliation{$^{1}$Department of Physics, The University of Tokyo,
Tokyo 113-0033, Japan\\
$^{2}$ERATO Macroscopic Quantum Project, JST, Tokyo 113-8656, Japan}
\title{Bogoliubov Theory and Lee-Huang-Yang Corrections
 in Spin-1 and Spin-2 Bose-Einstein Condensates 
in the Presence of the Quadratic Zeeman Effect} 
\begin{abstract}
We develop  
Bogoliubov theory of spin-1 and spin-2 Bose-Einstein condensates (BECs) 
in the presence of a quadratic Zeeman effect,
and derive the Lee-Huang-Yang (LHY) corrections to
the ground-state energy, pressure, sound velocity, and 
quantum depletion.
We investigate all the phases of spin-1 and spin-2
BECs that can be realized experimentally.
We also examine the stability of each phase
against quantum fluctuations and the quadratic Zeeman effect. 
Furthermore, we discuss a relationship between 
the number of symmetry generators that are spontaneously broken and 
that of Nambu-Goldstone (NG) modes.
It is found that in the spin-2 nematic phase
 there are special Bogoliubov modes that have gapless linear dispersion
relations but do not belong to  
the NG modes.
\end{abstract}
\maketitle
%\tableofcontents
\section{Introduction}
The Bogoliubov theory of weakly-interacting Bose-Einstein condensates
(BECs) \cite{bogoliubov} has served as
an indispensable tool
in diverse subfields of physics.
For a scalar BEC of bosons with mass $M$, particle-number density
$n$, and $s$-wave scattering length $a$,
the ground-state energy (GSE) of the system with volume $V$ is given by
\beq
E_0=\frac{2\pi\hbar^2V n^2}{M}a\left( 1+\frac{128}{15}
\sqrt{\frac{na^3}{\pi }} +\cdots  \right),\label{scalar}
\eeq    
where the first term on the right-hand side is the mean-field energy,
the second term gives a nonperturbative correction to it which
was first derived by Lee, Huang, and Yang (LHY)
\cite{lee,lhy}, 
and the higher-order terms were
discussed in Refs. \cite{wu,hugenholtz,sawada}. 
In the present paper, we discuss Bogoliubov theory and LHY corrections
of BECs with
spin degrees of freedom in the presence of a quadratic Zeeman effect.

The Bogoliubov theory of spinor BECs has been discussed 
extensively over the past decade.
The spin-1 Bogoliubov spectra have been derived  
in Refs. \cite{ho, ohmi,huang,ueda2} up to the linear Zeeman effect.
In Ref. \cite{szirmai}, the same problem is discussed
from a field-theoretic point of view.
The effect of the quadratic Zeeman energy on the spin-1 BEC has been discussed
in Refs. \cite{murata,ruostekoski}. 
The spin-2 BEC has been examined in the absence of an external magnetic
field in
Ref. \cite{suominen} and
up to the linear Zeeman effect in Ref. \cite{ueda}.
However, little attention has been paid to the GSEs.

In this paper, we develop 
a systematic renormalization procedure and derive GSEs, pressure,
sound velocity, and quantum depletion up to the LHY corrections.
The LHY corrections have been measured  
for a scalar BEC \cite{xu,papp} and for  a two-component Fermi gas
\cite{grimm} by
using methods to enhance quantum fluctuations.
The experiment in Ref. \cite{xu} utilized a 
strongly correlated system in an optical lattice, while  
the experiments in Refs. \cite{papp,grimm} 
amplified the coupling constant by means of
a Feshbach resonance.
Our analysis takes into account 
the quadratic Zeeman effect that is of great importance 
under many experimental situations
in which the linear Zeeman effect can be ignored.
Because the sign of the quadratic Zeeman term $q$ 
can be manipulated experimentally \cite{leslie},
both cases of positive and negative $q$ are analyzed.
It is shown that except for the ferromagnetic phase
the LHY correction in a spinor BEC is affected by the quadratic 
Zeeman effect and that it can be measured by making strongly correlated systems
or by controlling the external magnetic
field.

The order parameter of a spin-2 nematic BEC in the absence of
an external magnetic field
depends on an
additional parameter, $\eta$, 
that is not related to the symmetry of the 
Hamiltonian but describes the degeneracy between the
uniaxial and biaxial nematic phases.
As pointed out in Refs. \cite{turner,song}, however,
quantum fluctuations induce
a quantum phase transition between the two phases, lifting the degeneracy. 
We show that  
the quadratic Zeeman effect with $q<0$
causes the dynamical
instability in the uniaxial nematic phase, whereas it leaves 
the biaxial nematic phase stable. 
That is, the uniaxial nematic phase is unstable against an
infinitesimal negative quadratic Zeeman effect in the thermodynamic limit.
Conversely, the quadratic Zeeman effect with $q>0$ makes
the biaxial nematic phase dynamically unstable while 
it leaves the uniaxial nematic phase  stable.
However, it is possible to stabilize both of these phases for nonzero $q$
in a finite system.
We will show this for the case of a
spin-2 BEC. 

The Bogoliubov theory
predicts massless modes, which
can be interpreted as Nambu-Goldstone (NG) modes associated with
spontaneous symmetry breaking.
To elucidate this point, we discuss a 
relationship between the number of
symmetry generators that are spontaneously broken and the number of 
NG modes \cite{nielsen}. 
We apply the relationship 
to spin-1 and spin-2 BECs,
and
point out that for the uniaxial and biaxial nematic phases
there exist the Bogoliubov modes 
that have  gapless linear dispersion relations but do not belong to
the NG modes.

This paper is organized as follows. Section I\hspace{-.1em}I formulates
 the problem, and describes 
the low-energy Hamiltonian and Hartree-Fock approximation of a spin-$f$ BEC.
Section I\hspace{-.1em}I\hspace{-.1em}I discusses the problem  
of divergence of the GSE
and 
how to remove the divergence by
renormalization of the coupling constant.
Sections I\hspace{-.1em}V and V
examine the mean-field phase diagrams and Bogoliubov theory of 
spin-1 and spin-2 BECs, respectively, in the 
thermodynamic limit, and derive the Bogoliubov spectra
and LHY corrections.  
Section V\hspace{-.1em}I discusses 
the relationship between the number of symmetry generators that are
spontaneously broken and the number of NG modes in spinor BECs.
Section V\hspace{-.1em}I\hspace{-.1em}I\hspace{-.1em} provides the summary and concluding remarks.
The detailed derivations of the GSEs are described in Appendix A, and
the properties and equation numbers of the physical quantities
in each phase
and notations are listed in 
Appendix B.
\section{Formulation of the problem}
We consider a system of $N$ spin-$f$ identical bosons with mass $M$
that undergo
an $s$-wave
scattering subject to periodic boundary conditions.  
As in most experiments done in spinor BECs,
we consider the case in which the linear Zeeman effect can be ignored.
Let  $\displaystyle\hat{\Psi}_m(\mathbf{x})$ $(m=-f,-f+1,...,f)$ be
the field operator of a boson at position 
$\displaystyle\mathbf{x}$ with magnetic quantum number $m$, where
we  assume that an external magnetic field $B$ is 
applied in the $z$ direction.
Then,
the low-energy effective Hamiltonian of a spin-$f$ BEC is given by 
\beq
\hat{H}=\hat{H}_{\text{KE}}+\hat{H}_{\text{QZ}}+\hat{V},\label{secondint}
\eeq
where 
\beq
\hat{H}_{\text{KE}}=\int d\mathbf{x} \ \hat{\Psi}^{\dagger}_m(\mathbf{x})
\Big( -\frac{\hbar^{2}\nabla^2}{2M}\Big)\hat{\Psi}_m (\mathbf{x})
\eeq
is the kinetic energy, 
\beq
\hat{H}_{\text{QZ}}=qm^2\int d \mathbf{x} \ \hat{\Psi}^{\dagger}_m(\mathbf{x})
\hat{\Psi}_m (\mathbf{x})
\eeq
is the quadratic Zeeman term, and
\beq
\hat{V}=\sum_{F=0}^{2f}\frac{\bar{g}_F}{2} \sum_{M=-F}^{F}
\langle fmfm'|FM\rangle\langle FM|f\mu f\mu '\rangle
\int d\mathbf{x} \
\hat{\Psi}_m^{\dagger}(\mathbf{x}) 
\hat{\Psi}_{m'}^{\dagger}(\mathbf{x}) \hat{\Psi}_{\mu}(\mathbf{x}) 
\hat{\Psi}_{\mu '}(\mathbf{x})
\label{secondqint}
\eeq
is the interaction energy. The strength of the quadratic Zeeman term is
given by
$q=(g\mu_{B}B)^2/E_{\text{hf}}$, where $g$ is the Land\'{e} $g$-factor,
$\mu_{B}$ is the Bohr magneton, and $E_{\text{hf}}$ is the hyperfine energy
splitting.
In Eq. (\ref{secondqint}),
$\bar{g}_F$ is a bare coupling constant in the total spin $F$
channel and 
$\displaystyle \langle fmfn|FM\rangle$ is the Clebsch-Gordan coefficient.  
Here and henceforth,
repeated indices  such as $m, m', \mu ,\ \text{and}\ \mu '$ are
assumed to be summed over $f,f-1,...,-f$ unless otherwise stated.
Bose symmetry requires that the total spin $F$ in the $s$-wave channel
is even. In fact, it follows from the canonical 
commutation relations of bosons
and the properties of the 
Clebsch-Gordan coefficients that the terms in Eq. (\ref{secondqint}) 
with odd $F$
vanish identically.

We expand the field operator as 
\beq
\hat{\Psi}_m(\mathbf{x})=\frac{1}{\sqrt{V}}\sum_{\mathbf{k}}\hat{a}_{\mathbf{k},m}\text{e}^{i\mathbf{k\cdot
x}}, \label{eq:expandphi}
\eeq
where $V$ is the volume of the system and $\ah_{\mk,m}$ is the 
annihilation operator of a spin-$f$ boson with wave number $\mk$ and
magnetic quantum number $m$. Substituting Eq. (\ref{eq:expandphi})
in Eq. (\ref{secondint}), we obtain 
\beq
\hat{H}=\sum_{\mathbf{k}}(\epsilon_{\mathbf{k}}+qm^2)
\hat{a}^{\dagger}_{\mk,m}\hat{a}_{\mk,m} +\sum_{F=0}^{2f}\frac{\bar{g}_F}{2V} \sum_{M=-F}^{F} \sum_{\mathbf{k},\mathbf{p},\mathbf{q}}
\langle fmfm'|FM\rangle\langle FM|f\mu f\mu '\rangle
\hat{a}_{\mathbf{p},m}^{\dagger}\hat{a}_{\mathbf{q},m'}^{\dagger}
\hat{a}_{\mathbf{p+k},\mu}\hat{a}_{\mathbf{q-k},\mu '}, \label{secondint2}
\eeq
where $\epsilon_{\mathbf{k}}=\hbar^2\mathbf{k}^2/2M$.

In the mean-field or Hartree-Fock approximation, 
all bosons are assumed to occupy 
a single mode, which we label as $\mathbf{0}$: 
\beq
|\mathbf{\zeta}\rangle
=\frac{1}{\sqrt{N!}}\left(\sum_{m=-f}^{f}\zeta_m\hat{a}^{\dagger}_{\mathbf{0},m}
\right)^{N}|\text{vac}\rangle ,\label{coherents}
\eeq
where variational parameters $\zeta_m$ are assumed to satisfy 
the normalization condition 
$\sum_{m}|\zeta_{m}|^2=1$ and they are determined so as to
minimize the expectation value of the Hamiltonian.
By using the trial state (\ref{coherents}),
it is possible to classify the 
mean-field ground-state
phase in a number-conserving manner
\cite{ueda2,koashi,ueda}.

\section{Renormalization of the ground-state energy}  
If we take into account the effects of quantum fluctuations, 
physical quantities such as GSE
show ultraviolet divergence.
This divergence stems from use of the contact
interaction, which gives correct results if we only consider
the region $|\mathbf{x}|\gg r_0$, where $r_0$ is 
the range of the interaction. 
This implies 
that the effective Hamiltonian (\ref{secondint}) is 
only valid below a certain cutoff momentum. 
By introducing the cutoff,
the GSE no longer diverges but
depends explicitly on the cutoff.
The renormalization of the coupling constant eliminates  
the cutoff in favor of an observable, that is, an $s$-wave scattering length
in the present problem.  

To examine this problem of the coupling constant in detail, 
let us first consider the low-energy scattering between 
spin-$f$ identical bosons.
The scattering rate for each scattering channel is determined by
the $T$-matrix $\hat{T}$, 
which is the solution to the following equation \cite{pethick}:
\beq
\hat{T}=\hat{V}+\hat{V}\frac{1}{E-2\hat{H}_{\text{KE}}}\hat{T},
\eeq 
where $E$ is the total energy and $1/(E-2\hat{H}_{\text{KE}})$
is the two-particle Green's function.
We deal with the quadratic Zeeman term perturbatively by
assuming that it is at most of the order of
$\hat{V}$ -- the condition well satisfied in current experiments.
This assumption is implicitly made in classifying the
mean-field ground-state phases 
in Refs. \cite{murata,stenger,saito}. 

\begin{figure}[h]
 \begin{center}
  \includegraphics[width=0.5\linewidth]{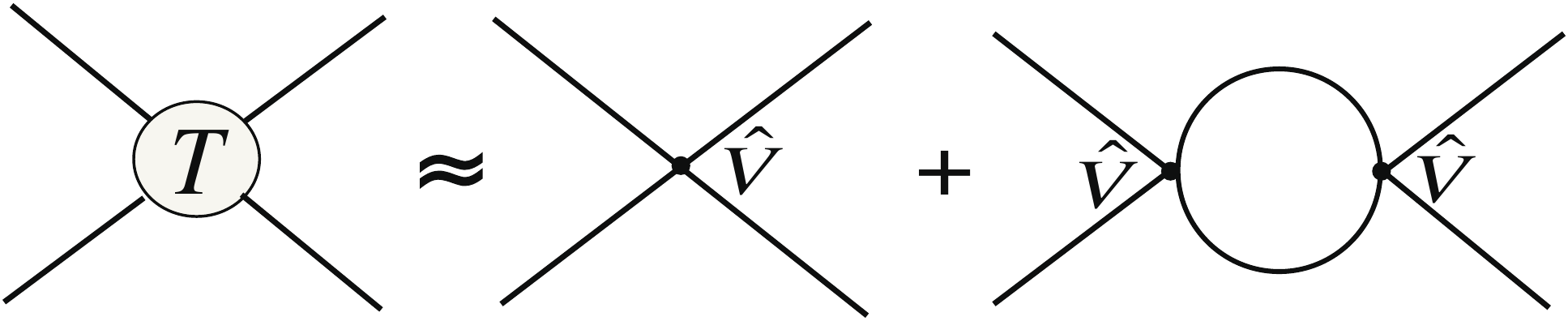}
  \caption{
Diagrammatic representation of the coupling-constant renormalization in
Eq. (\ref{renormalize}), where each solid line describes the one-particle
Green's function of the Hamiltonian $\hat{H}_{\text{KE}}$
and each vertex represents $\hat{V}$. 
The first diagram on the right-hand side represents
the Hartree-Fock term, and the second diagram represents the second-order term. }
  \label{feynman}
 \end{center}
\end{figure}

The $T$-matrix is related to the $s$-wave scattering lengths $a_F$
and 
renormalized coupling constants $g_F$ as follows:
\beq 
g_F\equiv\frac{4\pi\hbar^2 a_F}{M}=\lim_{k\to 0} \langle 
\mathbf{k'},F|\hat{T}|
\mathbf{k},F\rangle,
\label{renormalizeg}
\eeq
where 
$\mathbf{k}$ and $\mathbf{k'}$ are the incoming and outgoing wave
vectors, respectively, and $k\equiv |\mathbf{k}|=|\mathbf{k'}|$ because
of energy conservation. 
In the Hartree-Fock approximation, the $T$-matrix is 
approximated by $\hat{T}\approx\hat{V}$, and therefore
\beq
g_{F}\equiv\frac{4\pi\hbar^2 a_{F}}{M}=\bar{g}_{F}.
\eeq
In the Bogoliubov theory, however, this approximation does not 
remove the cutoff
dependence because the divergence occurs at the level of second order
in $\bar{g}_F$.
Thus, we must approximate the $T$-matrix up to second order 
in the coupling constants:  
\beq 
\hat{T}\approx\hat{V}+\hat{V}\frac{1}{E-2\hat{H}_{\text{KE}}}\hat{V}.
\label{bogoliubovt}
\eeq
The corresponding relation between the bare and renormalized coupling
constants is given by
\beq 
g_F\equiv\frac{4\pi\hbar^2 a_F}{M}
=\bar{g}_F-\frac{\bar{g}_F^2}{V}{\sum_{\mk}}^{'} \frac{1}{2\ek},
\label{renormalize}
\eeq
where the prime on the sum means to omit the term $\mk=\mathbf{0}$.
The diagrammatic representation of Eq. (\ref{renormalize}) is shown
in Fig. \ref{feynman}.

Next, we illustrate how this renormalization of the coupling constant 
eliminates the cut-off dependence. 
As we will show in subsequent sections, 
the GSE in the absence of an external magnetic field is expressed as
\beq 
E_0=E_0^{\text{MF}}+E_0^{\text{QF}}.
\label{eq:general-gse}
\eeq
Here $E_0^{\text{MF}}$ is the mean-field energy  
\beq
E_0^{\text{MF}}=\frac{Vn^2\bar{C}}{2},
\eeq 
where
$n=N/V$,
and $\bar{C}$ 
is a linear combination of the bare coupling constants.
The term $E_0^{\text{QF}}$ in Eq. \eqref{eq:general-gse} 
describes the contribution from
quantum fluctuations around the Hartree-Fock mean field
and
takes the following  form:
\beq
E_0^{\text{QF}}
=-\frac{\hbar^2}{4M}
\sum_{j}{\sk}^{'} 
 \Bigg[
k^2+\frac{2Mn\bar{C}_j}{\hbar^2}
-k\sqrt{k^2+\frac{4Mn\bar{C}_j}{\hbar^2}}\Bigg],
\label{eq:qfluc}
\eeq
where 
the sum over $j$ is taken over 
the Bogoliubov modes, each of which describes 
fluctuations such as density and spin fluctuations, and
$\bar{C}_j$ is a linear combination of the bare coupling
constants. 
 At the limit of $k\to \infty$, the integrand behaves as follows:  
\beq
k^2+\frac{2Mn\bar{C}_j}{\hbar^2}
-k\sqrt{k^2+\frac{4Mn\bar{C}_j}{\hbar^2}}
\xrightarrow[k\to \infty]{}\frac{1}{2}
\left(\frac{2MnC_j}{\hbar^2k}\right)^2 +O\left(\frac{1}{k^4}\right),
\eeq
where we substitute a renormalized coupling constant $C_j$ for
$\bar{C}_j$ on the right-hand side, 
which is correct up to the second order in the coupling 
constants.
On the other hand, from the $T$-matrix calculation 
of Eq. (\ref{renormalize}) (note that $g_F$ and $\bar{g}_F$ correspond
to $C$ and $\bar{C}$, respectively),
the mean-field energy is calculated to give
\beq
E_{0}^{\text{MF}}=\frac{Vn^2C^2}{2}+\frac{\hbar^2}{8M}\sum_{j}{\sk}^{'}
\left(\frac{2MnC_j}{\hbar^2k}\right)^2.
\eeq 
Here the second term on the right-hand side cancels
the ultraviolet part of Eq. (\ref{eq:qfluc}).
Thus, the cutoff dependence of $E_0^{\text{QF}}$  is removed.
Therefore, the GSE is given by
\beq
E_0&=&\frac{Vn^2C}{2}-\frac{\hbar^2}{4M}\sum_{j}{\sk}^{'}
\Bigg[k^2+\frac{2MnC_j}{\hbar^2}-k\sqrt{k^2+\frac{4MnC_j}{\hbar^2}}
-\frac{1}{2}\left(\frac{2MnC_j}{\hbar^2k}\right)^2\Bigg]\nonumber\\
&=&\frac{Vn^2C}{2}
-\frac{\hbar^2}{8\pi^2 M}\sum_j\left(\frac{2MnC_j}{\hbar^2}\right)^{\frac{5}{2}}
\int_{x_j}^{\infty}dx\
x^2\left(x^2+1-x\sqrt{x^2+2}-\frac{1}{2x^2}\right)
\nonumber\\
&\simeq&\frac{Vn^2C}{2}\Bigg(1+\frac{16\sqrt{M^3}}{15\hbar^3\pi^2}
\sum_j\frac{C_j}{C}\sqrt{nC_j^3}\Bigg),
\eeq
where $x_j\equiv 2\pi\hbar/(V^{1/3}\sqrt{2MnC_j})$, and
we take the thermodynamic limit $x_j=0$ to obtain the last expression.
Even if we incorporate the quadratic Zeeman effect, 
the above cancellation mechanism holds as shown in Appendix A.

\section{Spin-1 BEC}
For a spin-1 BEC, the total spin $F$
of two interacting bosons 
must be 0 or 2, and therefore,
Eq. (\ref{secondint2}) reduces to
\beq
\hat{H}=\sum_{\mathbf{k}}(\epsilon_{\mathbf{k}}+qm^2)
\hat{a}^{\dagger}_{\mk,m}\hat{a}_{\mk,m} 
+\frac{1}{2V}\sum_{\mathbf{k},\mathbf{p},\mathbf{q}}
\Bigg(\bar{c}_0^{(1)}\hat{a}_{\mathbf{p},m}^{\dagger}\hat{a}_{\mathbf{q},m'}^{\dagger}\hat{a}_{\mathbf{p+k},{m}}\hat{a}_{\mathbf{q-k},{m'}}\nonumber\\
+\bar{c}_1^{(1)}\mathbf{f}_{mm^{'}}\cdot\mathbf{f}_{\mu\mu '}\hat{a}_{\mathbf{p},m}^{\dagger}\hat{a}_{\mathbf{q},\mu}^{\dagger}
\hat{a}_{\mathbf{p+k},{m^{'}}}\hat{a}_{\mathbf{q-k},\mu '}\Bigg), 
 \label{spin1}
\eeq
where 
$\bar{c}_0^{(1)}=(\bar{g}_0+2\bar{g}_2)/3,$ 
$\bar{c}_1^{(1)}=(\bar{g}_2-\bar{g}_0)/3,$
and
$\mathbf{f}_{mm'}=(f^x_{mm'},f^y_{mm'},f^z_{mm'})$ represents a set
of the spin-1 matrices given by
\begin{eqnarray}
{ f}^x =\frac{1}{\sqrt{2}}
\begin{pmatrix}
	0 & 1 & 0 \\
	1 & 0 & 1 \\
	0 & 1 & 0 
\end{pmatrix}, \ \ 
{ f}^y =\frac{i}{\sqrt{2}}
\begin{pmatrix}
	0 & -1 & 0 \\
	1 & 0 & -1 \\
	0 & 1 & 0
\end{pmatrix},\ \ 
{ f}^z= \begin{pmatrix}
	1 & 0 & 0 \\
	0 & 0 & 0 \\
	0 & 0 & -1
\end{pmatrix}.
\label{spin1matrices}
\end{eqnarray}

\begin{figure}[h]
 \begin{center}
  \includegraphics[width=0.4\linewidth]{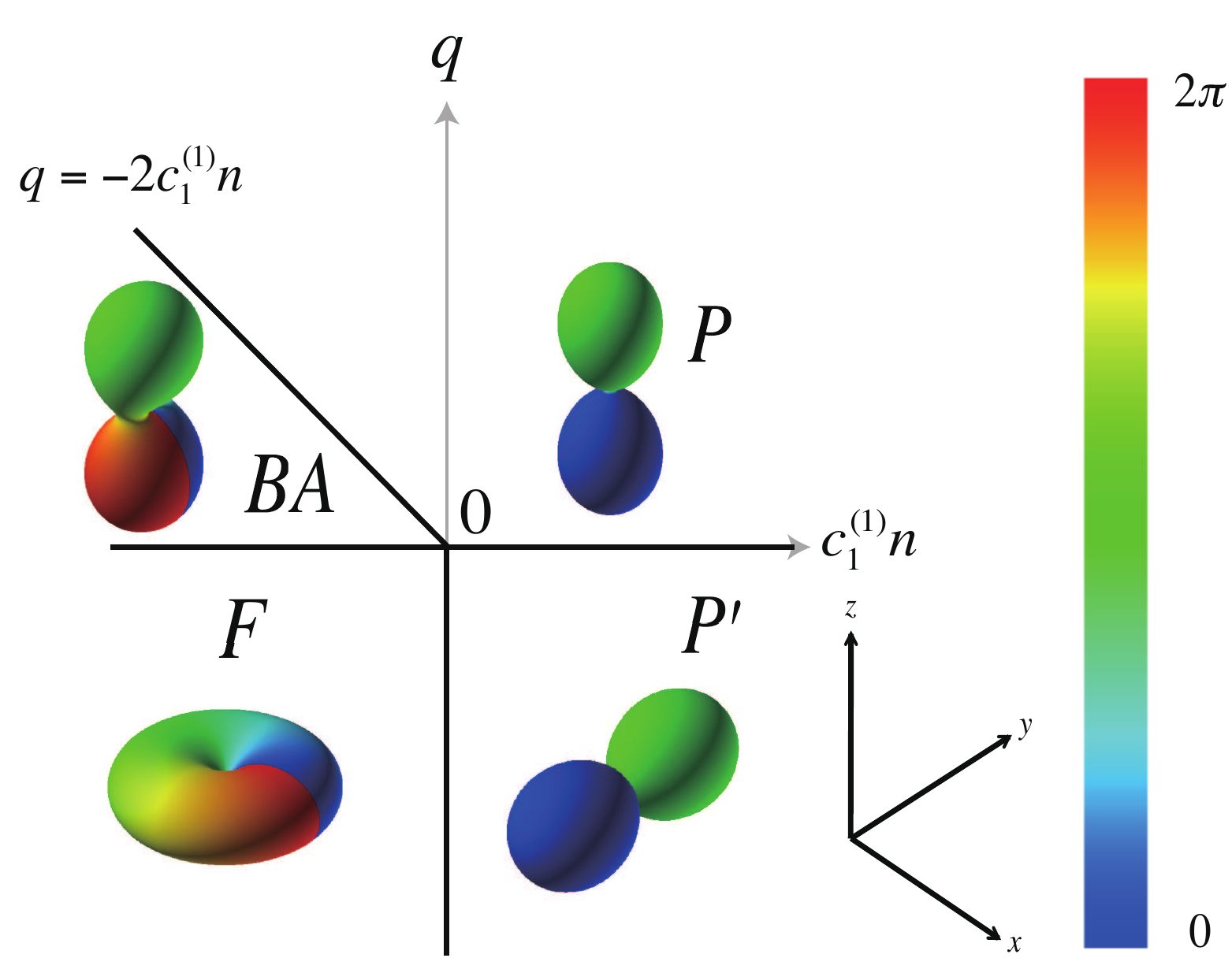}
  \caption{(Color online)
Phase diagram of spin-1 BECs, where F, P (P'), and BA stand for
the ferromagnetic, polar, and broken-axisymmetry phases, respectively.
The spinor order parameters $(\zeta_1,\zeta_0,\zeta_{-1})$ 
of P and P' are given by $(0,1,0)$ and
$(1/\sqrt{2},0,1/\sqrt{2})$, respectively,
which are transformed each other by a $\pi/2$ rotation about the
$x$ axis. The degeneracy of these states are lifted by the
quadratic Zeeman effect.
The thick lines represent the phase boundaries.
The shape of the wave function in each phase is
calculated in terms of polar angle $\theta$ and azimuthal angle $\phi$
as
$|\Psi(\theta,\phi)|^2\equiv|\sum_m\zeta_m\text{Y}_1^m(\theta,\phi)|^2$,
where $Y^m_1(\theta,\phi)$ $(m=1,0,-1)$ are the spherical harmonics of
rank 1 and
color represents the phase of $\Psi(\theta,\phi)$
(see the right color gauge).
The axes of the spin coordinates are shown at the bottom right
of the phase diagram. }
  \label{spin-1-phase}
 \end{center}
\end{figure} 
The possible phases and phase boundaries
of spin-1 BEC
are shown in Fig. \ref{spin-1-phase} 
and described as follows
\cite{stenger,murata}:
\begin{gather}
\text{ferromagnetic phase:} \ \ \ q< 0,  \ \ 
c_1^{(1)}<0,
\ \   \text{and} \ \  
\mathbf{\zeta}^{F}=(1,0,0),\label{ferro}\\
 \text{polar phase:} \ \ \ q>0,  \ \ q>-2c_1^{(1)}n,
\ \   \text{and}   \ \ 
\mathbf{\zeta}^{P}=(0,1,0),   \label{polar}\\ \ \ \ \ \ \ \ \ \ \ \ \ \
 \ 
 \ \ \ \ \ \ q<0,  \ \  
c_1^{(1)}>0, \  \ \ \text{and} \ \  
\mathbf{\zeta}^{P'}=\frac{1}{\sqrt{2}}(1,0,1), \label{polar'} \\
\text{broken-axisymmetry phase:} \ \ \ q>0, \ \  c_1^{(1)}<0, \ \ 
q<-2c_1^{(1)}n, \ \ \text{and} \ \ 
\mathbf{\zeta}^{BA}=\left(\frac{\sin\theta}{\sqrt{2}},\cos\theta,
\frac{\sin\theta}{\sqrt{2}}\right), \label{ba}
\end{gather}
where $c_0^{(1)}=(g_0+2g_2)/3$,
$c_1^{(1)}=(g_2-g_0)/3$, $\sin\theta=\sqrt{1/2+q/(4nc_1^{(1)})}$. 
In Fig. \ref{spin-1-phase}, the shape and color of the wave function
in each phase represent the symmetry of the order
parameter.
For example, the spinor of the polar phase, $\zeta^P$ has a rotational
symmetry about the $z$ axis, since the shape and color are 
symmetric about the same axis.
The ferromagnetic phase apparently does not have an axisymmetry because
the color changes around the $z$ axis.
However, the $U(1)$ gauge transformation can make up for
the variation of the color so that due to the spin-gauge symmetry, 
the ferromagnetic phase
maintains the $U(1)$ symmetry.
The broken-axisymmetry phase does not possess any continuous symmetry
because any $U(1)\times SO(2)$ transformations cannot 
compensate for the variation of the color shown. 

The ferromagnetic phase has a longitudinal magnetization, 
whereas the broken-axisymmetry phase has
a transverse one that depends on $q$. 
Moreover, the broken-axisymmetry phase has a finite
spin-singlet pair amplitude that also depends on $q$.
In contrast, both spinors of P and P' 
have no magnetization but a finite spin-singlet
pair amplitude that is independent of $q$.
These properties in each phase are summarized in Appendix B.
Since  $c_1^{(1)}>0$, 
the spin-1 $^{23}$Na condensate is in the polar
phase. On the other hand, since $c_1^{(1)}<0$, 
the spin-1 $^{87}$Rb condensates can be in any of
the ferromagnetic,
broken-axisymmetry, and polar phases, depending on the
sign and magnitude of $q$.
 
In the Bogoliubov theory, we replace operators $\hat{a}_{\mathbf{0},m}$ 
by c-numbers $\sqrt{N_0}\zeta_m$
and keep $\ahd_{\mk\ne 0,m}$
and $\ah_{\mk\ne 0,m}$ up to the second order in the Hamiltonian, where
$N_0$ is the number of condensate bosons, which, together with 
$\displaystyle\hat{n}_{\mk,m}\equiv\ahd_{\mk,m}\ah_{\mk,m}$,
satisfies 
\beq
N_0+{\sk}^{'}\sum_m \hat{n}_{\mathbf{k},m}=N,
\eeq
where $N$ is the total number of bosons.
The Bogoliubov Hamiltonian of a spin-1 BEC is thus given by \cite{ueda2,murata}
\beq
\hat{H}_{\text{eff}}&=&\frac{Vn^2}{2}\left(\bar{c}_0^{(1)}
+\bar{c}_1^{(1)}\langle\mathbf{f}\rangle^2\right)+qN\langle(f^z)^2\rangle
+{\sk}^{'}\Bigg[\left(\epsilon_{\mathbf{k}}
-nc_1^{(1)}\langle\mathbf{f}\rangle^2
+qm^2-q\langle(f^z)^2\rangle\right)\hat{a}^{\dagger}_{\mathbf{k},m}\hat{a}_{\mathbf{k},m}\nonumber\\
&&+nc_1^{(1)}\langle\mathbf{f}\rangle\cdot\mathbf{f}_{mm'}
\hat{a}^{\dagger}_{\mathbf{k},m}\hat{a}_{\mathbf{k},m'}
+\frac{nc_0^{(1)}}{2}
\left(2\hat{D}_{\mathbf{k}}^{\dagger}\hat{D}_{\mathbf{k}}+\hat{D}_{\mathbf{k}}\hat{D}_{\mathbf{-k}}+\hat{D}_{\mathbf{k}}^{\dagger}\hat{D}_{\mathbf{-k}}^{\dagger}\right)\nonumber\\
&&+\frac{nc_1^{(1)}}{2}\left(2\mathbf{\hat{F}}_{\mathbf{k}}^{\dagger}\cdot\mathbf{\hat{F}}_{\mathbf{k}}+\mathbf{\hat{F}}_{\mathbf{k}}\cdot\mathbf{\hat{F}}_{\mathbf{-k}}+
\mathbf{\hat{F}}_{\mathbf{k}}^{\dagger}\cdot\mathbf{\hat{F}}_{\mathbf{-k}}^{\dagger}\right)\Bigg], \label{bogoliubov1}
\eeq
where
\begin{gather}
\langle\mathbf{f}\rangle\equiv\sum_{m,m'}\mathbf{f}_{mm'}\zeta_m^{*}
\zeta_{m'} ,\\
\hat{D}_{\mathbf{k}}\equiv \sum_m \zeta^{*}_{m}\hat{a}_{\mathbf{k},m}, \label{density}\\
\mathbf{\hat{F}}_{k}\equiv \sum_{m,m'}\mathbf{f}_{mm'}
\zeta^{*}_{m}\hat{a}_{\mathbf{k},m'}. \label{spin}
\end{gather}
Here, $\hat{D}_{\mk}$ and $\hat{\mathbf{F}}_{\mk}$ denote the density and
spin fluctuation operators of the condensate, respectively.
In Eq. (\ref{bogoliubov1}), 
we substitute $c_i^{(1)}$ for $\bar{c}_i^{(1)}$ $(i=0,1)$
in  the sum over the momentum because
the Bogoliubov approximation is correct up to the second order in
the coupling constants.

\subsection{Ferromagnetic phase}
For the ferromagnetic phase
(\ref{ferro}),
Eq. (\ref{bogoliubov1}) reduces to
\beq
\hat{H}_{\text{eff}}^{F}&=&\frac{Vn^2(\bar{c}_0^{(1)}+\bar{c}_1^{(1)})}{2}+qN
+{\sk}^{'}\Bigg[
(\epsilon_{\mathbf{k}}-q)\hat{a}^{\dagger}_{\mathbf{k},0}
\hat{a}_{\mathbf{k},0}+
\left(\ek -2nc_1^{(1)}\right)
\hat{a}^{\dagger}_{\mathbf{k},-1}\hat{a}_{\mathbf{k},-1}
\nonumber\\
&&+
\left(\epsilon_{\mathbf{k}}+n(c_0^{(1)}+c_1^{(1)})\right)
\hat{a}^{\dagger}_{\mathbf{k},1}\hat{a}_{\mathbf{k},1}+
\frac{n(c_0^{(1)}+c_1^{(1)})}{2}(\hat{a}^{\dagger}_{\mathbf{k},1}\hat{a}_{\mathbf{-k},1}^{\dagger}+\hat{a}_{\mathbf{k},1}\hat{a}_{\mathbf{-k},1})
\Bigg].
\label{ferroh}
\eeq
Here, the $m=0$ and $-1$ modes are already diagonal, and 
the $m=1$ mode can be diagonalized by the standard Bogoliubov
transformation \cite{ho,ohmi}:
\beq
\bh_{\mk ,1}=\sqrt{\frac{\ek +n(c_0^{(1)}
+c_1^{(1)})+E_{\mk ,1}}{2E_{\mk ,1}}}\ah_{\mk ,1}
+\sqrt{\frac{\ek +n(c_0^{(1)}+c_1^{(1)})
-E_{\mk ,1}}{2E_{\mk ,1}}}\ahd_{\mmk ,1},
\eeq 
where $E_{\mk,1}$ is the Bogoliubov spectrum given by
\beq
E_{\mk ,1}=\sqrt{\ek \left[\ek +2n
(c_0^{(1)}+c_1^{(1)})\right]}.\label{eq:spin-1-f-b}
\eeq
The diagonalized Hamiltonian is 
\beq
\hat{H}_{\text{eff}}^{F}=E_{0}^{F}
+{\sk}^{'}\Bigg[ E_{\mk ,1}\bhd_{\mk ,1}\bh_{\mk ,1} 
+(\ek-q) \ahd_{\mk ,0}\ah_{\mk ,0}
+(\ek -2nc_1^{(1)})\ahd_{\mk ,-1}\ah_{\mk ,-1}\Bigg],
\label{bogohf}
\eeq
where
\beq
E_{0}^{F}=\frac{Vn^2(\bar{c}_0^{(1)}+\bar{c}_1^{(1)})}{2}+qN
-\frac{1}{2}{\sk}^{'}\left[\ek +n(c_0^{(1)}+c_1^{(1)})-E_{\mk,1}\right]
\eeq
is the GSE in the ferromagnetic phase. 
As can be seen from Eqs. (\ref{eq:spin-1-f-b}) and (\ref{bogohf}),
the $m=1$ mode is massless, and in the absence of the external magnetic
field, the $m=0$ mode is also massless. 
For the excitation energies with $m=0$ and $m=-1$ to be positive,
we must have $q<0$ and $c_1^{(1)}<0$.
For the $m=1$ Bogoliubov mode to be stable, $c_0^{(1)}+c_1^{(1)}>0$ is
required.
This condition ensures the mechanical stability of the mean-field ground
state; otherwise the compressibility would not be positive definite and
the system would become unstable against collapse. 
These requirements are consistent with the stability criteria of
the mean-field theory.
Conversely, if we prepare a spin-polarized state with 
$q>0$, $c_1^{(1)}>0$, or $c_0^{(1)}+c_1^{(1)}<0$, 
the state would undergo
the Landau instability for the $m=0$ and $-1$ modes with
quadratic spectra and the
dynamical instability for the $m=1$ mode with a linear spectrum.
The GSE per a volume $V$ in the ferromagnetic phase is calculated to give
\beq
\frac{E_0^{F}}{V}=qn+\frac{2\pi\hbar^2
n^2}{M}a_2\left(1+\frac{128}{15}\sqrt{\frac{na_2^3}{\pi}}\right),
\label{eq:spin-1-f-gse}
\eeq
where the last term  is the LHY
correction. The pressure is obtained from
Eq. (\ref{eq:spin-1-f-gse}) as
\beq
P=-\frac{\partial E_{0}^{F}}{\partial V}
       =\frac{2\pi\hbar^2 n^2a_2}{M}
\left(1+\frac{64}{5}\sqrt{\frac{na_2^3}{\pi}}\right) ,\label{eq:spin-1-f-p}
\eeq 
and the sound velocity is given by  
\beq
c=\sqrt{\frac{1}{M}\frac{\partial P}{\partial n}}
=\sqrt{\frac{4\pi\hbar^2 n a_2}{M^2}}
\left(1+8\sqrt{\frac{na_2^3}{\pi}}\right).
\label{eq:spin-1-f-sv}
\eeq
The quantum depletion is calculated to be
\beq
\frac{N-N_0}{N}=\frac{1}{N}{\sk}^{'}\sum_m n_{\mk,m} 
=\frac{8}{3}\sqrt{\frac{na_2^3}{\pi}}.
\label{eq:spin-1-f-dep}
\eeq
We note that the LHY corrections in the ferromagnetic phase
are not affected by the quadratic Zeeman effect.

\subsection{Polar phase}
The polar phase has two spinor configurations (\ref{polar})
and (\ref{polar'}), which are degenerate and connected each other by 
a $U(1)\times SO(3)$ transformation for $q=0$.
However, for nonzero $q$, 
the degeneracy is lifted, and they should be 
considered as different phases.
This is because the phase of $\eqref{polar}$ has a remaining $SO(2)$
symmetry, whereas the phase of (\ref{polar'})
is not invariant under any continuous transformation.
As a consequence, it is expected that
the number of the NG modes is different in each phase, and that
the low-energy behavior is also different.

\subsubsection{Case of $q>0$}
In the case of (\ref{polar}),
we introduce the following density fluctuation operator
$\ah_{\mk,d}$ 
and spin fluctuation operators $\ah_{\mk,f_x}$ and $\ah_{\mk,f_y}$
around the $x$ and $y$ axes:
\begin{gather}
\ah_{\mk ,d} =  \ah_{\mk ,0}, \label{densityo}\\
\ah_{\mk ,f_x}=\frac{1}{\sqrt{2}}(\ah_{\mk ,1}+\ah_{\mk ,-1}), \label{spinxo}\\
\ah_{\mk ,f_y}  =\frac{i}{\sqrt{2}}(\ah_{\mk ,1}-\ah_{\mk ,-1}).\label{spinyo}
\end{gather}
In terms of them, the total Hamiltonian is expressed as 
\beq
\hat{H}_{\text{eff}}^{P}&=&\frac{Vn^2\bar{c}_0^{(1)}}{2}
+{\sk}^{'}\Bigg\{\left(\epsilon_{\mathbf{k}}
+nc_0^{(1)}\right)
\ahd_{\mathbf{k},d}\ah_{\mathbf{k},d}
+\frac{nc_0^{(1)}}{2}
(\ahd_{\mathbf{k},d}\ahd_{\mathbf{-k},d}
+\ah_{\mathbf{k},d}\ah_{\mathbf{-k},d})\nonumber\\
&&+\sum_{j=f_x,f_y}\Big[\left(\epsilon_{\mathbf{k}}+q
+nc_1^{(1)}\right)
\ahd_{\mk ,j}\ah_{\mk ,j}
+\frac{nc_1^{(1)}}{2} (\ahd_{\mk ,j}\ahd_{\mmk ,j}+\ah_{\mk
,j}\ah_{\mmk ,j})\Big]\Bigg\}.
\nonumber\\ \label{eq:eff-ham-p}
\eeq
This effective Hamiltonian can be diagonalized by means of
the following Bogoliubov transformations
\begin{gather}
\bh_{\mk,d}=\sqrt{\frac{\ek+nc_0^{(1)}
+E_{\mk,d}}{2E_{\mk,d}}}
\ah_{\mk,d}
+\sqrt{\frac{\ek+nc_0^{(1)}-E_{\mk,d}}
{2E_{\mk,d}}}\ahd_{\mmk,d},\\
\bh_{\mk,j}=\sqrt{\frac{\ek+q+nc_1^{(1)}
+E_{\mk,f_{t}}}{2E_{\mk,f_t}}}\ah_{\mk,j}
+\sqrt{\frac{\ek+q+nc_1^{(1)}-E_{\mk,f_{t}}}
{2E_{\mk,f_t}}}
\ahd_{\mmk,j},
\end{gather}
with the result
\beq
\hat{H}_{\text{eff}}^{P}=E_{0}^{P}
+{\sk}^{'}\big[  E_{\mk ,d}\bhd_{\mk ,d}\bh_{\mk ,d}
+E_{\mk ,f_{t}}(\bhd_{\mk ,f_x}\bh_{\mk ,f_x}+\bhd_{\mk ,f_y}\bh_{\mk
,f_y})
\big],
\eeq
where the Bogoliubov spectra are given by
\begin{gather}
E_{\mk,d}=\sqrt{\ek (\ek +2nc_0^{(1)})},
\label{eq:spin-1-p-b1}
\\
E_{\mk,f_{t}}=\sqrt{(\ek+q) (\ek+q +2nc_1^{(1)}
)},
\label{eq:spin-1-p-b2}
\end{gather}
and the GSE by
\beq 
E_{0}^{P}=\frac{Vn^2\bar{c}_0^{(1)}}{2}-\frac{1}{2}{\sk}^{'} 
\Bigg[\left( \ek +nc_0^{(1)}- E_{\mk ,d}\right) 
+2\left(  \ek+q +nc_1^{(1)} 
-E_{\mk ,f_t}\right) \Bigg].
\eeq
The density mode ($d$) is massless because 
the $U(1)$ gauge symmetry is spontaneously broken in the
mean-field ground state,
while the transverse magnetization
($f_x$, $f_y$) modes are massive
for nonzero $q$, since the rotational degeneracies about the
$x$ and $y$ axes are lifted by the external magnetic field.

In the limit of $q\to 0$, 
the Bogoliubov spectra \eqref{eq:spin-1-p-b1} and \eqref{eq:spin-1-p-b2} 
reduce to those obtained in Refs. \cite{ho,ohmi}, and 
the transverse magnetization modes become massless since the symmetry of the
Hamiltonian becomes $U(1)\times SO(3)$ and the rotational symmetry
around the $x$ and $y$ axes are spontaneously broken.
The conditions $q>0$, $c_0^{(1)}>0$, and 
$q>-2c_1^{(1)}n$
are required for the Bogoliubov spectra to be positive semidefinite, 
and they are consistent
with the stability conditions of the mean-field ground state.
 
\begin{figure}[h]
\begin{center}
  \includegraphics[width=0.40\linewidth]{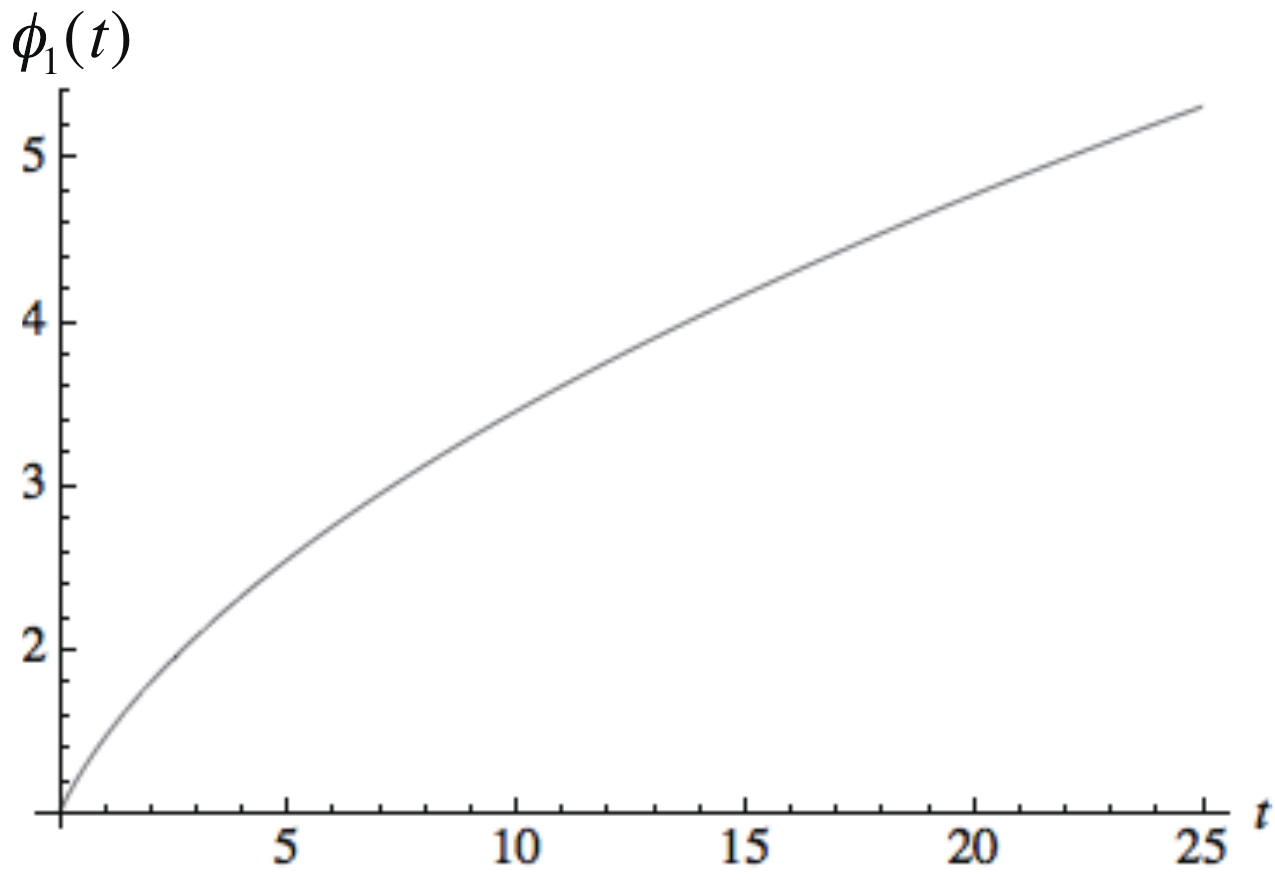}
 \caption{Plot of $\phi_1(t)$ in Eq. (\ref{eq:phi1}) which 
describes the contribution to
the ground-state energy (\ref{eq:spin-1-p-gse})
from spin fluctuations due to 
the quadratic Zeeman effect, where $t=q/(n|c_1^{(1)}|)-1$.}
  \label{fig:phi1}
\end{center}
\end{figure}

Using the renormalization procedure of the coupling constant shown 
in Eq. (\ref{renormalize}), 
we find that the GSE per volume $V$ is given by
(see Appendix A for derivation)
\beq
\frac{E_0^P}{V}
=
\frac{n^2c_0^{(1)}}{2}\Bigg[1+\frac{16\sqrt{M^3}}{15\pi^2\hbar^3}
\sqrt{n(c_0^{(1)})^3}\Bigg]
+\frac{16\sqrt{M^3}n^2|c_1^{(1)}|}{15\pi^2\hbar^3}
\sqrt{n(|c_1^{(1)}|)^3}\phi_1(t_1+\text{sgn}(c_1^{(1)}))
\label{eq:spin-1-p-gse}
\eeq
where $t_1=q/(n|c_1^{(1)}|)-1$, and
\beq
\phi_1(t)\equiv -\frac{15}{8\sqrt{2}}\int_{0}^{\infty}dx\ 
x^2\left( x^2+t+1-\sqrt{(x^2+t)(x^2+t+2)}
-\frac{1}{2x^2}\right)
\label{eq:phi1}
\eeq
with $\phi_1(0)=1$.
The GSE of the polar phase 
consists of two parts corresponding to density and spin fluctuation, and
the latter part depends on $\phi_1(t)$.
The behavior of $\phi_1(t)$  for positive $t$ 
 is plotted in FIG. \ref{fig:phi1}, which shows that
the GSE increases with increasing the
quadratic Zeeman effect.

\begin{figure}[h]
\begin{center}
  \includegraphics[width=0.40\linewidth]{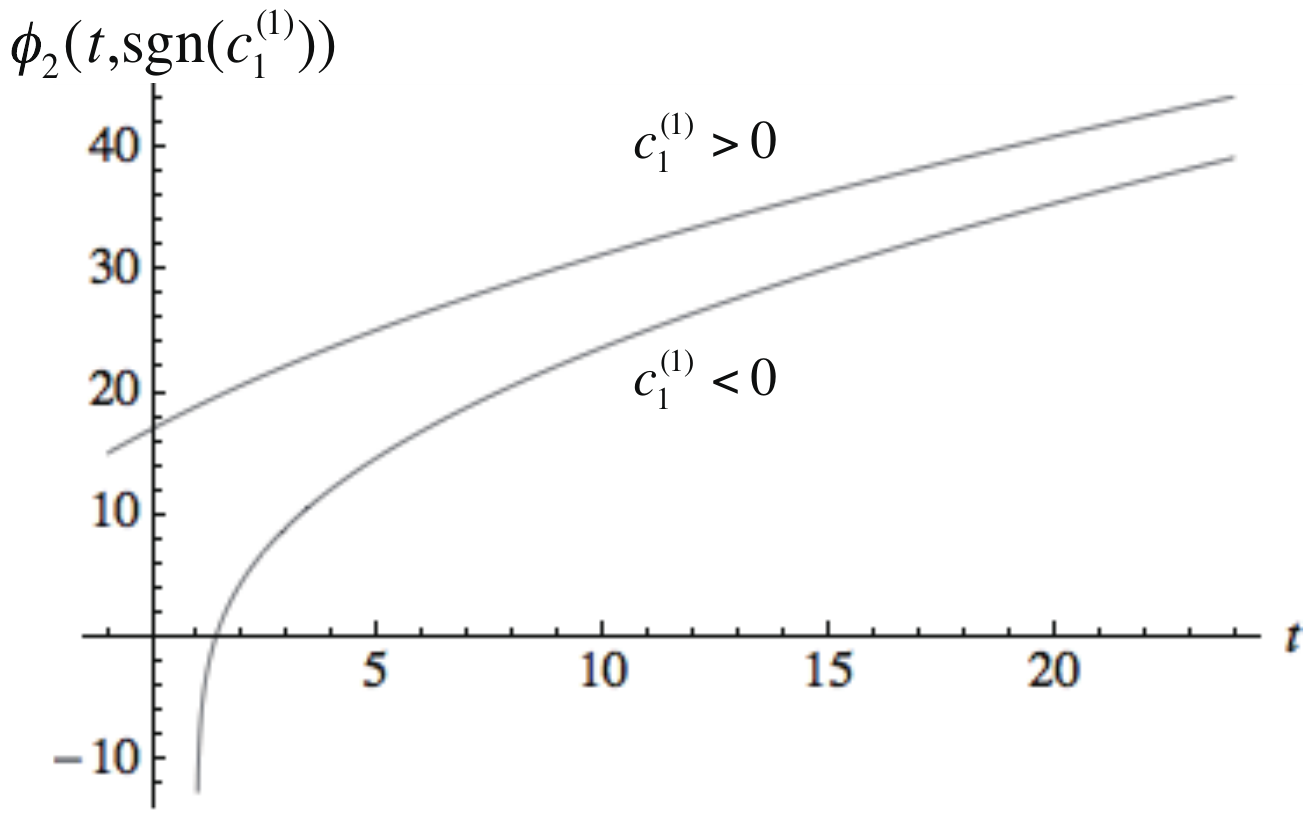}
 \caption{Plot of $\phi_2(t,\text{sgn}(c_1^{(1)}))$
 which gives the correction to the sound
velocity (\ref{eq:spin-1-p-sv}) due to the quadratic Zeeman effect.}
  \label{phi2}
\end{center}
\end{figure}

\begin{figure}
\begin{center}
  \includegraphics[width=0.40\linewidth]{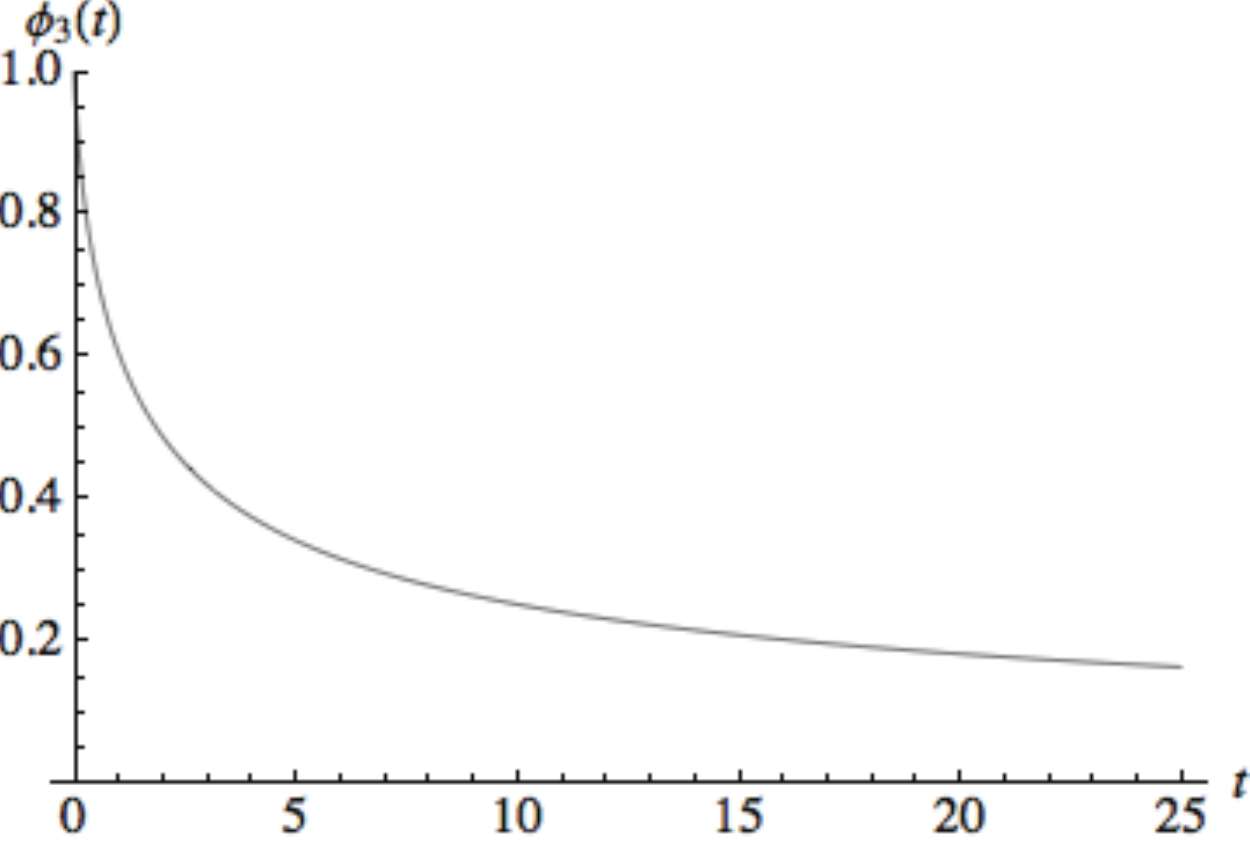}
 \caption{Plot of $\phi_3(t)$ which describes the quantum depletion
through Eq. (\ref{eq:spin-1-p-dep}).}
  \label{phi3}
\end{center}
\end{figure}

The pressure and sound
velocity are calculated as 
\beq
P=
\frac{n^2c_0^{(1)}}{2}\left(1+\frac{8\sqrt{M^3}}{5\pi^2\hbar^3}
\sqrt{n(c_0^{(1)})^3}\right)
+\frac{8\sqrt{M^3}n^2|c_1^{(1)}|}{15\pi^2\hbar^3}
\sqrt{n(|c_1^{(1)}|)^3}
\Big[3\phi_1(t_1+\text{sgn}(c_1^{(1)}))\nonumber\\
-2(t_1+1)\phi'_1(t_1+\text{sgn}(c_1^{(1)}))\Big], 
\label{eq:spin-1-p-p}
\eeq 
and  
\beq
c=
\sqrt{\frac{nc_0^{(1)}}{M}}\Bigg[1+\frac{\sqrt{M^3}}{\pi^2\hbar^3}
\sqrt{n(c_0^{(1)})^3}
+\frac{2\sqrt{M^3}}{15\pi^2\hbar^3}\left(\frac{|c_1^{(1)}|}{c_0^{(1)}}\right)
\sqrt{n(|c_1^{(1)}|)^3}\phi_2(t_1,\text{sgn}(c_1^{(1)}))\Bigg], 
\label{eq:spin-1-p-sv}
\eeq
where
\beq
\phi_2(t,\text{sgn}(c_1^{(1)}))
\equiv 15\phi_1(t+\text{sgn}(c_1^{(1)}))-12(t+1)
\phi'_1(t+\text{sgn}(c_1^{(1)}))
+4(t+1)^2\phi''_1(t+\text{sgn}(c_1^{(1)})),
\eeq
which is plotted in
Fig. \ref{phi2}.
We note that for the case of $c_1^{(1)}<0$, the quantum correction to 
the sound velocity due to spin fluctuations vanishes at $t\simeq 1.4$.
On the other hand, the quantum depletion is calculated to be
\beq
\frac{N-N_0}{N}&=&\frac{1}{N}{\sk}^{'}\langle\ahd_{\mk,d}\ah_{\mk,d}
+\ahd_{\mk,f_x}\ah_{\mk,f_x}+\ahd_{\mk,f_y}\ah_{\mk,f_y}\rangle
\nonumber\\
&=&
\frac{\sqrt{M^3}}{3\pi^2\hbar^3}\left(
\sqrt{n(c_0^{(1)})^3}+2\sqrt{n(|c_1^{(1)}|)^3}
\phi_3(t_1+\text{sgn}(c_1^{(1)}))\right),
\label{eq:spin-1-p-dep}
\eeq
where 
\beq
\phi_3(t)\equiv\frac{3}{\sqrt{2}}\int_0^{\infty}dx\ x^2
\left(\frac{x^2+t+1}{\sqrt{(x^2+t)(x^2+t+2)}}-1\right).
\eeq
The plot of $\phi_3(t)$ is shown in
Fig. \ref{phi3}, which shows that 
the quantum depletion is suppressed by the quadratic Zeeman effect.
In contrast, the quantum corrections to the GSE and sound velocity
are enhanced by it.
Since the variations of $\phi_1(t)$, $\phi_2(t,\pm 1)$,
and $\phi_3(t)$ in the typical region of $t$ are of the order of
$1\sim 10$, the magnetic-field dependence of the
LHY corrections is significant.
However, because $c_0^{(1)}\gg c_1^{(1)}$ in the alkali species,
the magnitudes 
of the LHY corrections are small and the quantum corrections 
arise mainly from density fluctuations.
\subsubsection{Case of $q<0$}
To analyze the case of (\ref{polar'}),
we introduce the following fluctuation operators:
\begin{gather}
\ah_{\mk ,d}=\frac{1}{\sqrt{2}}(\ah_{\mk ,1}+\ah_{\mk ,-1}), \\
\ah_{\mk ,f_x} =  \ah_{\mk ,0}, \\
\ah_{\mk ,f_z}  =\frac{1}{\sqrt{2}}(\ah_{\mk ,1}-\ah_{\mk ,-1}),
\end{gather}
where $\ah_{\mk,d}$, $\ah_{\mk,f_x}$, and $\ah_{\mk,f_{z}}$
describe the density fluctuation $(d)$
and spin fluctuations $(f_x,f_z)$ around the $x$ and $z$ axes, respectively.
The Hamiltonian is diagonalized by the following
Bogoliubov transformations
\begin{gather}
\bh_{\mk,d}=\sqrt{\frac{\ek+nc_0^{(1)}
+E_{\mk,d}}{2E_{\mk,d}}}
\ah_{\mk,d}
+\sqrt{\frac{\ek+nc_0^{(1)}-E_{\mk,d}}
{2E_{\mk,d}}}\ahd_{\mmk,d},\\
\bh_{\mk,f_x}=\sqrt{\frac{\ek-q+nc_1^{(1)}
+E_{\mk,f_{x}}}{2E_{\mk,f_x}}}\ah_{\mk,f_x}
+\sqrt{\frac{\ek-q+nc_1^{(1)}-E_{\mk,f_{x}}}
{2E_{\mk,f_x}}}
\ahd_{\mmk,f_x}, \\
\bh_{\mk,f_z}=\sqrt{\frac{\ek+nc_1^{(1)}
+E_{\mk,f_{z}}}{2E_{\mk,f_z}}}\ah_{\mk,f_z}
+\sqrt{\frac{\ek+nc_1^{(1)}-E_{\mk,f_{z}}}
{2E_{\mk,f_z}}}
\ahd_{\mmk,f_z},
\end{gather}
with the result
\beq
\hat{H}_{\text{eff}}^{P'}=E_{0}^{P'}
+{\sk}^{'} \big(E_{\mk ,d}\bhd_{\mk ,d}\bh_{\mk ,d}
+ E_{\mk ,f_{x}}\bhd_{\mk ,f_x}\bh_{\mk ,f_x}
+ E_{\mk, f_z}\bhd_{\mk ,f_z}\bh_{\mk ,f_z}
\big),
\eeq
where 
\beq
E_{0}^{P'}=qN+\frac{Vn^2\bar{c}_0^{(1)}}{2}-\frac{1}{2}{\sk}^{'} 
\Bigg[\left( \ek +nc_0^{(1)}- E_{\mk ,d}\right) 
+\left(  \ek-q +nc_1^{(1)} 
-E_{\mk ,f_x}\right) \nonumber\\
+\left( \ek +nc_1^{(1)} 
-E_{\mk, f_x}\right)
\Bigg]
\eeq
is the GSE, and
the Bogoliubov spectra are given by
\begin{gather}
E_{\mk,d}=\sqrt{\ek \left(\ek +2nc_0^{(1)}\right)},
\label{eq:spin-1-p'-b1}\\
E_{\mk,f_{x}}=\sqrt{(\ek-q) \left(\ek-q
 +2nc_1^{(1)}\right)}, \label{eq:spin-1-p'-b2}\\
E_{\mk,f_{z}}=\sqrt{\ek \left(\ek +2nc_1^{(1)}
\right)}.
\label{eq:spin-1-p'-b3}
\end{gather}
In contrast to the case of $q>0$, one of the spin fluctuation modes,
(\ref{eq:spin-1-p'-b3}), 
becomes massless.
This is because all the continuous symmetries of the Hamiltonian are
spontaneously broken for $q<0$.
On the other hand, the transverse spin mode is massive because
the rotational symmetry about the $x$ axis is  broken by
an external magnetic field.
For the Bogoliubov spectra to be real,
the conditions $q<0$, $c_0^{(1)}>0$, and $c_1^{(1)}>0$ must be
satisfied;
otherwise,
the state in Eq. (\ref{polar'})
would be dynamically unstable.
The GSE $E_{0}^{P'}$, 
pressure $P$, sound velocity $c$, and quantum depletion $(N-N_0)/N$ 
are given by
\beq
\frac{E_0^{P'}}{V}=qn+\frac{n^2c_0^{(1)}}{2}
\left(1+\frac{16\sqrt{M^3}}{15\pi^2\hbar^3}
\sqrt{n(c_0^{(1)})^3}\right)
+\frac{8\sqrt{M^3}n^2c_1^{(1)}}{15\pi^2\hbar^3}
\sqrt{n(c_1^{(1)})^3}
(1+\phi_1(t_2)),
\label{eq:spin-1-p'-gse}
\eeq
\beq
P=\frac{n^2c_0^{(1)}}{2}\left(1+\frac{8\sqrt{M^3}}{5\pi^2\hbar^3}
\sqrt{n(c_0^{(1)})^3}\right)
+\frac{4\sqrt{M^3}n^2c_1^{(1)}}{15\pi^2\hbar^3}
\sqrt{n(c_1^{(1)})^3}
(3+3\phi_1(t_2)-2t_2\phi'_1(t_2)),
\label{eq:spin-1-p'-p}
\eeq
\beq
c=\sqrt{\frac{nc_0^{(1)}}{M}}\Bigg[1+\frac{\sqrt{M^3}}{\pi^2\hbar^3}
\sqrt{n(c_0^{(1)})^3}
+\frac{\sqrt{M^3}}{15\pi^2\hbar^3}\left(\frac{c_1^{(1)}}{c_0^{(1)}}\right)
\sqrt{n(c_1^{(1)})^3}(15+\phi_2(t_2-1,+1))\Bigg],
\label{eq:spin-1-p'-sv}
\eeq
\beq
\frac{N-N_0}{N}=\frac{\sqrt{M^3}}{3\pi^2\hbar^3}\left(
\sqrt{n(c_0^{(1)})^3}+\sqrt{n(c_1^{(1)})^3}(1+\phi_3(t_2))\right),
\label{eq:spin-1-p'-dep}
\eeq
where $t_2=-q/(nc_1^{(1)})$. Reflecting the difference in the
Bogoliubov spectra, the LHY corrections for Eq. (\ref{polar'}) are
also different from those for Eq. (\ref{polar}).

\subsection{Broken-axisymmetry phase}
\begin{figure}[h]
 \begin{center}
\subfigure{
  \includegraphics[width=0.4\linewidth]{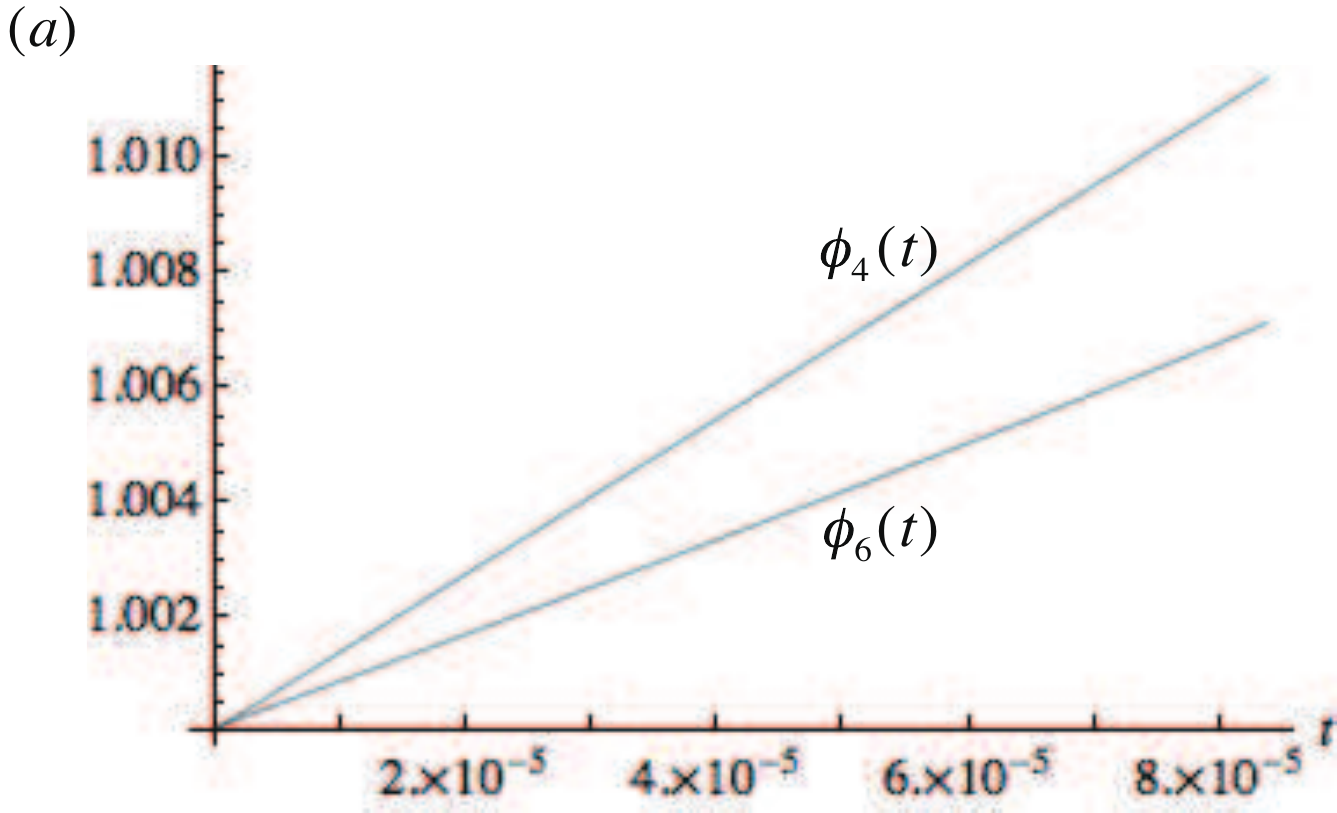}
}
\subfigure{
 \includegraphics[width=0.4\linewidth]{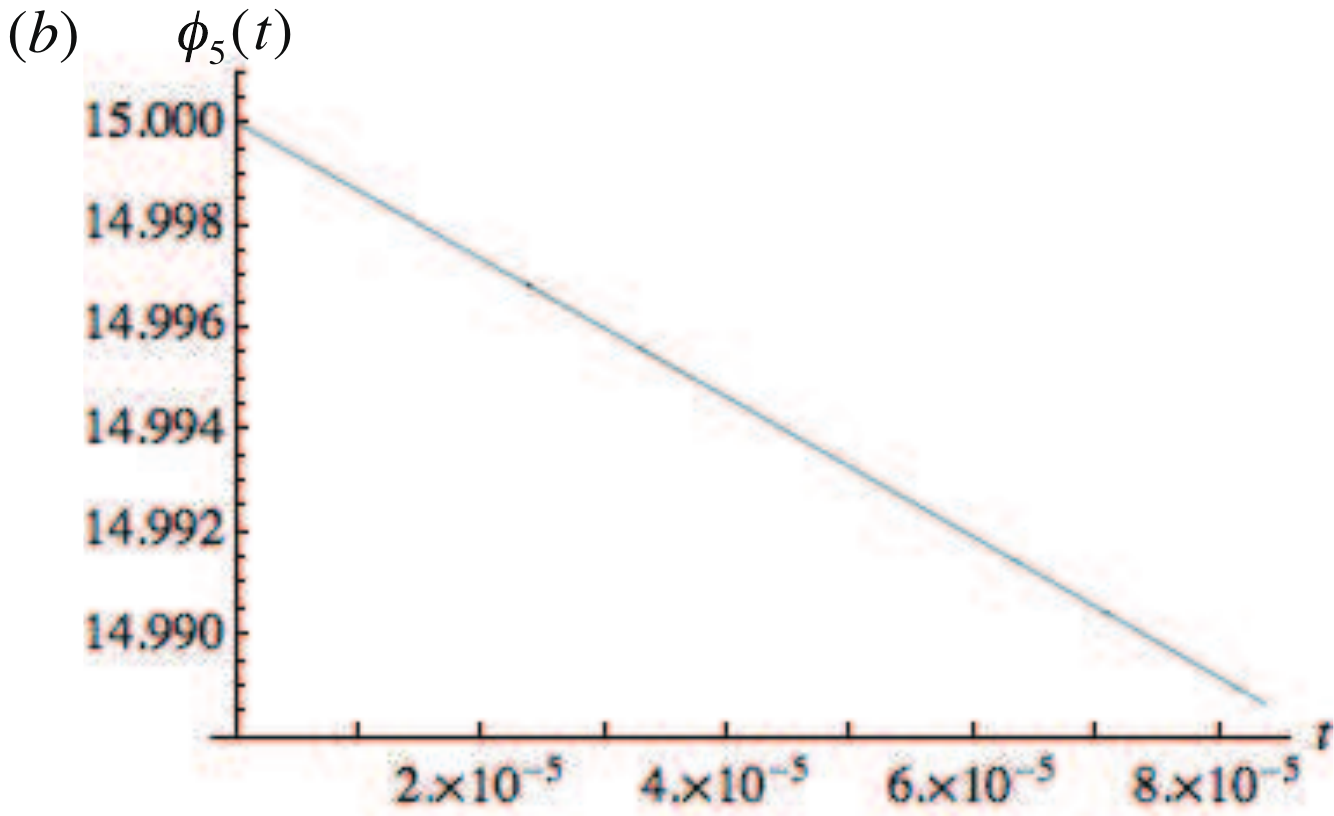}
}
  \caption{Plots of (a) $\phi_4(t)$ and $\phi_6(t)$, and
(b) $\phi_5(t)$.
Over the ranges of $t$ shown,
they are well approximated as
$\phi_4(t)\simeq 1+135t$ and $\phi_6(t)\simeq 1+80t+50000t^2$.
($\phi_5(t)$ is given by (\ref{eq:phi5}))}
  \label{phi4}
 \end{center}
\end{figure}

The mean-field ground state of the broken-axisymmetry phase is
parametrized by Eq. (\ref{ba}).
We define the following fluctuation operators:
\beq
\ah_{\mk,d}=\frac{\sin\theta}{\sqrt{2}}\left(
\ah_{\mk,1}+\ah_{\mk,-1}\right)+\cos\theta\ah_{\mk,0},
\eeq
\beq
\ah_{\mk,f_z}=\frac{1}{\sqrt{2}}\left(\ah_{\mk,1}-\ah_{\mk,-1}\right).
\eeq
In addition, we introduce the following operator that is
independent of $\ah_{\mk,d}$ and $\ah_{\mk,f_z}$:
\beq
\ah_{\mk,\theta}=\frac{\cos\theta}{\sqrt{2}}\left(
\ah_{\mk,1}+\ah_{\mk,-1}\right)-\sin\theta\ah_{\mk,0}.
\label{eq:ba-additional}
\eeq
In terms of them, the total Hamiltonian is expressed as
\beq
\hat{H}_{\text{eff}}^{BA}&=&\frac{Nq}{2}+\frac{Vn^2
(\bar{c}_0^{(1)}+\bar{c}_1^{(1)})}{2}+\frac{Nq^2}{8n\bar{c}_1^{(1)}}
+{\sk}^{'}\Bigg[\left(\ek+\frac{q}{2}\right)\ahd_{\mk,f_z}\ah_{\mk,f_z}
+\frac{q}{4}\left(\ah_{\mk,f_z}\ah_{\mmk,f_z}
+\ahd_{\mk,f_z}\ahd_{\mmk,f_z}\right)\nonumber\\
&&+\left(\ek+nc_0^{(1)}+nc_1^{(1)}
-nc_q^{(1)}\right)\ahd_{\mk,d}\ah_{\mk,d}
+\left(\ek-2nc_1^{(1)}+nc_q^{(1)}\right)
\ahd_{\mk,\theta}\ah_{\mk,\theta}
\nonumber\\
&&+\frac{n(c_0^{(1)}+c_1^{(1)}
-c_q^{(1)})}{2}\left(
\ah_{\mk,d}\ah_{\mmk,d}+\ahd_{\mk,d}\ahd_{\mmk,d}\right)
+\frac{nc_q^{(1)}}{2}\left(\ah_{\mk,\theta}\ah_{\mmk,\theta}
+\ahd_{\mk,\theta}\ahd_{\mmk,\theta}\right)\nonumber\\
&&-\frac{q\sin 2\theta}{2}\left(\ahd_{\mk,d}\ah_{\mk,\theta}+
\ahd_{\mk,\theta}\ah_{\mk,d}+\ah_{\mk,d}\ah_{\mmk,\theta}
+\ahd_{\mk,d}\ahd_{\mmk,\theta}
\right)
\Bigg],
\eeq
where 
\beq
c_q^{(1)}\equiv\frac{q^2}{4n^2c_1^{(1)}}.
\eeq
To diagonalize the sub-Hamiltonian
for the spin fluctuation mode around the $z$ axis,
we perform the following transformation:
\beq
\bh_{\mk,f_z}=\sqrt{\frac{\ek+q/2+E_{\mk,f_z}}{2E_{\mk,f_z}}}
\ah_{\mk,f_z}+\sqrt{\frac{\ek+q/2-E_{\mk,f_z}}{2E_{\mk,f_z}}}
\ahd_{\mmk,f_z},
\eeq
where the Bogoliubov spectrum is given by
\beq
E_{\mk,f_z}=\sqrt{\ek\left(\ek+q\right)}.
\label{eq:spin-1-ba-b1}
\eeq
On the other hand, for the density fluctuation mode and the $\theta$
mode
in Eq. \eqref{eq:ba-additional},
we consider the following transformations:
\beq
\hat{\mathbf{B}}_{\mk}=U(k)\hat{\mathbf{A}}_{\mk,d\theta}
+V(k)\hat{\mathbf{A}^{\dagger}}_{\mmk,d\theta},
\label{eq:baspec}
\eeq
where
the bold symbols represent the following set of the operators
\beq
\hat{\mathbf{B}}_{\mk}= ^{t}(\bh_{\mk,-},\bh_{\mk,+}),\ \
\hat{\mathbf{A}}_{\mk,d\theta}= ^{t}(\ah_{\mk,d},\ah_{\mk,\theta}),
\eeq
and
$U(k)$ and $V(k)$ are $2\times 2$ real matrices,
\beq 
U(k)=\frac{1}{2}
\left( 
\begin{array}{cc}
\frac{1}{2C_{1-}(k)E_1(k)}
+C_{1-}(k)X_{1+}(k)
 & q\ek C_{1-}(k)\sin 2\theta -\frac{X_{1-}(k)}{2q\ek C_{1-}(k)
E_1(k)\sin 2\theta}\\
-\frac{1}{2C_{1+}(k)E_1(k)}
+C_{1+}(k)X_{1-}(k)
 &   q\ek C_{1+}(k)\sin 2\theta +\frac{X_{1+}(k)}{2q\ek 
C_{1+}(k)E_1(k)\sin 2\theta}
\end{array} 
\right) ,
\eeq
\beq 
V(k)=\frac{1}{2}
\left( 
\begin{array}{cc}
-\frac{1}{2C_{1-}(k)E_{1}(k)}
+C_{1-}(k)X_{1+}(k)
 & q\ek C_{1-}(k)\sin 2\theta +\frac{X_{1-}(k)}{2q\ek C_{1-}(k)
E_1(k)\sin 2\theta}\\
\frac{1}{2C_{1+}(k)E_1(k)}
+C_{1+}(k)X_{1-}(k)
 &   -q\ek C_{1+}(k)\sin 2\theta -\frac{X_{1+}(k)}{2q
\ek C_{1+}(k)E_1(k)\sin 2\theta}
\end{array} 
\right) ,
\label{eq:ba-82}
\eeq
with
\begin{gather}
X_{1\pm}(k)=-n(c_0^{(1)}+3c_1^{(1)}-2c_q^{(1)})\ek
+2n^2c_1^{(1)}(c_1^{(1)}-c_q^{(1)})
\pm E_1(k),\\
E_1(k)=\Bigg\{[(n^2(c_0^{(1)}+3c_1^{(1)})^2-4n^2c_q^{(1)}
(c_0^{(1)}+2c_1^{(1)})]\ek^2
-4n^3c_1^{(1)}(c_0^{(1)}+3c_1^{(1)})
(c_1^{(1)}-c_q^{(1)})\ek\nonumber\\
+[2n^2c_1^{(1)}(c_1^{(1)}-c_q^{(1)})
]^2
\Bigg\}^{1/2},\\
C_{1\pm}(k)=\sqrt{\frac{E_{\mk,\pm}}{X^2_{1\mp}(k)\ek+q^2\ek^2
(\ek-2nc_1^{(1)})\sin^22\theta 
}}.
\label{eq:ba-85}
\end{gather}
In Eq. \eqref{eq:ba-85}, $E_{\mk,\pm}$ are the Bogoliubov spectra given by
\beq
E_{\mk,\pm}&=&\sqrt{\ek^2+n(c_0^{(1)}-c_1^{(1)})\ek
+2n^2c_1^{(1)}(c_1^{(1)}-c_q^{(1)})
\pm E_1(k)}.
\label{eq:spin-1-ba-b2}
\eeq
Using the transformations (\ref{eq:baspec}), 
the total Hamiltonian is 
diagonalized as 
\beq
\hat{H}_{\text{eff}}^{BA}=E_0^{BA}+
{\sk}^{'}\Bigg[E_{\mk,f_z}\bhd_{\mk,f_z}\bh_{\mk,f_z}+
E_{\mk,-}\bhd_{\mk,-}\bh_{\mk,-}
+E_{\mk,+}\bhd_{\mk,+}\bh_{\mk,+}
\Bigg],
\eeq
where the GSE is given by
\beq
E_0^{BA}=\frac{Nq}{2}+\frac{Vn^2
(\bar{c}_0^{(1)}+\bar{c}_1^{(1)})}{2}+\frac{Nq^2}{8n\bar{c}_1^{(1)}}
-\frac{1}{2}{\sk}^{'}\Big[
\left(\ek+q/2-E_{\mk,f_z}\right)\nonumber\\
+(2\ek+nc_0^{(1)}-nc_1^{(1)}-E_{\mk,+}
-E_{\mk,-})
\Big].
\eeq
Since 
$E_{\mk,-}^2\simeq 2n(c_0^{(1)}+c_1^{(1)})\ek$ in the long-wavelength limit, 
$E_{\mk,-}$ is linear and massless.
Furthermore, the $f_z$ mode is also linear and massless  for nonzero $q$. 
This is because
the $U(1)\times SO(2)$ symmetry is completely broken in the
broken-axisymmetry phase.
The positivity of the density and spin modes is ensured if
$q>0$, $c_1^{(1)}<0$, $c_0^{(1)}+c_1^{(1)}>0$, and $q<-2c_1^{(1)}n$.
Even though each mode is still massless for $q=0$, in the long-wavelength limit,
the spectrum of $f_z$ changes from linear with $q\ne 0$ to quadratic 
with $q = 0$. 
The physical origin of this change is discussed in Sec. V. 
Using the renormalization procedure discussed in Appendix A, 
we find that the GSE per volume $V$ 
in the broken-axisymmetry phase is given by
\beq
\frac{E_0^{BA}}{V}&=&\frac{nq}{2}+\frac{n^2(c_0^{(1)}+c_1^{(1)})}{2}
+\frac{q^2}{8c_1^{(1)}}+\frac{\sqrt{2M^3q^5}}{15\pi^2\hbar^3}
+\frac{8\sqrt{M^3}[n(c_0^{(1)}+c_1^{(1)})]^{\frac{5}{2}}}{15\pi^2\hbar^3}
\phi_4(t_3),
\label{eq:spin-1-ba-gse}
\eeq
with  $t_3=q^2/(nc_0^{(1)}+nc_1^{(1)})^2$. Here
\beq
\phi_4(t)&\equiv&-\frac{15}{8\sqrt{2}}\int_0^{\infty}
dx x^2\left(2x^2+y_1-\phi_{4}^{(+)}(t)-\phi_{4}^{(-)}(t)-\frac{1}{2x^2}
+\frac{c_0^{(1)}t}{4c_1^{(1)}x^2}\right),
\eeq
\beq
\phi_{4}^{(\pm)}(t)&\equiv&\sqrt{x^4+y_1x^2+y_2(t)
\pm\sqrt{y_3(t)x^4+y_4(t)x^2+y_2^2(t)}},
\eeq 
with
$y_1=(c_0^{(1)}-c_1^{(1)})/(c_0^{(1)}+c_1^{(1)})$,
$y_2(t)=(2c_1^{(1)})^2/(c_0^{(1)}+c_1^{(1)})^2-t/2$,
$y_3(t)=(c_0^{(1)}+3c_1^{(1)})^2/(c_0^{(1)}+c_1^{(1)})^2
-(c_0^{(1)}+2c_1^{(1)})t/c_1^{(1)}$, and
$y_4(t)=-4(c_1^{(1)})^2(c_0^{(1)}+3c_1^{(1)})/(c_0^{(1)}+c_1^{(1)})^3
+(c_0^{(1)}+3c_1^{(1)})t/(c_0^{(1)}+c_1^{(1)})$.
The behavior of $\phi_4(t)$ is shown in Fig. \ref{phi4}, where
the typical values of $t$ for $^{87}$Rb is $10^{-5}$.
As can be seen from Fig. \ref{phi4}(a), the density component of the GSE 
increases slightly due to the quadratic Zeeman effect. 
The pressure and the sound velocity are
\beq
P=\frac{n^2(c_0^{(1)}+c_1^{(1)})}{2}-\frac{q^2}{8c_1^{(1)}}
-\frac{\sqrt{2M^3q^5}}{15\pi^2\hbar^3}
+\frac{4\sqrt{M^3}[n(c_0^{(1)}+c_1^{(1)})]^{\frac{5}{2}}}{15\pi^2\hbar^3}
(3\phi_4(t_3)-4t_3\phi'_4(t_3)),
\label{eq:spin-1-ba-p}
\eeq
\beq
c=\sqrt{\frac{n(c_0^{(1)}+c_1^{(1)})}{M}}\left(1+
\frac{\sqrt{M^3}}{15\pi^2\hbar^3}\sqrt{n(c_0^{(1)}+c_1^{(1)})^3}
\phi_5(t_3)\right),
\label{eq:spin-1-ba-sv}
\eeq
where
\beq
\phi_5(t)&\equiv&15\phi_4(t)-16t\phi'_4(t)+16t^2\phi''_4(t).
\label{eq:phi5}
\eeq
The behavior of $\phi_5(t)$ in Fig. \ref{phi4}(b) shows 
that the LHY correction of the sound velocity
decreases slightly as the external magnetic field increases.

The quantum depletion is 
\beq
\frac{N-N_0}{N}&=&\frac{1}{N}{\sk}^{'}\langle\ahd_{\mk,f_z}\ah_{\mk,f_z}
+\ahd_{\mk,d}\ah_{\mk,d}+\ahd_{\mk,\theta}\ah_{\mk,\theta}
\rangle\nonumber\\
&=&\frac{\sqrt{M^3}}{3\pi^2\hbar^3}
\left(\sqrt{\frac{q^3}{8n^2}}
+\sqrt{n(c_0^{(1)}+c_1^{(1)})^3}\phi_6(t_3)\right),
\label{eq:spin-1-ba-dep}
\eeq
with
\beq
\phi_6(t)\equiv\frac{6}{\sqrt{2}}\int_0^{\infty}dxx^2(V_{11}^2+V_{12}^2
+V_{21}^2+V_{22}^2),
\eeq
where $V_{ij}$'s are the the components of $V(k)$ in Eq. \eqref{eq:ba-82}.
The behavior of $\phi_6$ in Fig. \ref{phi4}(a) shows that
the quantum depletion increases with increasing
$t_3=q^2/(nc_0^{(1)}+nc_1^{(1)})^2$.
However, the magnetic susceptibility of the quantum corrections in the
broken-axisymmetry phase is small because the variations of 
$\phi_4(t)$, $\phi_5(t)$, and $\phi_6(t)$ over the typical range of
$t$ are of the order of $10^{-2}$.
    
\section{Spin-2 BEC}
\begin{figure}[h]
  \includegraphics[width=0.60\linewidth]{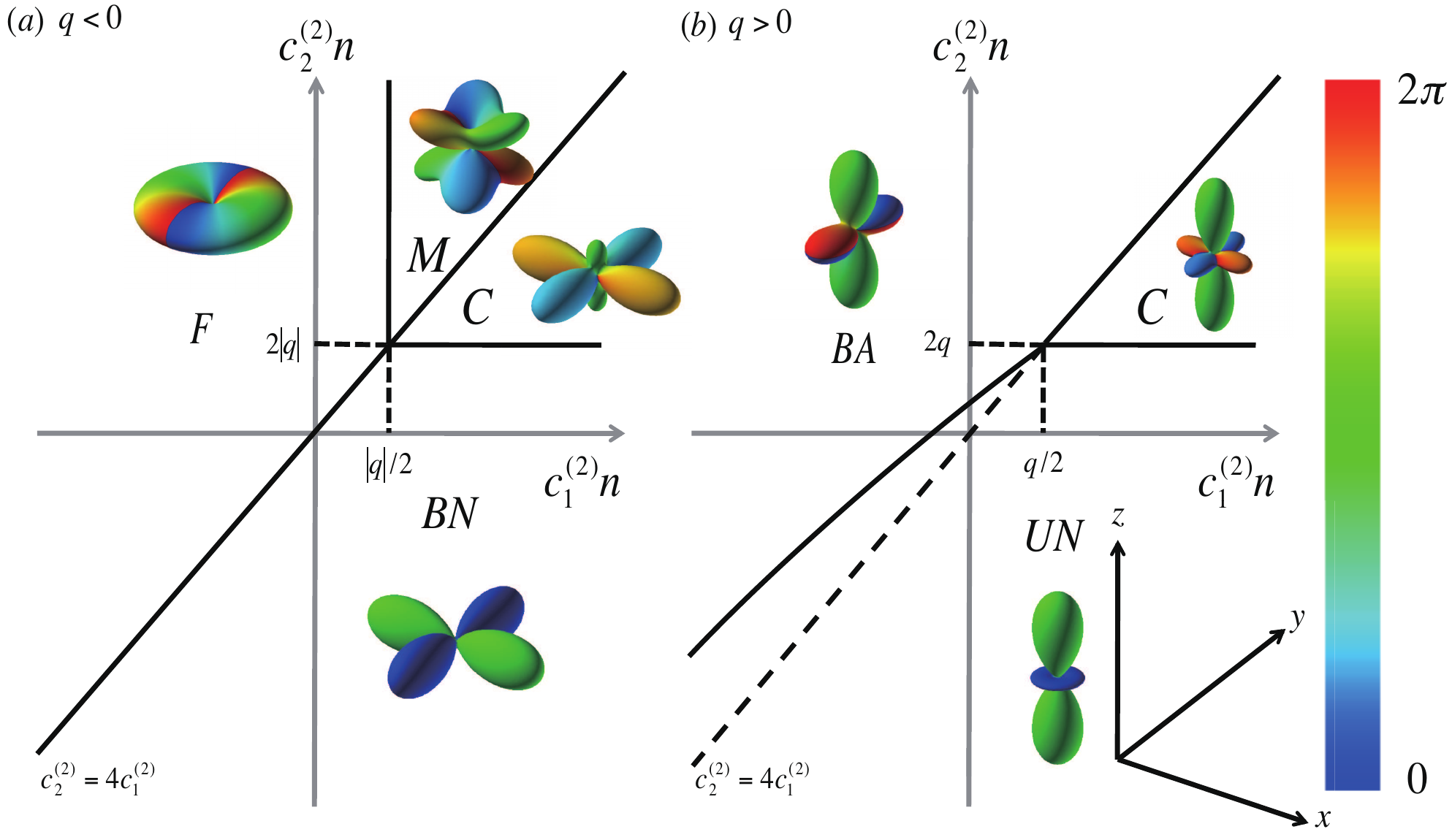}
 \caption{(Color online)
Phase diagrams of spin-2 BECs for nonzero $q$. 
(a) Case of $q<0$, where 
BN, C, F, and M stand for the biaxial nematic, cyclic,
ferromagnetic, and mixed phases, respectively.
In the absence of an external magnetic field ($q=0$), the 
C and M states can be transformed to each other by a rotation in space.
(b) Case of $q>0$, where  
the UN and BA stand for the uniaxial nematic and
broken-axisymmetry phases, respectively.
The thick lines show the phase boundaries.}
\label{spin-2-phase-q}
\end{figure}

For the case of a spin-2 BEC,
the total spin in the two-body interaction $F$ 
takes on the values of 0, 2, or 4, and
Eq. (\ref{secondint2}) reduces to \cite{ueda}
\beq
\hat{H}&=&\sum_{\mathbf{k}}
(\epsilon_{\mathbf{k}}+qm^2)
\hat{a}^{\dagger}_{\mk,m}\hat{a}_{\mk,m} 
+\frac{1}{2V}\sum_{\mathbf{k},\mathbf{p},\mathbf{q}}
\Big( \bar{c}_0^{(2)}\hat{a}_{\mathbf{p},m}^{\dagger}\hat{a}_{\mathbf{q},m'}^{\dagger}\hat{a}_{\mathbf{p+k},{m}}\hat{a}_{\mathbf{q-k},{m'}}\nonumber\\
&&+\bar{c}_1^{(2)}\mathbf{f}_{mm^{'}}\cdot\mathbf{f}_{\mu\mu '}\hat{a}_{\mathbf{p},m}^{\dagger}\hat{a}_{\mathbf{q},\mu}^{\dagger}
\hat{a}_{\mathbf{p+k},{m^{'}}}\hat{a}_{\mathbf{q-k},\mu '}
+\bar{c}_2^{(2)}(-1)^{m+m'}\hat{a}_{\mathbf{p},m}^{\dagger}
\hat{a}_{\mathbf{q},-m}^{\dagger}
\hat{a}_{\mathbf{p+k},{m'}}\hat{a}_{\mathbf{q-k},{-m'}}\Big) ,  \label{spin2}
\eeq
where 
$\bar{c}_0^{(2)}=(4\bar{g}_2+3\bar{g}_4)/7,$
$\bar{c}_1^{(2)}=(\bar{g}_4-\bar{g}_2)/7,$
$\bar{c}_2^{(2)}=(7\bar{g}_0-10\bar{g}_2+3\bar{g}_4)/35,$
and the spin-2 matrices are given by
\begin{eqnarray}
{f}^x &=& \begin{pmatrix}
       0 & 1 & 0 & 0 & 0\\
       1 & 0 & \sqrt{3\over 2} & 0 & 0\\
       0 & \sqrt{3 \over 2} & 0 & \sqrt{3\over 2} & 0\\
       0 & 0 & \sqrt{3\over 2} & 0 & 1\\
       0 & 0 & 0 & 1 & 0 \end{pmatrix},\ \ \ 
{ f}^y = \begin{pmatrix}
       0 & -i & 0 & 0 & 0\\
       i & 0 & -i\sqrt{3\over 2} & 0 & 0\\
       0 & i\sqrt{3 \over 2} & 0 & -i\sqrt{3\over 2} & 0\\
       0 & 0 & i\sqrt{3\over 2} & 0 & -i\\
       0 & 0 & 0 & i & 0 \end{pmatrix}, \ \ 
{ f}^z = \begin{pmatrix} 
       2 & 0 & 0 & 0 & 0 \\ 
       0 & 1 & 0 & 0 & 0 \\   
       0 & 0 & 0 & 0 & 0 \\ 
       0 & 0 & 0 & -1 & 0 \\
       0 & 0 & 0 & 0 & -2 \\
\end{pmatrix}.
\label{spin-2Matrices}
\end{eqnarray}
The order parameter in a spin-2 BEC can be expanded in terms of
the spherical harmonics of rank 2 as
\beq
\Psi =\sum_m \zeta_m \text{Y}^m_2(\mathbf{n})
\equiv\frac{1}{2}\sqrt{\frac{15}{8\pi}}\mathbf{n}^{T}Q\mathbf{n},
\eeq
where $\mathbf{n}=^t(n_x,n_y,n_z)$ and
\beq
Q=
\left( 
\begin{array}{ccc}
\zeta_2+\zeta_{-2}-\sqrt{\frac{2}{3}}\zeta_{0} & i(\zeta_2-\zeta_{-2}) & -\zeta_1+\zeta_{-1} \\
i(\zeta_2-\zeta_{-2}) & -\zeta_2-\zeta_{-2}-\sqrt{\frac{2}{3}}\zeta_{0}  & -i(\zeta_1+\zeta_{-1})  \\
-\zeta_1+\zeta_{-1} & -i(\zeta_1+\zeta_{-1}) & 2\sqrt{\frac{2}{3}}\zeta_0 \\
\end{array} 
\right) \label{tensor}
\eeq 
is a traceless symmetric tensor \cite{mermin,ueda}.
In the absence of an external magnetic field, the mean-field energy of 
a spin-2 BEC can be completely characterized by $Q$, and
the ground-state phases are
ferromagnetic, nematic, or cyclic, depending on the values of 
the coupling constants.
\begin{figure}[h]
 \begin{center}
  \includegraphics[width=0.4\linewidth]{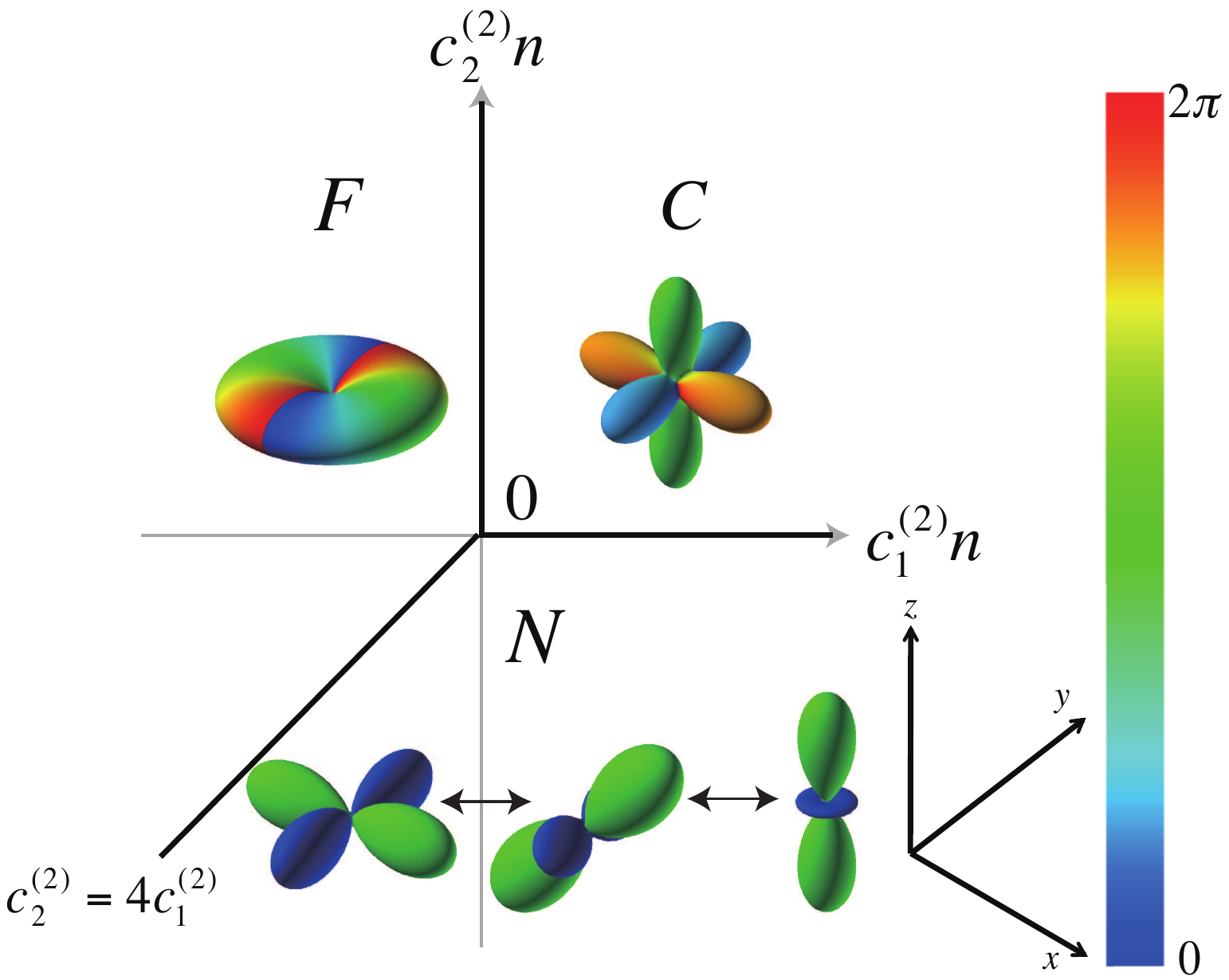}
  \caption{(Color online)
Phase diagram of spin-2 BECs in the absence of an external magnetic
  field, where C, F, and N show the cyclic, ferromagnetic, and nematic 
phases, respectively.
The thick lines represent the phase boundaries.
The shape of the wave function is
  determined from $|\Psi(\theta,\phi)|^2\equiv 
|\sum_m\zeta_m\text{Y}_2^{m}(\theta,\phi)|^2$ and the color indicates
the phase of $\Psi(\theta,\phi)$. 
The axes of the spin coordinate are depicted at the bottom right
of the phase diagram.
Note that the uniaxial (right)  and biaxial (left)
nematic phases are degenerate at the mean-field level. }
  \label{spin-2-phase}
 \end{center}
\end{figure} 
The parameter regimes of these phases and the spinor order parameters 
are  given by (see Fig. \ref{spin-2-phase})
\cite{ciobanu,koashi,ueda,barnett,mermin}
\begin{gather}
\text{ferromagnetic phase:} \ \ \  
c_1^{(2)} < 0, \ \ 
c_2^{(2)}>4c_1^{(2)}, \ \ \   \text{and} \ \ \
\mathbf{\zeta}^{F}=(1,0,0,0,0),\label{ferro2}\\
\text{nematic phase:} \
\ \   c_2^{(2)} < 0,  \ \  
c_2^{(2)}<4c_1^{(2)}, \
\text{and} \ \  \mathbf{\zeta}^{N}=
\left(\frac{\sin\eta}{\sqrt{2}},0,\cos\eta,0,\frac{\sin\eta}{\sqrt{2}}\right),  \label{polar2}\\
\text{cyclic phase:} \ \ \   
c_1^{(2)}>0, \ \ c_2^{(2)} >0,  \ \text{and} \ \mathbf{\zeta}^{C}=\frac{1}{2}(1,0,\sqrt{2},0,-1),\label{cyclic}
\end{gather}
where $c_0^{(2)}=(4g_2+3g_4)/7$, $c_1^{(2)}=(g_4-g_2)/7$,
$c_2^{(2)}=(7g_0-10g_2+3g_4)/35$.
The nematic phase has no magnetization but features a finite
spin-singlet
pair amplitude,
whereas the cyclic phase in the absence of an external magnetic field
has neither magnetization nor spin-singlet
amplitude.
It is predicted \cite{ciobanu,widera,tojo} that the ground states of
$^{23}$Na is nematic and 
that of 
$^{87}$Rb lies in the vicinity of the phase boundary between
the nematic and cyclic phases.
($^{85}$Rb is unstable at zero magnetic field in the thermodynamic limit 
because $c_0^{(2)}<0$.)
The nematic phase has an additional continuous parameter $\eta$
(see $\zeta^N$ in Eq. (\ref{polar2})),
which is not related to the symmetry of the Hamiltonian but
represents the degeneracy of the uniaxial and biaxial spin nematic
phases \cite{barnett,turner,song}. 
However, this additional degeneracy arises only when
the external magnetic field vanishes.
In fact, for $q<0$, the biaxial nematic phase has a lower ground-state
energy than the uniaxial nematic phase, and
the boundaries and spinor order parameters are 
given as follows \cite{saito}:
\begin{gather}
\text{ferromagnetic phase:} \ \ \ 
c_2^{(2)}>4c_1^{(2)}, \ \
 \ c_1^{(2)} < \frac{|q|}{2n}, 
 \ \ \   \text{and} \ \ \
\mathbf{\zeta}^{F}=(1,0,0,0,0),\label{qferro2}\\
\text{biaxial nematic phase:} \ \ \ 
c_2^{(2)}<4c_1^{(2)},
\ \   c_2^{(2)} < \frac{2|q|}{n}, \ \  
\text{and} \ \  \mathbf{\zeta}^{BN}=
\left(\frac{1}{\sqrt{2}},0,0,0,\frac{1}{\sqrt{2}}\right),  \label{qpolar2}\\
\text{cyclic phase:} \ \ \ 
 c_2^{(2)}<4c_1^{(2)}, \ \ 
c_1^{(2)}>\frac{|q|}{2n}, 
\ \ c_2^{(2)} >\frac{2|q|}{n},
 \ \text{and} \ \mathbf{\zeta}^{C}=\left(\frac{\sin\theta}{\sqrt{2}},
0,\cos\theta,0,-\frac{\sin\theta}{\sqrt{2}}\right),\label{qcyclic}\\
\text{mixed phase:} \ \ \
c_2^{(2)}>4c_1^{(2)}, \ \ 
c_1^{(2)}>\frac{|q|}{2n}, 
\ \ c_2^{(2)} >\frac{2|q|}{n},
 \ \text{and} \ \mathbf{\zeta}^{M}=\left(\cos\theta',
0,0,\sin\theta',0\right) \ \nonumber\\ 
\text{or} \  (0, \sin\theta',0,0,\cos\theta'),
\label{qm}
\end{gather}
where $\cos\theta=\sqrt{1/2+q/nc_2^{(2)}}$ and
$\cos\theta'=\sqrt{1/3-q/3nc_1^{(2)}}$.
The phase diagram is shown in Fig. \ref{spin-2-phase-q}(a).
The cyclic phase for $q<0$ does not have any magnetization but
has a finite spin-singlet pair amplitude that depends on $q$.
The mixed phase does not have a spin-singlet pair amplitude but
has a finite longitudinal magnetization that depends on $q$.

On the other hand, for $q>0$,
based on a numerical calculation, we find that
the ground-state phases are given as follows:
\begin{gather}
\text{broken-axisymmetry phase:} 
\ \ \ \zeta^{BA}=(\pm a,b,c,b,\pm a), \ \ +\  (-) \ 
\text{sign for}\  c_1^{(2)}<0\ (>0) \\
\text{uniaxial nematic phase:} 
\ \ \ \zeta^{UN}=(0,0,1,0,0), \label{uncon}\\
\text{cyclic phase:}
\ \ \ \zeta^{C}=\left(\frac{\sin\theta}{\sqrt{2}},0,\cos\theta,0,
-\frac{\sin\theta}{\sqrt{2}}\right),
\end{gather}
where 
$a$, $b$, and $c$ are positive  except for
the case of $c_1^{(2)}=0$ and $c_2^{(2)}>0$, where 
$a=0$.
The broken-axisymmetry phase has a transverse magnetization and
a spin-singlet pair amplitude, both of which depend on $q$,
as in the case of the spin-1 
broken-axisymmetry phase.
The phase boundaries between the broken-axisymmetry and
cyclic phases and those between the cyclic and uniaxial nematic phases
correspond to those between the mixed and cyclic phases and between
the cyclic and biaxial nematic phases, respectively.
The boundary between the broken-axisymmetry and uniaxial nematic
phases is determined numerically.
The phase diagram is shown
in Fig. \ref{spin-2-phase-q}(b).  
 
The Bogoliubov Hamiltonian in a spin-2 BEC is given by \cite{ueda}
\beq
\hat{H}_{\text{eff}}&=&\frac{Vn^2}{2}\left(
\bar{c}_0^{(2)}+\bar{c}_1^{(2)}\langle\mathbf{f}\rangle^2
+4\bar{c}_2^{(2)}|\langle
s_{-}\rangle|^2\right)+
qN\langle(f^z)^2\rangle\nonumber\\
&&+{\sk}^{'}\Bigg[\left(\epsilon_{\mathbf{k}}-nc_1^{(2)}
\langle\mathbf{f}\rangle^2-4nc_2^{(2)}|\langle
s_{-}\rangle |^2+qm^2-q\langle(f^z)^2\rangle
\right)\hat{a}^{\dagger}_{\mathbf{k},m}\hat{a}_{\mathbf{k},m}\nonumber\\
&&+\frac{nc_0^{(2)}}{2}
\left(2\hat{D}_{\mathbf{k}}^{\dagger}\hat{D}_{\mathbf{k}}
+\hat{D}_{\mathbf{k}}\hat{D}_{\mathbf{-k}}
+\hat{D}_{\mathbf{k}}^{\dagger}\hat{D}_{\mathbf{-k}}^{\dagger}\right)
+\frac{nc_1^{(2)}}{2}
\left(2\mathbf{\hat{F}}_{\mathbf{k}}^{\dagger}
\cdot\mathbf{\hat{F}}_{\mathbf{k}}
+\mathbf{\hat{F}}_{\mathbf{k}}\cdot\mathbf{\hat{F}}_{\mathbf{-k}}+
\mathbf{\hat{F}}_{\mathbf{k}}^{\dagger}
\cdot\mathbf{\hat{F}}_{\mathbf{-k}}^{\dagger}\right)\nonumber\\
&&+nc_1^{(2)}
\langle\mathbf{f}\rangle\cdot\mathbf{f}_{mm'}
 \hat{a}^{\dagger}_{\mathbf{k},m}\hat{a}_{\mathbf{k},m'}
+2nc_2^{(2)}
 (\hat{S}^{-}_{\mk})^{\dagger}\hat{S}^{-}_{\mk} \nonumber\\
&&+nc_2^{(2)}
(-1)^m\left(\langle s_{-}
\rangle\hat{a}^{\dagger}_{\mathbf{k},m}\hat{a}_{\mathbf{-k},-m}^{\dagger}+
\langle
s_{-}\rangle^{*}\hat{a}_{\mathbf{k},m}\hat{a}_{\mathbf{-k},-m}\right)\Bigg],
 \label{bogoliubov2}
\eeq
where 
\beq
\langle s_- \rangle =\frac{1}{2}\sum_m (-1)^{m} 
\zeta_{m}\zeta_{-m}
\eeq
is the spin-singlet pair amplitude, and
\beq
\hat{S}^{-}_{\mk}=\sum_{m}(-1)^{m}\zeta_{m}\ah_{\mk ,-m}, \label{pair}
\eeq
is the pair fluctuation operator of the condensate.
As in the case of a spin-1 BEC, 
we substitute $c_i^{(2)}$ for $\bar{c}_i^{(2)}$ $(i=0,1,2)$ in
the sum over the momentum in Eq. (\ref{bogoliubov2}).

\subsection{Ferromagnetic phase}
For the ferromagnetic phase (\ref{ferro2}),
the sub-Hamiltonian with respect to $\ah_{\mk,2}$ 
is diagonalized by the  Bogoliubov transformation \cite{ueda}
\beq
\bh_{\mk,2}=\sqrt{\frac{\ek+n(c_0^{(2)}+4c_1^{(2)})+E_{\mk,2}}
{2E_{\mk,2}}}\ah_{\mk,2}
+\sqrt{\frac{\ek+n(c_0^{(2)}+4c_1^{(2)})-E_{\mk,2}}
{2E_{\mk,2}}}\ahd_{\mmk,2},
\eeq
with the Bogoliubov spectrum
\beq
E_{\mk ,2}=\sqrt{\ek \left[\ek +2n
(c_0^{(2)}+4c_1^{(2)})\right]}.
\label{eq:spin-2-f-b}
\eeq 
The diagonalized Hamiltonian is given by
\beq
\hat{H}_{\text{eff}}^{F}&=&E_{0}^F
+{\sk}^{'}\Bigg[ E_{\mk ,2}\bhd_{\mk ,2}\bh_{\mk ,2} 
+(\ek-3q) \ahd_{\mk ,1}\ah_{\mk ,1}
+\left(\ek-4q -4nc_1^{(2)}\right)\ahd_{\mk ,0}\ah_{\mk ,0}\nonumber\\
&&+\left(\ek-3q -6nc_1^{(2)}\right)
\ahd_{\mk ,-1}\ah_{\mk ,-1}+\left(\ek
-8nc_1^{(2)}+2nc_2^{(2)}\right)
\ahd_{\mk ,-2}\ah_{\mk ,-2}\Bigg],
\eeq
where
\beq
E_{0}^F=4qN+\frac{Vn^2(\bar{c}_0^{(2)}+4\bar{c}_1^{(2)})}{2}
-\frac{1}{2}{\sk}^{'}\left(\ek
+n(c_0^{(2)}+4c_1^{(2)})
-E_{\mk,2}\right)
\label{eq:spin-2-f-117}
\eeq
is the GSE.
The $m=2$ mode is massless, 
and in the absence of an external magnetic field, the $m=1$ mode also
becomes massless.
For the excitation energies to be positive, 
we must have $c_0^{(2)}+4c_1^{(2)}>0$,
$q<0$, $c_1^{(2)}<|q|/(2n)$, and $c_2^{(2)}>4c_1^{(2)}$.
The GSE $E_0^F$, pressure $P$, sound velocity $c$, and quantum depletion
$(N-N_0)/N$ can be  
calculated as
\begin{gather}
\frac{E_0^{F}}{V}
=4qn+\frac{2\pi\hbar^2 n^2a_4}{M}\left(1+\frac{128}{15}
\sqrt{\frac{na_4^3}{\pi }}\right),
\label{eq:spin-2-f-gse}\\
P=\frac{2\pi\hbar^2 n^2a_4}{M}
\left(1+\frac{64}{5}\sqrt{\frac{na_4^3}{\pi}}\right) ,
\label{eq:spin-2-f-p}\\
c=\sqrt{\frac{4\pi\hbar^2 n a_4}{M^2}}
\left(1+8\sqrt{\frac{na_4^3}{\pi}}\right),
\label{eq:spin-2-f-sv}\\
\frac{N-N_0}{N}=\frac{8}{3}\sqrt{\frac{na_4^3}{\pi}},
\label{eq:spin-2-f-dep}
\end{gather}
where the renormalization of $\bar{g}_4$ is carried out to obtain the finite GSE.

\subsection{Nematic phase}
Let us next discuss the nematic phase.
We discuss the cases of $q=0$ and $q\ne 0$ separately, since
the stationary solutions  are different.

\subsubsection{Case of $q=0$}
In the absence of an external magnetic field, the nematic phase is 
characterized by Eq. (\ref{polar2}). 
We introduce the following four independent fluctuation operators:
\begin{gather}
\ah_{\mk ,d}=\frac{\sin\eta}{\sqrt{2}}(\ah_{\mk, 2}+\ah_{\mk ,-2})+\cos\eta\ah_{\mk ,0},\label{d2}\\
\ah_{\mk ,f_x}=\frac{1}{\sqrt{2}}(\ah_{\mk ,1}+\ah_{\mk, -1}),\\
\ah_{\mk ,f_y}=\frac{i}{\sqrt{2}}(-\ah_{\mk ,1}+\ah_{\mk, -1}),\\
\ah_{\mk ,f_z}=\frac{1}{\sqrt{2}}(\ah_{\mk 2}-\ah_{\mk ,-2}), \label{fluc-ne-z}
\end{gather}
where $\ah_{\mk,d}$  
and $\ah_{\mk,f_{x,y,z}}$ describe density and spin fluctuations, 
respectively \cite{note:nematic}.
Furthermore, we introduce the following nematic
fluctuation operator:
\beq
\ah_{\mk ,\eta}=\frac{\cos\eta}{\sqrt{2}}(\ah_{\mk ,2}+\ah_{\mk ,-2})-\sin\eta \ah_{\mk ,0}.\label{no}
\eeq
This operator describes fluctuations of the nematic order parameter; 
in fact, the coefficients in Eq. (\ref{no}) are related to the
nematic order parameter (\ref{polar2}) by
\beq
\frac{\partial
\mathbf{\zeta}^N}{\partial\eta}=\left(\frac{\cos\eta}{\sqrt{2}},0,-\sin\eta,0,\frac{\cos\eta}{\sqrt{2}}\right).
\eeq 
In terms of the operators
in Eqs. (\ref{d2})-(\ref{no}),
the total Hamiltonian can be decomposed into sub-Hamiltonians as
\beq
\hat{H}_{\text{eff}}^{N}&=&\frac{Vn^2( \bar{c}_0^{(2)}+\bar{c}_2^{(2)})}{2}
+{\sk}^{'}\Bigg\{\left[\ek+n(c_0^{(2)}
+c_2^{(2)})\right]
\ahd_{\mk ,d}\ah_{\mk ,d}
+\frac{n(c_0^{(2)}
+c_2^{(2)})}{2}
(\ahd_{\mk ,d}\ahd_{\mmk ,d}+\ah_{\mk ,d}\ah_{\mmk ,d})
\nonumber\\
&&+\left[\ek+nc_3^{(2)}(\eta+\pi/3)\right]
\ahd_{\mk ,f_x}\ah_{\mk ,f_x}
+\frac{nc_3^{(2)}(\eta+\pi/3)}{2}
(\ahd_{\mk ,f_x}\ahd_{\mmk ,f_x}+\ah_{\mk ,f_x}\ah_{\mmk ,f_x}) 
\nonumber\\
&&+\left[\ek +nc_3^{(2)}(\eta-\pi/3)
\right]\ahd_{\mk ,f_y}\ah_{\mk ,f_y}
-\frac{nc_3^{(2)}(\eta-\pi/3)}{2}
(\ahd_{\mk ,f_y}\ahd_{\mmk ,f_y}+\ah_{\mk ,f_y}\ah_{\mmk ,f_y})
\nonumber\\
&&+\left[\ek+nc_3^{(2)}(\eta)\right]\ahd_{\mk ,f_z}\ah_{\mk ,f_z}
+\frac{nc_3^{(2)}(\eta)}{2}
(\ahd_{\mk ,f_z}\ahd_{\mmk ,f_z}+\ah_{\mk ,f_z}\ah_{\mmk ,f_z})\nonumber\\
&&+\left(\ek 
-nc_2^{(2)}\right)\ahd_{\mk ,\eta}\ah_{\mk ,\eta}
+\frac{nc_2^{(2)}}{2}
(\ahd_{\mk ,\eta}\ahd_{\mmk ,\eta}+\ah_{\mk ,\eta}\ah_{\mmk ,\eta})\Bigg\},
\eeq
where $\displaystyle c_3^{(2)}(\eta)\equiv 4c_1^{(2)}\sin^2\eta-c_2^{(2)}$.
To diagonalize this Hamiltonian, we consider the following Bogoliubov
transformations:
\begin{gather}
\bh_{\mk,d}=\sqrt{\frac{\ek+n(c_0^{(2)}
+c_2^{(2)})+E_{\mk,d}}{E_{\mk,d}}}\ah_{\mk,d}
+\sqrt{\frac{\ek+n(c_0^{(2)}
+c_2^{(2)})-E_{\mk,d}}{E_{\mk,d}}}\ahd_{\mmk,d},\\
\bh_{\mk,f_x}=\sqrt{\frac{\ek+nc_3^{(2)}(\eta+\pi/3) 
+E_{\mk,f_x}}{E_{\mk,f_x}}}\ah_{\mk,f_x}
+\sqrt{\frac{\ek+nc_3^{(2)}(\eta+\pi/3)
-E_{\mk,f_x}}{E_{\mk,f_x}}}\ahd_{\mmk,f_x},\\
\bh_{\mk,f_y}=\sqrt{\frac{\ek+nc_3^{(2)}(\eta-\pi/3)
+E_{\mk,f_y}}{E_{\mk,f_y}}}\ah_{\mk,f_y}
-\sqrt{\frac{\ek+nc_3^{(2)}(\eta-\pi/3)
-E_{\mk,f_y}}{E_{\mk,f_y}}}\ahd_{\mmk,f_y},\\
\bh_{\mk,f_z}=\sqrt{\frac{\ek+nc_3^{(2)}(\eta)
+E_{\mk,f_z}}{E_{\mk,f_z}}}\ah_{\mk,f_z}
+\sqrt{\frac{\ek+nc_3^{(2)}(\eta)
-E_{\mk,f_z}}{E_{\mk,f_z}}}\ahd_{\mmk,f_z},\\
\bh_{\mk,\eta}=\sqrt{\frac{\ek-nc_2^{(2)}
+E_{\mk,\eta}}{E_{\mk,\eta}}}\ah_{\mk,\eta}
-\sqrt{\frac{\ek-nc_2^{(2)}-E_{\mk,\eta}}{E_{\mk,\eta}}}\ahd_{\mmk,\eta},
\end{gather}
where the Bogoliubov spectra are given by\cite{turner,song}
\begin{gather}
E_{\mk ,d}=\sqrt{\ek[\ek+2n(c_0^{(2)}+c_2^{(2)})]},
\label{eq:spin-2-n-b1}\\
E_{\mk
,f_x}=\sqrt{\ek[\ek+2nc_3^{(2)}(\eta+\pi/3
) ]},\label{eq:spin-2-n-b2}\\
E_{\mk
,f_y}=\sqrt{\ek[\ek+2nc_3^{(2)}(\eta-\pi/3
) ]},\label{eq:spin-2-n-b3}\\
E_{\mk ,f_z}=\sqrt{\ek[\ek+2nc_3^{(2)}(\eta)]},\label{eq:spin-2-n-b4}\\
E_{\mk ,\eta}=\sqrt{\ek(\ek-2nc_2^{(2)})} \label{eq:spin-2-n-b5}.
\end{gather}
The diagonalized Hamiltonian is then given by
\beq
\hat{H}_{\text{eff}}^{N}&=&E_0^N(\eta)
+{\sk}^{'}\Bigg[E_{\mk, d}\bhd_{\mk,d}\bh_{\mk,d}+E_{\mk, f_x}\bhd_{\mk,f_x}\bh_{\mk,f_x}+E_{\mk, f_y}\bhd_{\mk,f_y}\bh_{\mk,f_y}+E_{\mk, f_z}\bhd_{\mk,f_z}\bh_{\mk,f_z}
+E_{\mk, \eta}\bhd_{\mk,\eta}\bh_{\mk,\eta} \Bigg],
\nonumber\\
\eeq 
where
\beq
E_0^N(\eta)&=&\frac{Vn^2(\bar{c}_0^{(2)}
+\bar{c}_2^{(2)})}{2}
-\frac{1}{2}{\sk}^{'}\Bigg[ \left(\ek+n
(c_0^{(2)}+c_2^{(2)})-E_{\mk ,d} \right)+  
\left(\ek+nc_3^{(2)}(\eta+\pi/3)
-E_{\mk ,f_x} \right) \nonumber\\
&&+\left(\ek+nc_3^{(2)}(\eta-\pi/3)
-E_{\mk ,f_y} \right)  
+\left(\ek+nc_3^{(2)}(\eta)-E_{\mk ,f_z} \right)  
+\left(\ek-nc_2^{(2)} -E_{\mk ,\eta} \right)
\Bigg].
\nonumber\\
\eeq
Although the total number of the symmetry generators of the
Hamiltonian is $4$, the five Bogoliubov massless modes emerge.
The physical origin of this result is discussed in Sec. V.
We also note that regardless of the value of $\eta$,
the above Bogoliubov spectra are real  
if $c_0^{(2)}+c_2^{(2)}>0$, $c_2^{(2)}<0$, and
$c_2^{(2)}<4c_1^{(2)}$, in agreement with the stability criteria of
the mean-field ground state. 
As shown in Appendix A,
the GSE $E_0^N(\eta)$, pressure, sound velocity, quantum depletion are
respectively given by
\beq
\frac{E_0^N(\eta)}{V}&=&\frac{n^2(c_{0}^{(2)}+c_2^{(2)})}{2}
\left(1+\frac{16\sqrt{M^3}}{15\pi^2\hbar^3}\sqrt{n
(c_{0}^{(2)}+c_2^{(2)})^{3}}
\right)
+\frac{8\sqrt{M^3}}{15\pi^2\hbar^3}\Bigg\{
(n|c_2^{(2)}|)^{\frac{5}{2}}\nonumber\\
&&+[n(2c_1^{(2)}-c_2^{(2)})]^{\frac{5}{2}}
\sum_{j=0}^{2}[1+X\cos(2\eta+2\pi j/3)]^{\frac{5}{2}}
\Bigg\},
\label{eq:spin-2-n-gse}
\eeq
\beq
P(\eta)
&=&
\frac{n^2(c_{0}^{(2)}+c_2^{(2)})}{2}
\left(1+\frac{8\sqrt{M^3}}{5\pi^2\hbar^3}
\sqrt{n(c_{0}^{(2)}+c_2^{(2)})^3}\right)
+\frac{4\sqrt{M^3}}{5\pi^2\hbar^3}\Bigg\{
(n|c_2^{(2)}|)^{\frac{5}{2}}\nonumber\\
&&+[n(2c_1^{(2)}-c_2^{(2)})]^{\frac{5}{2}}
\sum_{j=0}^{2}[1+X\cos(2\eta+2\pi j/3)]^{\frac{5}{2}}
\Bigg\},
\label{eq:spin-2-n-p}
\eeq
\beq
c(\eta)
&=&\sqrt{\frac{n(c_{0}^{(2)}+c_2^{(2)})}{M}}
\Bigg\{1+\frac{\sqrt{M^3}}{\pi^2\hbar^3}\Bigg[
\sqrt{n(c_{0}^{(2)}+c_2^{(2)})^3}
-\frac{\sqrt{n(-c_2^{(2)})^5}}
{c_{0}^{(2)}+c_{2}^{(2)}}\nonumber\\
&&+\frac{\sqrt{n(2c_1^{(2)}
-c_2^{(2)})^5}}
{c_0^{(2)}+c_2^{(2)}}
\sum_{j=0}^{2}(1+X\cos(2\eta+2\pi j/3))^{\frac{5}{2}}
\Bigg]
\Bigg\},
\label{eq:spin-2-n-sv}
\eeq
\beq
\frac{N-N_0}{N}=\frac{\sqrt{M^3}}{3\pi^2\hbar^3}
\Bigg[\sqrt{n(c_0^{(2)}+c_2^{(2)})^3}
+\sqrt{n\left(c_3^{(2)}
(\eta+\pi/3)\right)^3}
+\sqrt{n\left(c_3^{(2)}
(\eta-\pi/3)\right)^3}\nonumber\\
+\sqrt{n\left(c_3^{(2)}
\left(\eta\right)\right)^3}
+\sqrt{n(|c_2^{(2)}|)^3}
\Bigg],\label{eq:spin-2-n-dep}
\eeq
where $X=-2c_1^{(2)}/(2c_1^{(2)}-c_2^{(2)})$ 
and $2c_1^{(2)}-c_2^{(2)} \ge 0$
in the nematic phase. 

\begin{figure}[h]
 \begin{center}
  \includegraphics[width=0.4\linewidth]{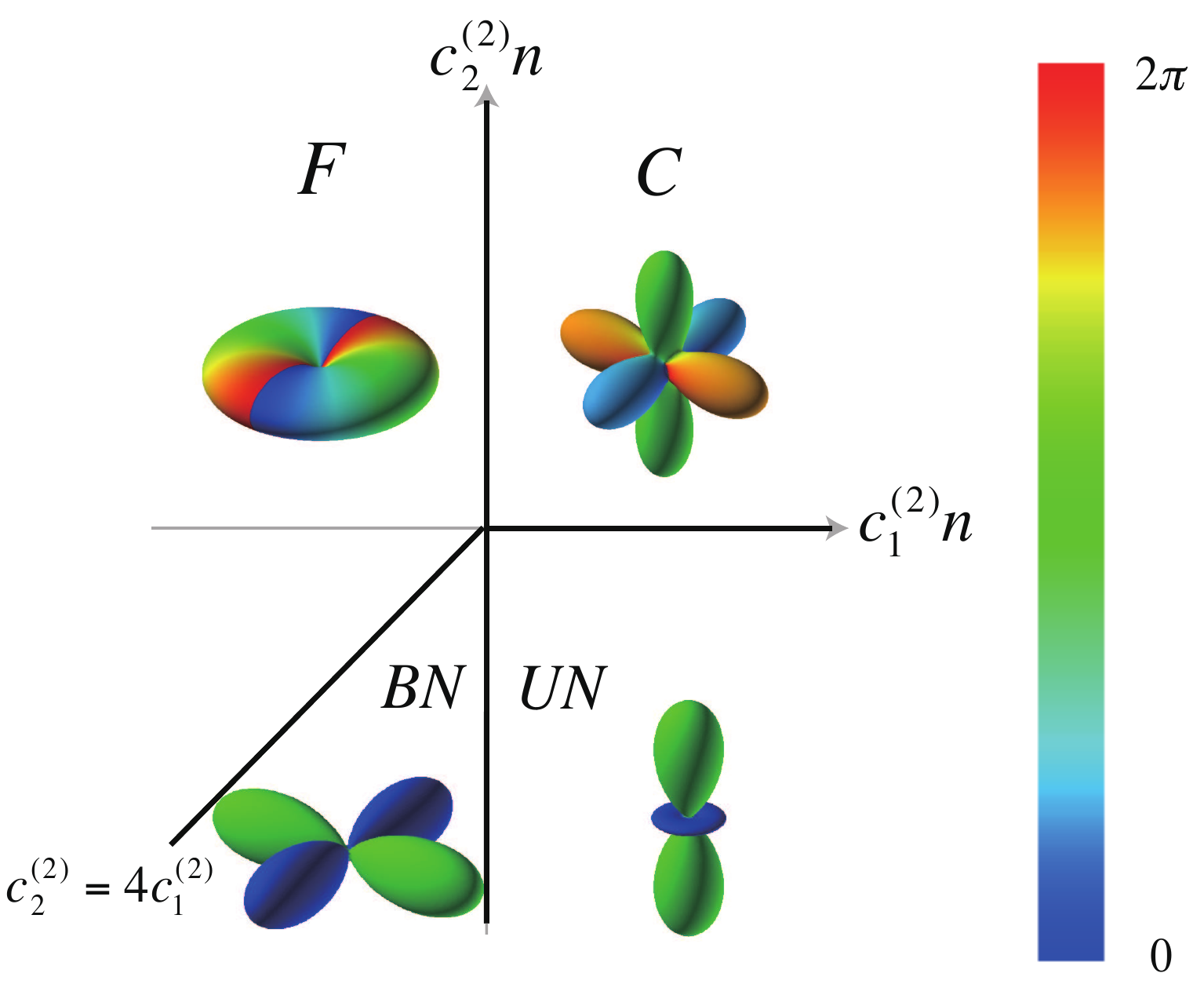}
  \caption{(Color online)
Beyond-mean-field-theory phase diagram of spin-2 BECs 
in the absence of an external magnetic
  field obtained by the Bogoliubov theory, 
where C, F, UN, and BN stand for the cyclic, ferromagnetic,
uniaxial nematic, and biaxial nematic 
phases, respectively.
The thick lines show the phase boundaries.
The phase diagram is consistent with that obtained in Refs. \cite{turner,song}.}
  \label{spin-2-phase-bogoliubov}
 \end{center}
\end{figure}

We note that 
the GSE satisfies the following symmetries:
\beq
E_0^N(\eta+\pi/3)=E_0^N(\eta) \ \ \text{and} \ \  
E_0^N(-\eta)=E_0^N(\eta).
\eeq
Therefore, other physical properties derived from it such as the
pressure, 
sound velocity,
and quantum depletion also satisfy similar relations:
we may therefore restrict the domain of the definition of $\eta$ to 
$\displaystyle 0\le \eta <\pi/3$ to find the ground state.
We define the part of the GSE that depends on $\eta$ as
\beq
E/\omega= \sum_{j=0}^{2}[1+X\cos(2\eta+2\pi j/3)]^{\frac{5}{2}},
\label{eeta}
\eeq
where $
\omega = 8\sqrt{M^3}V[n(2c_1^{(2)}-c_2^{(2)})]^{\frac{5}{2}}/(15
\pi^2\hbar^3).$
Since $\eta=0\ \text{and} \ \pi/6$ 
satisfy $\partial E/\partial\eta=0$,
these points are the candidates of the ground state.
\begin{figure}[h]
 \begin{center}
  \includegraphics[width=0.40\linewidth]{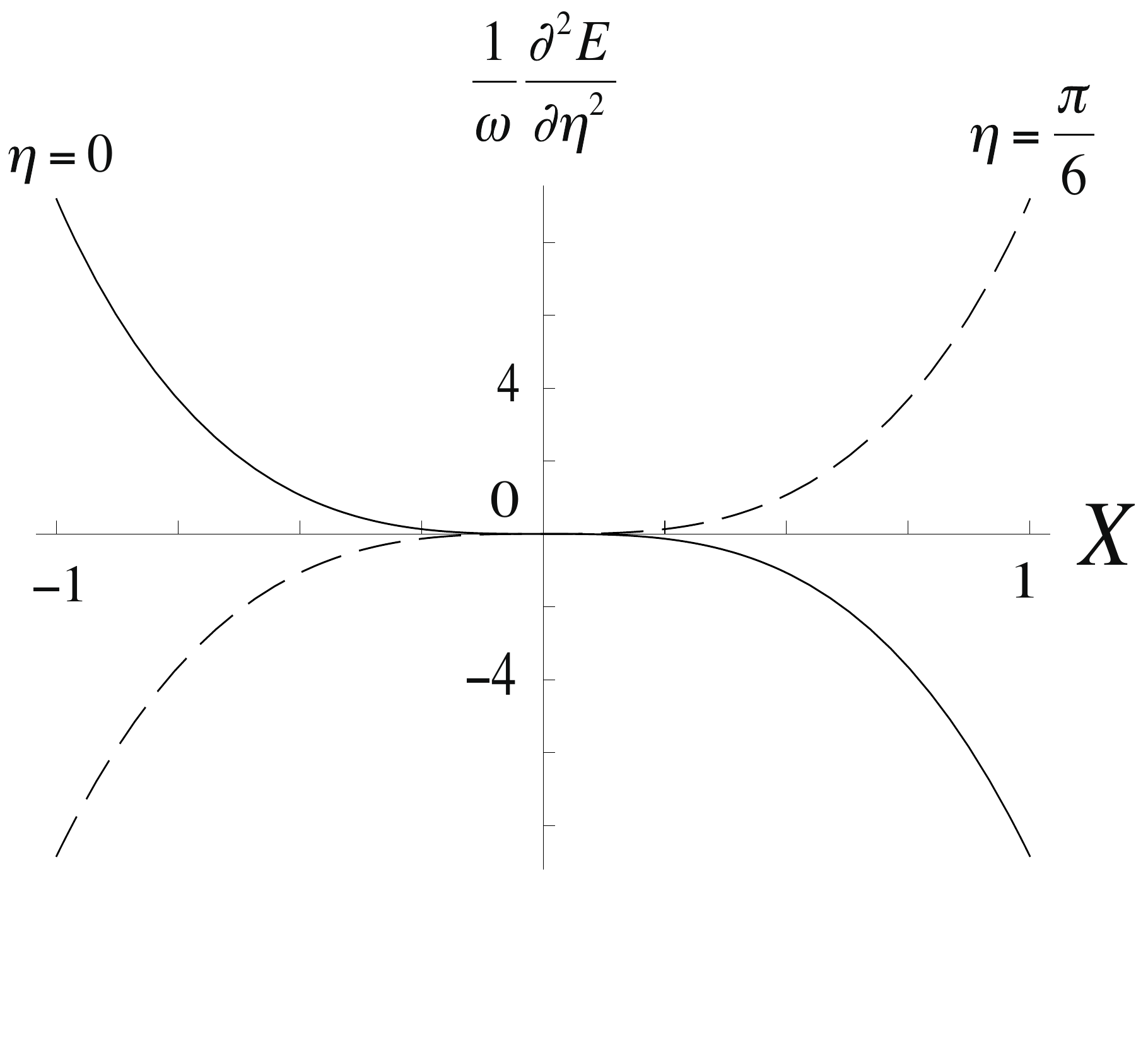}
  \caption{Second derivative of E with respect to $\eta$. 
The solid and dashed curves correspond to
$\eta=0\ \text{and} \ \pi/6$, 
respectively.  }
  \label{min}
 \end{center}
\end{figure} 
The second derivative with respect to $\eta$ is plotted in 
FIG. \ref{min}, which shows that
$\eta=0$ and $\eta=\pi/6$ correspond to
the ground state for
$X<0$ and $X>0$, respectively \cite{turner,song}. 
The phase transition between
uniaxial and biaxial nematic phases occurs at $X=0$ because 
the diagonalized order-parameter matrices (\ref{tensor}) for $X<0$ and $X>0$ 
given respectively by
\beq 
Q_{X<0}=
\left( 
\begin{array}{ccc}
-\sqrt{\frac{2}{3}} & 0 & 0 \\
0 & -\sqrt{\frac{2}{3}} & 0 \\
0 & 0 & 2\sqrt{\frac{2}{3}} \\
\end{array} 
\right) \ \ \text{and}\ \ 
Q_{X>0}=
\left( 
\begin{array}{ccc}
0 & 0 & 0 \\
0 & -\sqrt{2} & 0 \\
0 & 0 & \sqrt{2} \\
\end{array} 
\right) ,
\eeq
cannot be connected by any transformation 
that belongs to $U(1)\times SO(3)$ because
Det$(Q_{X>0})=0$ and Det$(Q_{X<0})\ne 0$.
The phase diagram that incorporates the quantum fluctuations arising from
the Bogoliubov theory is depicted in
Fig. \ref{spin-2-phase-bogoliubov}. 
Here, we note that the many-body analysis
in the single-mode approximation (SMA) \cite{koashi,ueda,uchino}
does not predict such a phase transition
because the SMA ignores
quantum fluctuations of  momentum that
play a crucial role in the phase transition between the uniaxial and
biaxial nematic phases.

We briefly comment on another stationary state 
$\displaystyle\zeta'=\frac{1}{\sqrt{2}}(0,1,0,1,0)$.
This state does not provide any new piece of 
information
because it can be transformed into a biaxial nematic
state by an element of $U(1)\times SO(3)$; in fact,
\beq
\zeta'=e^{i\theta}e^{-i\gamma f^z}e^{-i\beta f^y}e^{-i\alpha
f^z}\zeta^{BN},
\eeq
where
$\displaystyle\zeta^{BN}=\frac{1}{\sqrt{2}}(1,0,0,0,1)$,
$\theta=\pi/2$,
$\alpha=\pi/4$,
$\beta=\pi/2$,
and $\displaystyle\gamma=0$.
Therefore, this state belongs to the biaxial nematic phase
and it has the same spectra as 
the biaxial nematic state in the absence of an external magnetic field. 

There is 
a hidden symmetry in the mean-field stationary solution 
of the nematic phase.
The nematic phase is characterized as 
$|\langle s_-\rangle |=1/2$ and 
$\langle \mathbf{f}\rangle =\mathbf{0}$, 
both of which remain invariant under the
$U(1)\times SO(3)$ transformations and are independent on $\eta$.
In fact, we can show that the full symmetry group that leaves
the nematic invariant
is  $U(1)\times SO(5)$,
which includes the $U(1)\times SO(3)$ as a subgroup.
This is confirmed by using the fact that
the pair-singlet interaction term, which is proportional to
$\bar{c}_2^{(2)}$ in the Hamiltonian, has the $SO(5)$ symmetry \cite{uchino}, 
and that $\langle\mathbf{f} \rangle =\mathbf{0}$ if 
$|\langle s_- \rangle| =1/2$ \cite{ueda,ciobanu}.
In Sec.  V\hspace{-.1em}I, we discuss a role of the $SO(5)$ symmetry 
and relationship with the massless modes in the nematic phase.

\subsubsection{Biaxial nematic phase}
The effective Hamiltonian of the biaxial 
nematic phase defined in Eq. \eqref{qpolar2} 
can be decomposed by the transformations of Eqs. (\ref{d2})-
(\ref{no}):
\beq
\hat{H}_{\text{eff}}^{BN}&=&\frac{Vn^2(\bar{c}_0^{(2)}+\bar{c}_2^{(2)})}{2}
+4qN+{\sk}^{'}\Bigg\{(\ek+nc_0^{(2)}+nc_2^{(2)})\ahd_{\mk,d}\ah_{\mk,d}\nonumber\\
&&+\frac{n(c_0^{(2)}+c_2^{(2)})}{2}
(\ah_{\mk,d}\ah_{\mmk,d}+\ahd_{\mk,d}\ahd_{\mmk,d})\nonumber\\
&&+\sum_{j=f_x,f_y}\Big[(\ek-3q+nc_1^{(2)}-nc_2^{(2)})
\ahd_{\mk,j}\ah_{\mk,j}+\frac{n(c_1^{(2)}-c_2^{(2)})}{2}
(\ah_{\mk,j}\ah_{\mmk,j}
+\ahd_{\mk,j}\ahd_{\mk,j})\Big]\nonumber\\
&&+(\ek+4nc_1^{(2)}-nc_2^{(2)})\ahd_{\mk,f_z}\ah_{\mk,f_z}
+\frac{n(4c_1^{(2)}-c_2^{(2)})}{2}
(\ah_{\mk,f_z}\ah_{\mmk,f_z}+\ahd_{\mk,f_z}\ahd_{\mmk,f_z})\nonumber\\
&&+(\ek-4q-nc_2^{(2)})\ahd_{\mk,\eta}\ah_{\mk,\eta}
+\frac{nc_2^{(2)}}{2}
(\ah_{\mk,\eta}\ah_{\mmk,\eta}+\ahd_{\mk,\eta}\ahd_{\mmk,\eta})
\Bigg\}.
\eeq
This Hamiltonian can be diagonalized by 
the following Bogoliubov transformations
\begin{gather}
\bh_{\mk,d}=\sqrt{\frac{\ek+n(c_0^{(2)}
+c_2^{(2)})+E_{\mk,d}}{E_{\mk,d}}}\ah_{\mk,d}
+\sqrt{\frac{\ek+n(c_0^{(2)}
+c_2^{(2)})-E_{\mk,d}}{E_{\mk,d}}}\ahd_{\mmk,d},\\
\bh_{\mk,j}=\sqrt{\frac{\ek-3q+n(c_1^{(2)}
-c_2^{(2)})
+E_{\mk,f_t}}{E_{\mk,f_t}}}\ah_{\mk,j}
+\sqrt{\frac{\ek-3q+n(c_1^{(2)}
-c_2^{(2)})
-E_{\mk,f_t}}{E_{\mk,f_t}}}\ahd_{\mmk,j}, \\
\bh_{\mk,f_z}=\sqrt{\frac{\ek+n(4c_1^{(2)}
-c_2^{(2)})
+E_{\mk,f_z}}{E_{\mk,f_z}}}\ah_{\mk,f_z}
+\sqrt{\frac{\ek+n(4c_1^{(2)}
-c_2^{(2)} )
-E_{\mk,f_z}}{E_{\mk,f_z}}}\ahd_{\mmk,f_z},\\
\bh_{\mk,\eta}=\sqrt{\frac{\ek-4q-nc_2^{(2)}
+E_{\mk,\eta}}{E_{\mk,\eta}}}\ah_{\mk,\eta}
-\sqrt{\frac{\ek-4q-nc_2^{(2)}
-E_{\mk,\eta}}{E_{\mk,\eta}}}\ahd_{\mmk,\eta},
\end{gather}
with the result
\beq
\hat{H}_{\text{eff}}^{BN}=E_0^{BN}
+{\sk}^{'}\big[ E_{\mk ,d}\bhd_{\mk ,d}\bh_{\mk ,d}
+ E_{\mk ,f_t}(\bhd_{\mk ,f_x}\bh_{\mk ,f_x}
+\bhd_{\mk ,f_y}\bh_{\mk ,f_y})
+ E_{\mk ,f_z}\bhd_{\mk ,f_z}\bh_{\mk ,f_z}
+ E_{\mk ,\eta}\bhd_{\mk ,\eta}\bh_{\mk ,\eta}
\big],
\nonumber\\
\eeq  
where
\beq
E_{0}^{BN}&=&\frac{Vn^2(\bar{c}_0^{(2)}
+\bar{c}_2^{(2)})}{2}
+4qN
-\frac{1}{2}{\sk}^{'} \Bigg[
\left(  \ek +nc_0^{(2)}
+nc_2^{(2)} -E_{\mk ,d}\right) \nonumber\\
&&+2\left(  \ek -3q +nc_1^{(2)}
-nc_2^{(2)} -E_{\mk ,f_t}\right) \nonumber\\
&&+\left(  \ek +4nc_1^{(2)} 
-nc_2^{(2)} -E_{\mk ,f_z}\right) 
+\left( \ek -4q -nc_2^{(2)}
- E_{\mk ,\eta}\right) \Bigg]
\eeq
is the GSE, and the Bogoliubov spectra are given by
\begin{gather}
E_{\mk ,d}=\sqrt{\ek (\ek +2n( c_0^{(2)}
+c_2^{(2)}))},\label{eq:spin-2-bn-b1}\\
E_{\mk,f_t}=\sqrt{(\ek -3q ) (\ek-3q 
+2n( c_1^{(2)}-c_2^{(2)}))},\label{eq:spin-2-bn-b2}\\
E_{\mk,f_z}=\sqrt{\ek (\ek +2n(4c_1^{(2)}
-c_2^{(2)}))}, \label{eq:spin-2-bn-b3}\\
E_{\mk ,\eta}=\sqrt{(\ek -4q) (\ek-4q 
-2nc_2^{(2)})}. \label{eq:spin-2-bn-b4}
\end{gather}
The density and 
spin fluctuations around the $z$ axis are massless and 
those of nematic and spin fluctuations around the $x$ and
$y$ axes are massive.
This is because the symmetry of the Hamiltonian is reduced to 
$U(1)\times SO(2)$ due to  the external magnetic field and the fact
that the isotropy group of the biaxial nematic phase does not
include any continuous group.
The Bogoliubov spectra in Eqs. (\ref{eq:spin-2-bn-b2}) and
(\ref{eq:spin-2-bn-b4}) are positive semidefinite 
if $q<0$, which is consistent with the stability criterion 
of the mean-field ground state. 

The GSE $E_{0}^{BN}$, pressure, sound velocity, and quantum depletion
up to the LHY corrections are given as
follows:
\beq
\frac{E_0^{BN}}{V}
&=&
4qn+\frac{n^2(c_0^{(2)}+c_2^{(2)})}{2}
\left(1+
\frac{16\sqrt{M^3}}{15\pi^2\hbar^3}
\sqrt{n(c_0^{(2)}+c_2^{(2)})^3}
\right)
+\frac{8\sqrt{M^3}}{15\pi^2\hbar^3}
\Big\{(n|c_2^{(2)}|)^{\frac{5}{2}}\phi_1(t_5-\text{sgn}(c_2^{(2)}))\nonumber\\
&&+2(n|c_1^{(2)}-c_2^{(2)}|)^{\frac{5}{2}}
\phi_1(t_4+\text{sgn}(c_1^{(2)}-c_2^{(2)}))
+[n(4c_1^{(2)}-c_2^{(2)})]^{\frac{5}{2}}
\Big\}, 
\label{eq:spin-2-bn-gse}
\eeq
\beq
P&=&
\frac{n^2(c_0^{(2)}+c_2^{(2)})}{2}
\left(1+
\frac{8\sqrt{M^3}}{5\pi^2\hbar^3}
\sqrt{n(c_0^{(2)}+c_2^{(2)})^3}
\right)\nonumber\\
&&+\frac{4\sqrt{M^3}}{15\pi^2\hbar^3}
\Big\{(n|c_2^{(2)}|)^{\frac{5}{2}}
\Big[3\phi_1(t_5-\text{sgn}(c_2^{(2)}))
-2(t_5+1)\phi'_1(t_5-\text{sgn}(c_2^{(2)}))\Big]\nonumber\\
&&+2(n|c_1^{(2)}-c_2^{(2)}|)^{\frac{5}{2}}
\Big[ 3\phi_1(t_4+\text{sgn}(c_1^{(2)}-c_2^{(2)}))
-2(t_4+1)\phi'_1(t_4+\text{sgn}(c_1^{(2)}-c_2^{(2)})\Big]\nonumber\\
&&+3[n(4c_1^{(2)}-c_2^{(2)})]^{\frac{5}{2}}
\Big\},
\label{eq:spin-2-bn-p}
\eeq
\beq
c&=&
\sqrt{\frac{n(c_0^{(2)}+c_2^{(2)})}{M}}
\Bigg\{1+\frac{\sqrt{M^3}}{\pi^2\hbar^3}
\sqrt{n(c_0^{(2)}+c_2^{(2)})^3}
+\frac{\sqrt{M^3}}{15\pi^2\hbar^3}
\frac{\sqrt{n|c_2^{(2)}|^{5}}}
{c_0^{(2)}+c_2^{(2)}}
\phi_2(t_5,-\text{sgn}(c_2^{(2)}))
\nonumber\\
&&+\frac{2\sqrt{M^3}}{15\pi^2\hbar^3}
\frac{\sqrt{n|c_1^{(2)}-c_2^{(2)}|^{5}}}
{c_0^{(2)}+c_2^{(2)}}
\phi_2(t_4,\text{sgn}(c_1^{(2)}-c_2^{(2)}))
+\frac{\sqrt{M^3}}{\pi^2\hbar^3}
\frac{\sqrt{n|4c_1^{(2)}-c_2^{(2)}|^5}}
{c_0^{(2)}+c_2^{(2)}}
\Bigg\}, 
\label{eq:spin-2-bn-sv}
\eeq
\beq
\frac{N-N_0}{N}&=&
\frac{\sqrt{M^3}}{3\pi^2\hbar^3}
\Big[\sqrt{n(c_0^{(2)}+c_2^{(2)})^3}
+2\sqrt{n|c_1^{(2)}-c_2^{(2)}|^3}
+\sqrt{n(|c_2^{(2)}|)^3}\phi_3(t_5-\text{sgn}(c_2^{(2)}))\nonumber\\
&&+\phi_3(t_4+\text{sgn}(c_1^{(2)}-c_2^{(2)}))
+\sqrt{n(4c_1^{(2)}-c_2^{(2)})^3}
\Big], 
\label{eq:spin-2-bn-dep}
\eeq
where 
$\displaystyle t_4=-3q/(n|c_1^{(2)}-c_2^{(2)}|)-1$ and
$\displaystyle t_5=-4q/(n|c_2^{(2)}|)-1$.
For $q<0$, all the LHY corrections are positive definite, 
indicating the robustness of the biaxial nematic phase against
the beyond-Bogoliubov effect.

\subsubsection{Uniaxial nematic phase}
The effective Hamiltonian of the uniaxial nematic state in
Eq. \eqref{uncon} is obtained by
using the canonical transformations of Eqs. (\ref{d2})- (\ref{no}) as
\beq
\hat{H}_{\text{eff}}^{UN}&=&\frac{Vn^2(\bar{c}_0^{(2)}+\bar{c}_2^{(2)})}{2}
+{\sk}^{'}\Bigg\{(\ek+nc_0^{(2)}+nc_2^{(2)})\ahd_{\mk,d}\ah_{\mk,d}
+\frac{n(c_0^{(2)}+c_2^{(2)})}{2}
(\ah_{\mk,d}\ah_{\mmk,d}+\ahd_{\mk,d}\ahd_{\mmk,d})\nonumber\\
&&+\sum_{j=f_x,f_y}\Big[(\ek+q+3nc_1^{(2)}-nc_2^{(2)})
\ahd_{\mk,j}\ah_{\mk,j}
+\frac{n(3c_1^{(2)}-c_2^{(2)})}{2}
(\ah_{\mk,j}\ah_{\mmk,j}+
\ahd_{\mk,j}\ahd_{\mk,j})\Big]\nonumber\\
&&+\sum_{l=f_z,\eta}\Big[(\ek+4q-nc_2^{(2)})\ahd_{\mk,l}\ah_{\mk,l}
+\frac{nc_2^{(2)}}{2}
(\ah_{\mk,l}\ah_{\mmk,l}+\ahd_{\mk,l}\ahd_{\mmk,l})\Big]
\Bigg\}. 
\eeq
This can be diagonalized by means of
the following Bogoliubov transformations,
\begin{gather}
\bh_{\mk,d}=\sqrt{\frac{\ek+n(c_0^{(2)}
+c_2^{(2)})+E_{\mk,d}}{E_{\mk,d}}}\ah_{\mk,d}
+\sqrt{\frac{\ek+n(c_0^{(2)}
+c_2^{(2)})-E_{\mk,d}}{E_{\mk,d}}}\ahd_{\mmk,d},\\
\bh_{\mk,j}=\sqrt{\frac{\ek+q+n(3c_1^{(2)}
-c_2^{(2)} )
+E_{\mk,f_t}}{E_{\mk,f_t}}}\ah_{\mk,j}
+\sqrt{\frac{\ek+q+n(3c_1^{(2)}
-c_2^{(2)} )
-E_{\mk,f_t}}{E_{\mk,f_t}}}\ahd_{\mmk,j}, \\
\bh_{\mk,l}=\sqrt{\frac{\ek+4q-nc_2^{(2)}
+E_{\mk,\eta}}{E_{\mk,\eta}}}\ah_{\mk,l}
-\sqrt{\frac{\ek+4q-nc_2^{(2)}
-E_{\mk,\eta}}{E_{\mk,\eta}}}\ahd_{\mmk,l}, 
\end{gather}
with the result
\beq
\hat{H}_{\text{eff}}^{UN}=E_0^{UN}
+{\sk}^{'}\big[ E_{\mk ,d}\bhd_{\mk ,d}\bh_{\mk ,d}
+ E_{\mk ,f_t}(\bhd_{\mk ,f_x}\bh_{\mk ,f_x}
+\bhd_{\mk ,f_y}\bh_{\mk ,f_y})
+ E_{\mk ,\eta}(\bhd_{\mk ,f_z}\bh_{\mk ,f_z}
+\bhd_{\mk ,\eta}\bh_{\mk ,\eta})
\big],
\nonumber\\
\eeq  
where
\beq
E_0^{UN}&=&\frac{Vn^2(\bar{c}_0^{(2)}
+\bar{c}_2^{(2)})}{2}
-\frac{1}{2}{\sk}^{'} \Bigg[\left( \ek +nc_0^{(2)}
+nc_2^{(2)} - E_{\mk ,d}\right)\nonumber\\
&&+2\left(  \ek+q +3nc_0^{(2)}
-nc_2^{(2)} -E_{\mk ,f_t}\right)
+2\left(  \ek+4q -nc_2^{(2)} -E_{\mk ,\eta}\right) 
\Bigg]
\eeq
is the GSE, and the Bogoliubov spectra are given by
\begin{gather}
E_{\mk ,d}=\sqrt{\ek (\ek +2n(c_0^{(2)}
+c_2^{(2)}))}, \label{eq:spin-2-un-b1}\\
E_{\mk ,f_t}=\sqrt{(\ek +q) (\ek +q 
+2n( 3c_1^{(2)}-c_2^{(2)}))},\label{eq:spin-2-un-b2}\\
E_{\mk ,\eta}=\sqrt{(\ek +4q) (\ek +4q -2nc_2^{(2)})}.
\label{eq:spin-2-un-b3}
\end{gather}
The density fluctuation is massless,
while the nematic and spin fluctuations
around all the axes are massive. This is because the isotropy 
group of the uniaxial nematic phase includes the 
continuous group $SO(2)$ and, therefore, one massless
mode should appear.
As can be seen from Eqs. 
(\ref{eq:spin-2-un-b2}) and (\ref{eq:spin-2-un-b3}),
the Bogoliubov spectra are positive definite only if $q>0$, which is
consistent with the mean-field analysis.

The GSE $E_0^{UN}$, pressure, sound velocity, and quantum depletion 
are given by
\beq
\frac{E_0^{UN}}{V}&=&
\frac{n^2(c_0^{(2)}+c_2^{(2)})}{2}\left(
1+\frac{16\sqrt{M^3}}{15\pi^2\hbar^3}
\sqrt{n(c_0^{(2)}+c_2^{(2)})^3}\right)
+\frac{16\sqrt{M^3}}{15\pi^2\hbar^3}
\Big[(n|c_2^{(2)}|)^{\frac{5}{2}}\phi_1(t_7-\text{sgn}(c_2^{(2)}))\nonumber\\
&&+(n|3c_1^{(2)}-c_2^{(2)}|)^{\frac{5}{2}}
\phi_1(t_6+\text{sgn}(3c_1^{(2)}-c_2^{(2)}))
\Big],
\label{eq:spin-2-un-gse}
\eeq
\beq
P&=&
\frac{n^2(c_0^{(2)}+c_2^{(2)})}{2}
\left(1+
\frac{8\sqrt{M^3}}{5\pi^2\hbar^3}
\sqrt{n(c_0^{(2)}+c_2^{(2)})^3}
\right)\nonumber\\
&&+\frac{8\sqrt{M^3}}{15\pi^2\hbar^3}
\Big\{(n|c_2^{(2)}|)^{\frac{5}{2}}
\Big[3\phi_1(t_7-\text{sgn}(c_2^{(2)}))
-2(t_7+1)\phi'_1(t_7-\text{sgn}(c_2^{(2)}))\Big]\nonumber\\
&&+(n|3c_1^{(2)}-c_2^{(2)}|)^{\frac{5}{2}}
\Big[3\phi_1(t_6+\text{sgn}(3c_1^{(2)}-c_2^{(2)}))
-2(t_6+1)\phi'_1(t_6+\text{sgn}(3c_1^{(2)}-c_2^{(2)}))\Big]
\Big\},
\nonumber\\
\label{eq:spin-2-un-p}
\eeq
\beq
c&=&
\sqrt{\frac{n(c_0^{(2)}+c_2^{(2)})}{M}}
\Bigg[1+\frac{\sqrt{M^3}}{\pi^2\hbar^3}
\sqrt{n(c_0^{(2)}+c_2^{(2)})^3}
+\frac{2\sqrt{M^3}}{15\pi^2\hbar^3}
\Bigg(\frac{\sqrt{n|c_2^{(2)}|^5}}
{c_0^{(2)}+c_2^{(2)}}
\phi_2(t_7,-\text{sgn}(c_2^{(2)}))\nonumber\\
&&+\frac{\sqrt{n|3c_1^{(2)}-c_2^{(2)}|^5}}
{c_0^{(2)}+c_2^{(2)}}
\phi_2(t_6,\text{sgn}(3c_1^{(2)}-c_2^{(2)}))
\Bigg)
\Bigg], 
\label{eq:spin-2-un-sv}
\eeq
\beq
\frac{N-N_0}{N}=
\frac{\sqrt{M^3}}{3\pi^2\hbar^3}
\Big[\sqrt{n(c_0^{(2)}+c_2^{(2)})^3}
+2\sqrt{n(|c_2^{(2)}|)^3}\phi_3(t_7-\text{sgn}(c_2^{(2)}))\nonumber\\
+2\sqrt{n(|3c_1^{(2)}-c_2^{(2)}|)^3}
\phi_3(t_6+\text{sgn}(3c_1^{(2)}-c_2^{(2)}))
\Big], 
\label{eq:spin-2-un-dep}
\eeq
where $t_6=q/(n|3c_1^{(2)}-c_2^{(2)}|)-1$, $t_7=q/(n|c_2^{(2)}|)-1$.
The LHY corrections become imaginary for $q<0$, which
implies that the system undergoes the dynamical instability, 
and that the uniaxial nematic phase is only stable for $q>0$.

\subsection{Cyclic phase}
In addition to the
stable configuration of Eq. (\ref{qcyclic}), 
we also examine the tetrahedral
configuration of Eq. (\ref{cyclic}), even though the latter configuration
is not a stationary solution of the mean-field theory for nonzero $q$.
This is because
the tetrahedral configuration has attracted considerable attention, since it
gives rise to nontrivial phenomena such as
the non-Abelian vortices \cite{zhou,makela,kobayashi}.

\subsubsection{Stable configuration for nonzero $q$}
\begin{figure}[!h]
\subfigure{ 
  \includegraphics[width=0.40\linewidth]{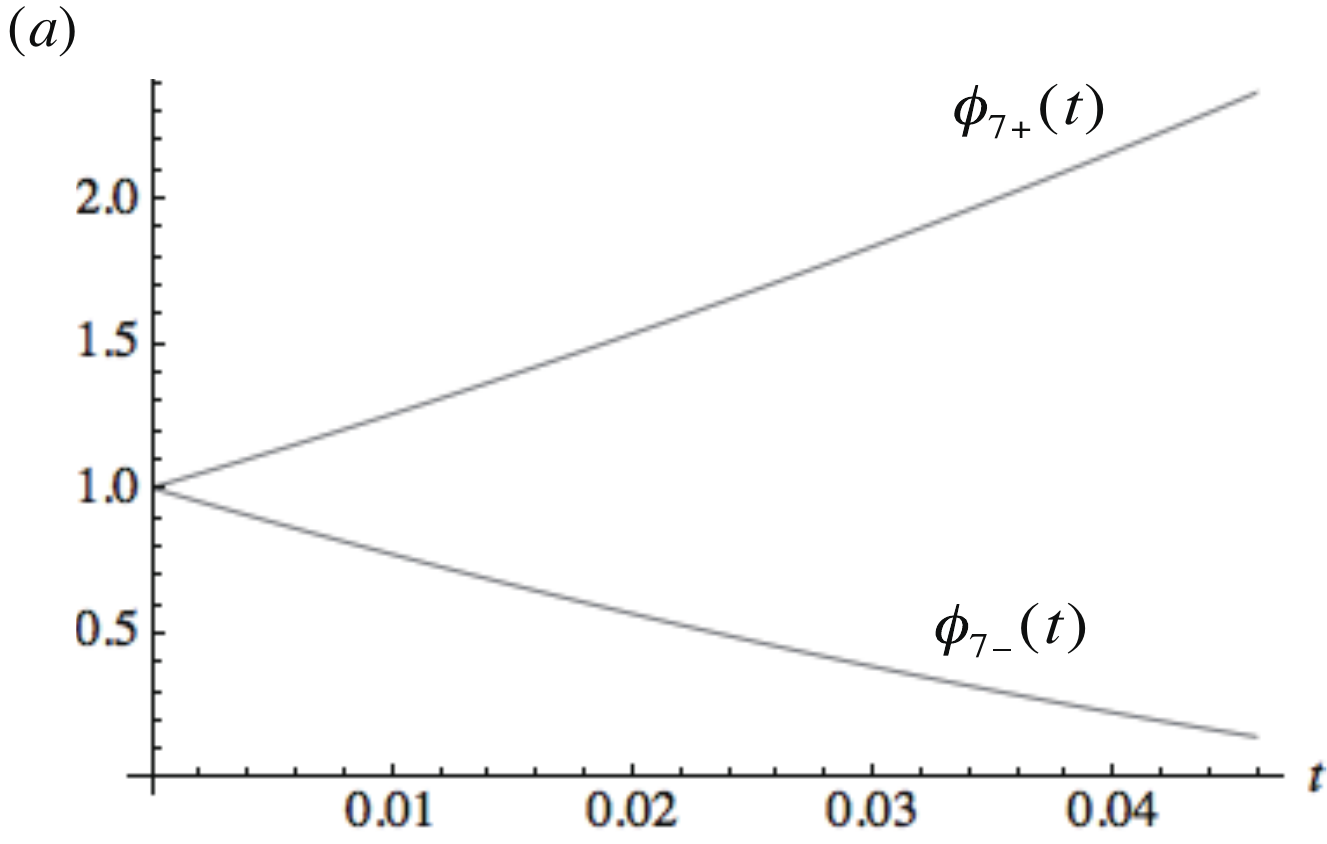}
 \label{phi9}
}
\subfigure{
 \includegraphics[width=0.40\linewidth]{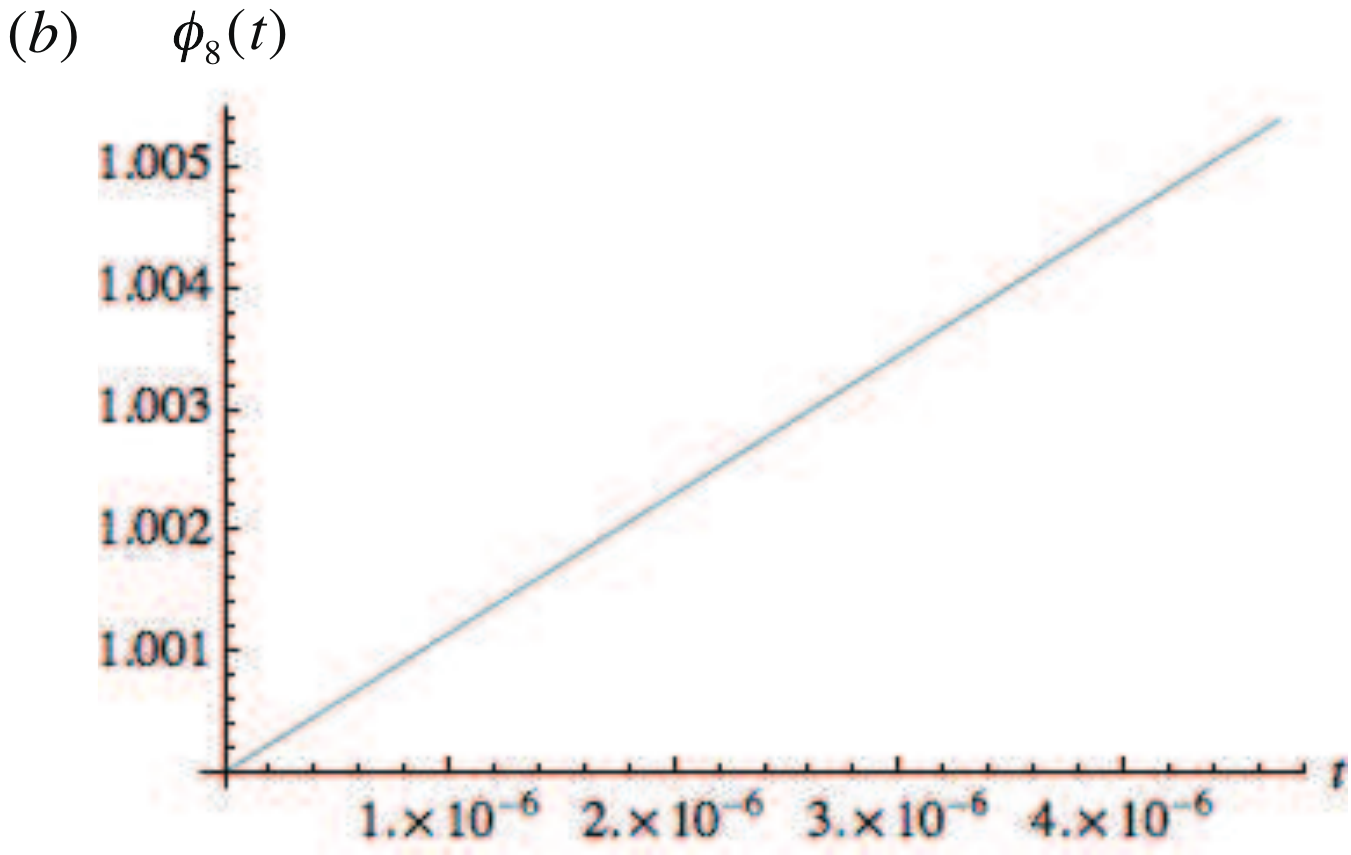}
 \label{phi10}
}
  \caption{Plots of (a) $\phi_{7\pm}(t)$ and (b) $\phi_{8}(t)$.
These functions are approximated as
$\phi_{7\pm}(t)\simeq 1\pm 24.1t+116t^2$
and $\phi_8(t)\simeq 1+1140t$. }
\end{figure}

For the stable cyclic phase in Eq. (\ref{qcyclic}), we
consider the following canonical transformations:
\beq
\ah_{\mk ,d}&=&\frac{\sin\theta}{\sqrt{2}}(\ah_{\mk ,2}-\ah_{\mk ,-2})+
\cos\theta\ah_{\mk ,0},\\
\ah_{\mk ,f_x}&=&\frac{\sin\theta+\sqrt{3}\cos\theta}
{\sqrt{2\sin^2\theta+6\cos^2\theta}}\ah_{\mk ,1}
+\frac{-\sin\theta+\sqrt{3}\cos\theta}
{\sqrt{2\sin^2\theta+6\cos^2\theta}}\ah_{\mk ,-1},\\
\ah_{\mk ,f_y}&=&i\frac{-\sin\theta+\sqrt{3}\cos\theta}
{\sqrt{2\sin^2\theta+6\cos^2\theta}}\ah_{\mk ,1}
-i\frac{\sin\theta+\sqrt{3}\cos\theta}
{\sqrt{2\sin^2\theta+6\cos^2\theta}}\ah_{\mk ,-1},\\
\ah_{\mk ,f_z}&=&\frac{1}{\sqrt{2}}(\ah_{\mk ,2}+\ah_{\mk ,-2}),\\
\ah_{\mk ,\theta}&=&\frac{\cos\theta}{\sqrt{2}}(-\ah_{\mk ,2}+\ah_{\mk
,-2})
-\cos\theta\ah_{\mk ,0},
\eeq
where $\ah_{\mk,d}$, $\ah_{\mk,f_x}$, $\ah_{\mk,f_y}$, and $\ah_{\mk,f_z}$
represent the density and spin fluctuations, 
and $\ah_{\mk,\theta}$ is the operator that commute with
the other four operators.
Then, the Hamiltonian is given by
\beq
\hat{H}_{\text{eff}}^{C}&=&2qN+\frac{Vn^2\bar{c}_0^{(2)}}{2}
-\frac{2Vq^2}{\bar{c}_{2}^{(2)}}
+{\sk}^{'}\Bigg\{\left[\ek-q+2nc_1^{(2)}
\left(1+\frac{q}{nc_2^{(2)}}\right)\right]
(\ahd_{\mk,f_x}\ah_{\mk,f_x}+\ahd_{\mk,f_y}\ah_{\mk,f_y})\nonumber\\
&&+\left[nc_1^{(2)}\left(1+\frac{q}{nc_2^{(2)}}\right)
-q\left(\frac{-\sin^2\theta+3\cos^2\theta}{\sin^2\theta+3\cos^2\theta}\right)
\right]\nonumber\\
&&\times(\ah_{\mk,f_x}\ah_{\mmk,f_x}+\ah_{\mk,f_y}\ah_{\mmk,f_y}
+\ahd_{\mk,f_x}\ahd_{\mmk,f_x}+\ahd_{\mk,f_y}\ahd_{\mmk,f_y})\nonumber\\
&&-\frac{2\sqrt{3}q\sin 2\theta}{\sin^2\theta+3\cos^2\theta}
(\ah_{\mk,f_x}\ah_{\mmk,f_y}-\ahd_{\mk,f_x}\ahd_{\mmk,f_y})
+\left(\ek+2nc_1^{(2)}+2q-\frac{4qc_1^{(2)}}{c_2^{(2)}}\right)
\ahd_{\mk,f_z}\ah_{\mk,f_z}\nonumber\\
&&+\frac{1}{2}\left(2nc_1^{(2)}+2q-
\frac{4qc_1^{(2)}}{c_2^{(2)}}\right)
(\ah_{\mk,f_z}\ah_{\mmk,f_z}+\ahd_{\mk,f_z}\ahd_{\mmk,f_z})\nonumber\\
&&+\left(\ek+nc_0^{(2)}+nc_q^{(2)}\right)
\ahd_{\mk,d}\ah_{\mk,d}+\frac{n(c_0^{(2)}
+c_q^{(2)})}{2}(\ah_{\mk,d}\ah_{\mmk,d}
+\ahd_{\mk,d}\ahd_{\mk,d})\nonumber\\
&&+\left(\ek+nc_0^{(2)}-nc_q^{(2)}\right)
\ahd_{\mk,\theta}\ah_{\mk,\theta}
-\frac{nc_q^{(2)}}{2}(\ah_{\mk,\theta}\ah_{\mmk,\theta}
+\ahd_{\mk,\theta}\ahd_{\mk,\theta})\nonumber\\
&&-2q\sin 2\theta
(\ahd_{\mk,d}\ah_{\mk,\theta}+\ahd_{\mk,\theta}\ah_{\mk,\theta}
+\ah_{\mk,d}\ah_{\mmk,\theta}+\ahd_{\mk,d}\ahd_{\mmk,\theta})
\Bigg\},
\eeq
where 
$c_q^{(2)}\equiv 4q^2/(n^2c_2^{(2)})$.
For nonzero $q$, the above Hamiltonian is not decomposed into
sub-Hamiltonians completely because all modes except for
$\ah_{\mk,f_z}$ couple.
For the $f_z$ mode, we consider the following Bogoliubov 
transformation:
\beq
\bh_{\mk,f_z}=\sqrt{\frac{\ek+2nc_1^{(2)}+2q
-4c_1^{(2)}q/c_2^{(2)}+E_{\mk,f_z}}{2E_{\mk,f_z}}}\ah_{\mk,f_z}
+\sqrt{\frac{\ek+2nc_1^{(2)}+2q
-4c_1^{(2)}q/c_2^{(2)}-E_{\mk,f_z}}{2E_{\mk,f_z}}}\ahd_{\mmk,f_z},
\nonumber\\
\eeq
where
\beq
E_{\mk,f_z}=\sqrt{\ek\left(\ek+4nc_1^{(2)}+4q
-8c_1^{(2)}q/c_2^{(2)}\right)}
\label{eq:spin-2-c-b1}
\eeq
is the Bogoliubov spectrum.
For the $f_x$ and $f_y$ modes, we consider the following Bogoliubov
transformations:
\beq
\bh_{\mk,f_{t1}}&=&\frac{1}{2}
\Bigg[\sqrt{\frac{\ek-q+2nc_1^{(2)}+2c_1^{(2)}q/c_2^{(2)}
+E_{\mk,f_t}}
{E_{\mk,f_t}}}
(\ah_{\mk,f_x}+
\ah_{\mk,f_y})\nonumber\\
&&+
e^{i\tau}\sqrt{\frac{\ek-q+2nc_1^{(2)}+2c_1^{(2)}q/c_2^{(2)}
-E_{\mk,f_t}}
{E_{\mk,f_t}}}
(\ahd_{\mmk,f_x}+\ahd_{\mmk,f_y})
\Bigg],
\eeq
\beq
\bh_{\mk,f_{t2}}&=&\frac{1}{2}\Bigg[
\sqrt{\frac{\ek-q+2nc_1^{(2)}+2c_1^{(2)}q/c_2^{(2)}
+E_{\mk,f_t}}
{E_{\mk,f_t}}}
(\ah_{\mk,f_x}-
\ah_{\mk,f_y})\nonumber\\
&&+
e^{-i\tau}\sqrt{\frac{\ek-q+2nc_1^{(2)}+2c_1^{(2)}q/c_2^{(2)}
-E_{\mk,f_t}}
{E_{\mk,f_t}}}
(\ahd_{\mmk,f_x}-\ahd_{\mmk,f_y})
\Bigg],
\eeq
\beq
\beta=nc_1^{(2)}\left(1+\frac{q}{nc_2^{(2)}}\right)-q\left(
\frac{-\sin^2\theta+3\cos^2\theta+2\sqrt{3}i\sin\theta\cos\theta
}{\sin^2\theta+3\cos^2\theta}
\right),
\eeq
where $\tau\equiv -\text{Im}\beta/\text{Re}\beta$ and
\beq
E_{\mk,f_t}=\sqrt{3n^2c_1^{(2)}c_q^{(2)}-3q^2
+\ek\left(\ek+4nc_1^{(2)}+4nc_1^{(2)}q/c_2^{(2)}-2q\right)},
\label{eq:spin-2-c-b2}
\eeq
is the doubly degenerate Bogoliubov spectrum.
Finally, for the density and $\theta$ modes, we consider the following
transformations:
\beq
\hat{\mathbf{B}}_{\mk }=U(k)\hat{\mathbf{A}}_{\mk,d\theta }
+V(k)\hat{\mathbf{A}^{\dagger}}_{\mmk,d\theta },
\eeq
where
$
\hat{\mathbf{B}}_{\mk}= ^{t}(\bh_{\mk,-},\bh_{\mk,+}),\ 
\hat{\mathbf{A}}_{\mk,d\theta}= ^{t}(\ah_{\mk,d},\ah_{\mk,\theta}),$
\beq 
U(k)=\frac{1}{2}
\left( 
\begin{array}{cc}
\frac{1}{2C_{2-}(k)E_2(k)}
+C_{2-}(k)X_{2+}(k)
 & 4q\ek C_{2-}(k)\sin 2\theta -\frac{X_{2-}(k)}{8q\ek C_{2-}(k)
E_2(k)\sin 2\theta}
 \\
-\frac{1}{2C_{2+}(k)E_2(k)}
+C_{2+}(k)X_{2-}(k)
 &   4q\ek C_{2+}(k)\sin 2\theta +\frac{X_{2+}(k)}{8q\ek C_{2+}(k)
E_2(k)\sin 2\theta}
\end{array} 
\right) ,
\eeq
\beq 
V(k)=\frac{1}{2}
\left( 
\begin{array}{cc}
-\frac{1}{2C_{2-}(k)E_2(k)}
+C_{2-}(k)X_{2+}(k)
 & 4q\ek C_{2-}(k)\sin 2\theta +\frac{X_{2-}(k)}{8q\ek C_{2-}(k)
E_2(k)\sin 2\theta}
 \\
\frac{1}{2C_{2+}(k)E_2(k)}
+C_{2+}(k)X_{2-}(k)
 &   4q\ek C_{2+}(k)\sin 2\theta -\frac{X_{2+}(k)}{8q\ek C_{2+}(k)
E_2(k)\sin 2\theta}
\end{array} 
\right) ,
\eeq
\begin{gather}
X_{2\pm}(k)=-n(c_0^{(2)}-2c_2^{(2)}+2c_q^{(2)})\ek
+2n^2c_2^{(2)}(c_2^{(2)}-c_q^{(2)})\pm E_2(k),\\
C_{2\pm}(k)=\sqrt{\frac{E_{\mk,\pm}}{X^2_{2\mp}(k)\ek
+16q^2\sin^2 2\theta\ek^2(\ek+2nc_2^{(2)})
}},
\end{gather}
\beq
E_2(k)=\Bigg\{\{ (nc_0^{(2)}-2nc_2^{(2)})^2
+4n^2c_q^{(2)}(c_0^{(2)}-c_2^{(2)})
\}\ek^2-4n^3c_2^{(2)}
(c_0^{(2)}-2c_2^{(2)}
)(c_2^{(2)}-c_q^{(2)}
)\ek\nonumber\\
+\{2n^2c_2^{(2)}(c_2^{(2)}-c_q^{(2)})
\}^2\Bigg\}^{1/2},
\eeq
and $E_{\mk,\pm}$ are the Bogoliubov spectra given by
\beq
E_{\mk,\pm}=\sqrt{\ek\left(\ek+nc_0^{(2)}+2nc_2^{(2)}\right)
+2n^2c_2^{(2)}\left(c_2^{(2)}
-c_q^{(2)}\right)\pm E_2(k)}.
\label{eq:spin-2-c-b3}
\eeq
Using the above transformations, the total Hamiltonian is
diagonalized as follows:
\beq
\hat{H}_{\text{eff}}^C=E_0^C+{\sk}^{'}\Big[E_{\mk,f_t}
(\bhd_{\mk,f_{t1}}\bh_{\mk,f_{t1}}+\bhd_{\mk,f_{t2}}\bh_{\mk,f_{t2}}
)+E_{\mk,f_z}\bhd_{\mk,f_z}\bh_{\mk,f_z}
+E_{\mk,+}\bhd_{\mk,+}\bh_{\mk,+}
+E_{\mk,-}\bhd_{\mk,-}\bh_{\mk,-}
\Big],
\nonumber\\
\eeq
where
\beq 
E_0^C&=&2qN+\frac{Vn^2\bar{c}_0^{(2)}}{2}-\frac{2Vq^2}{\bar{c}_2^{(2)}}
-\frac{1}{2}{\sk}^{'}\Bigg[2\left(\ek+2nc_1^{(2)}
+2c_1^{(2)}q/c_2^{(2)}
-q-E_{\mk,f_t}\right)\nonumber\\
&&+\left(\ek+2nc_1^{(2)}+2q-4c_1^{(2)}q/c_2^{(2)}
-E_{\mk,f_z}
\right)
+\left(2\ek+nc_0^{(2)}+2nc_2^{(2)}-E_{\mk,+}-E_{\mk,-}
\right)
\Bigg].
\nonumber\\
\eeq
In the long-wavelength limit,
$E_{\mk,-}^2\simeq 2nc_0^{(2)}\ek$;
therefore, $E_{\mk,-}$ is linear and massless. 
This is the NG mode associated with  
the spontaneous breaking of the $U(1)$ gauge symmetry.
The spectrum on the 
spin fluctuations around the $z$ axis is also linear and massless 
since the $SO(2)$
symmetry is spontaneously broken at the mean-field level.
The spectra around the transverse axes are massive because
the rotational symmetry about transverse directions
are explicitly broken by an external magnetic field.
We note that
this cyclic configuration is stable if $c_0^{(2)}>c_2^{(2)}$,
$c_2^{(2)}<4c_1^{(2)}$,
$c_1^{(2)}>|q|/(2n)$, and $c_2^{(2)}>2|q|/n$.
It is robust regardless the sign of $q$, which is 
consistent with the stability condition of the mean-field ground state.  
Using the renormalization procedure discussed in Appendix A, we find that
the GSE per  volume $V$ is given by
\beq
\frac{E_0^{C}}{V}=
2qn+\frac{n^2c_0^{(2)}}{2}-\frac{2q^2}{c_2^{(2)}}
+\frac{8\sqrt{M^3}}{15\pi^2\hbar^3}\Bigg[(
2nc_1^{(2)}+2q-4c_1^{(2)}q/c_2^{(2)})^{\frac{5}{2}}
+2(2nc_1^{(2)})^{\frac{5}{2}}\phi_{7\pm}(t_8)\nonumber\\
+(nc_0^{(2)})^{\frac{5}{2}}\phi_{8}(t_9)
\Bigg], 
\label{eq:spin-2-c-gse}
\eeq
where the $+$ $(-)$ sign in $\phi_{7\pm}(t)$ 
corresponds to the case of $q>0$ $(q<0)$,
\beq
\phi_{7\pm}(t)\equiv -\frac{15}{8\sqrt{2}}\int_0^{\infty} dx
x^2\left(x^2+z_{1\pm}(t)-\sqrt{z_2(t)+x^2(x^2+2z_{1\pm}(t))}
-\frac{z_{1\pm}^2(t)-z_2(t)}{2x^2}\right),\nonumber\\
\eeq
\beq
\phi_{8}(t)\equiv -\frac{15}{8\sqrt{2}}\int_0^{\infty} dx
x^2\left(2x^2+z_3-\phi_{8}^{(+)}(t)-\phi_{8}^{(-)}(t)-\frac{1}{2x^2}
-\frac{z_4(t)}{x^2}
\right),
\eeq
\beq
\phi_{8}^{(\pm)}(t)\equiv \sqrt{x^4+z_3x^2+z_5(t)\pm\sqrt{z_6(t)x^4
+z_7(t)x^2+z_5^2(t)}},
\eeq
with
$t_8=|q|/2nc_1^{(2)}$, $t_9=(2q)^2/(nc_0^{(2)})^2$,
$z_{1\pm}(t)=1\mp t(1-2c_1^{(2)}/c_2^{(2)})$,
$z_2(t)=t^2(12c_1^{(2)}/c_2^{(2)}-3)$,
$z_3=(c_0^{(2)}+2c_2^{(2)})/c_0^{(2)}$,
$z_4(t)=t(1+c_0^{(2)}/c_2^{(2)})$,
$z_5(t)=2(c_2^{(2)}/c_0^{(2)})^2-2t$,
$z_6(t)=(1-2c_2^{(2)}/c_0^{(2)})^2+4t(c_0^{(2)}/c_2^{(2)}-1)$,
and
$z_7(t)=-4(c_2^{(2)}/c_0^{(2)})(c_0^{(2)}-2c_2^{(2)})[c_2^{(2)}/(c_0^{(2)})^2
-t/c_2^{(2)}]$.
The plots of $\phi_{7\pm}(t)$ and $\phi_{8}(t)$  
in Fig. \ref{phi9} show
that the spin components of the GSE around the $x$ and $y$ axes for $q<0$ 
(for $q>0$) 
decrease (increase), while the density component of the GSE increases
as the external magnetic field increases. 
The typical values of $t_8$ and $t_9$ 
are of the order of $10^{-2}$ and $10^{-6}$, respectively,
for the parameters of $^{87}$Rb. 
\begin{figure}[!h]
\subfigure{
  \includegraphics[width=0.4\linewidth]{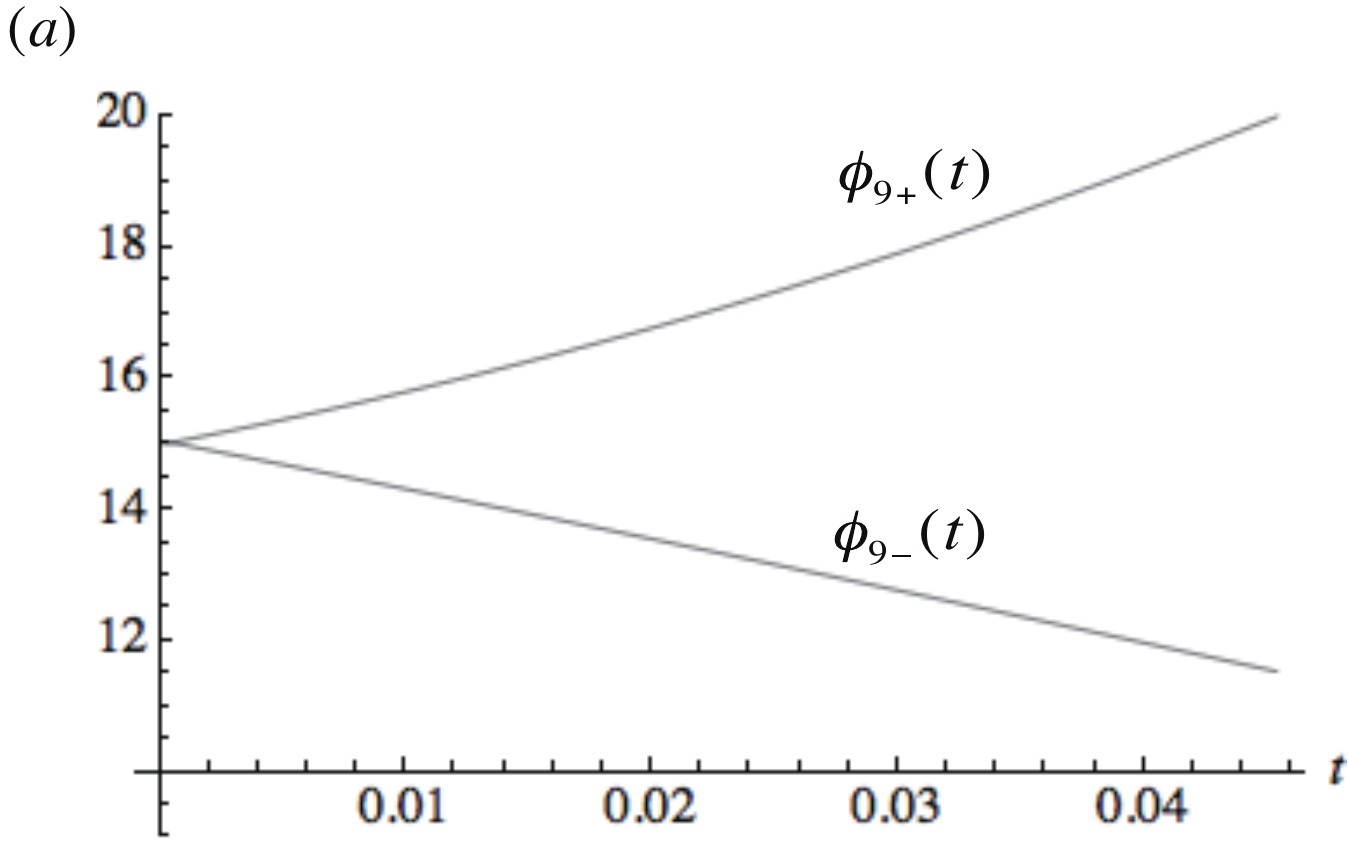}
}
\subfigure{
   \includegraphics[width=0.4\linewidth]{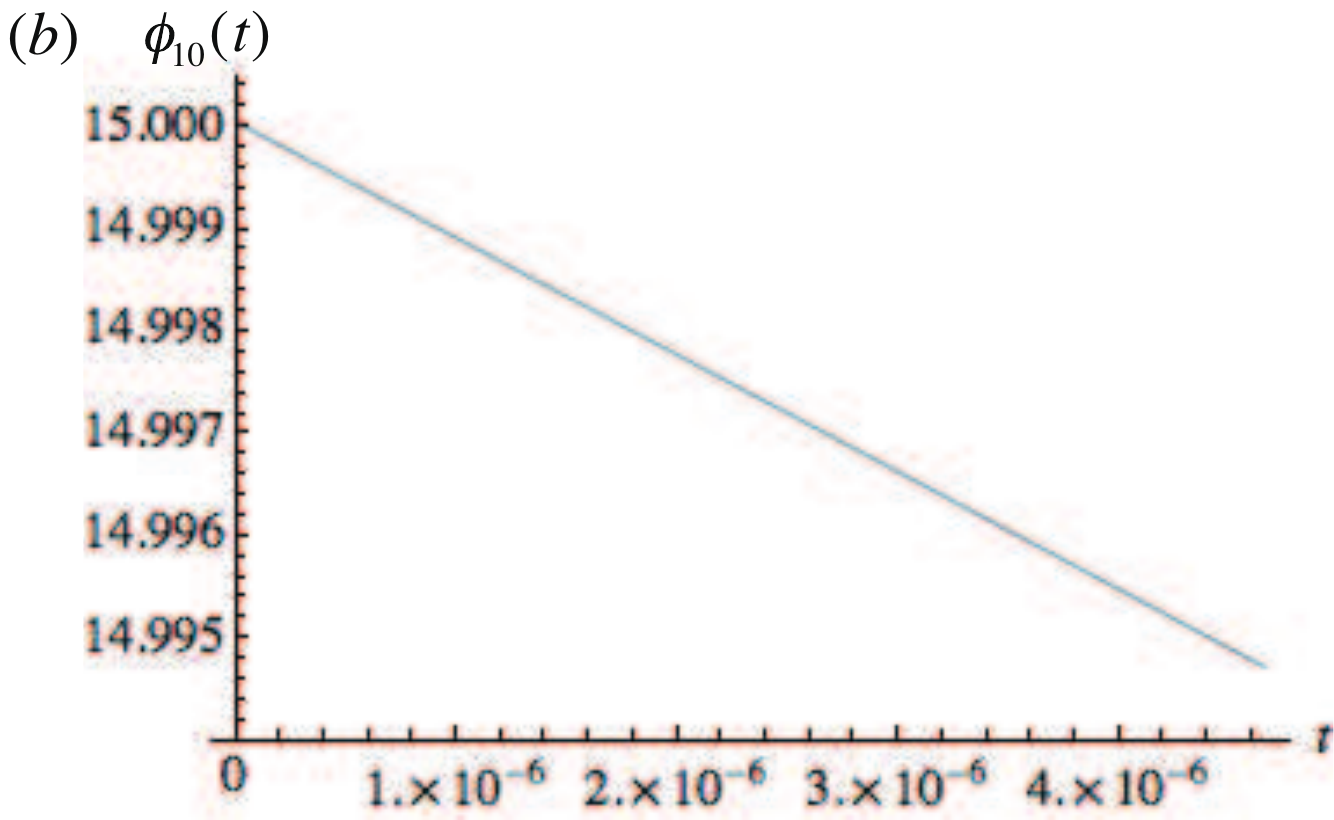}
}
  \caption{Plots of $\phi_{9\pm}(t)$ (a) and $\phi_{10}(t)$ (b).}
  \label{phi13}
\end{figure}
\begin{figure}[!h]
\subfigure{
  \includegraphics[width=0.4\linewidth]{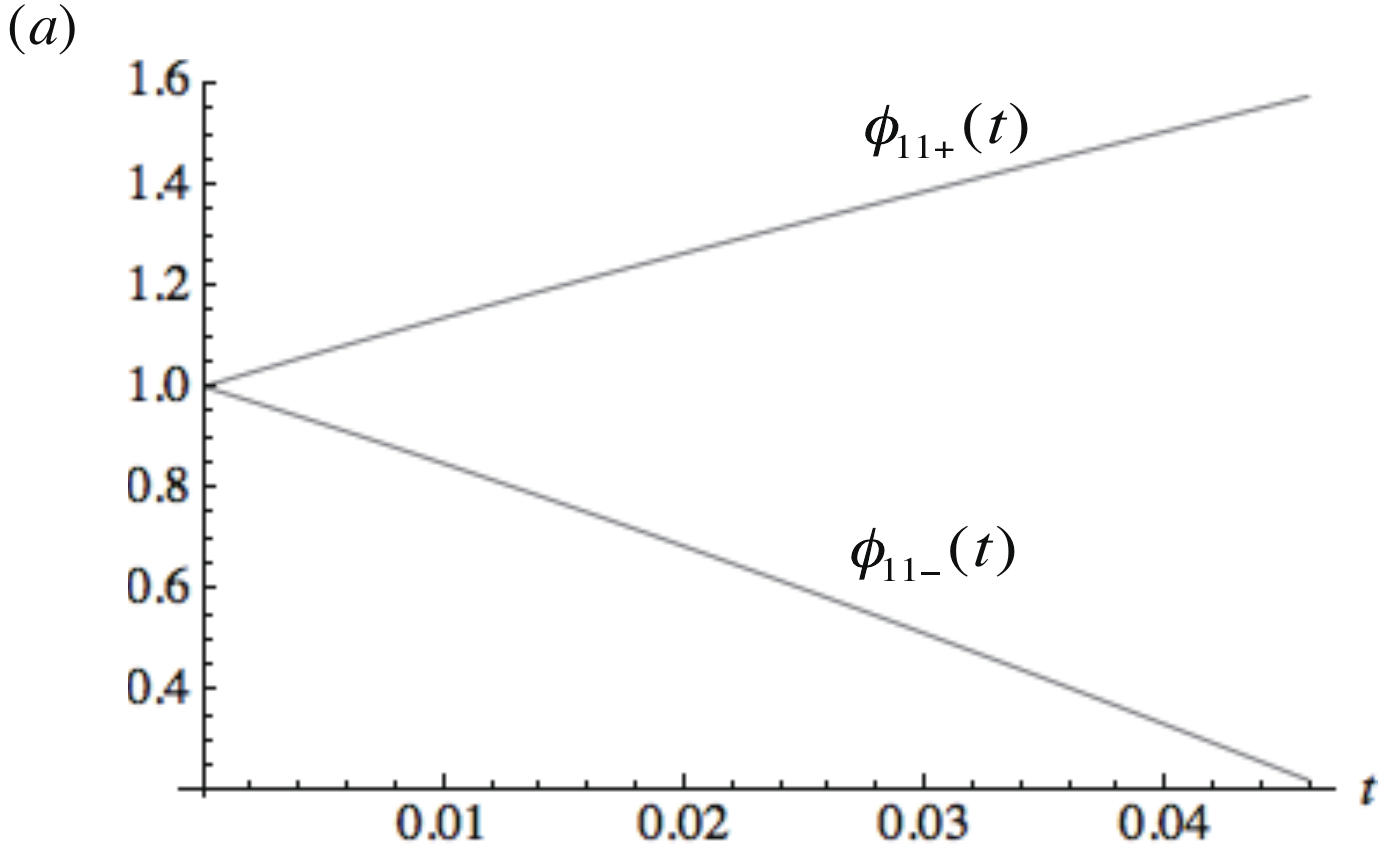}
} 
\subfigure{
 \includegraphics[width=0.4\linewidth]{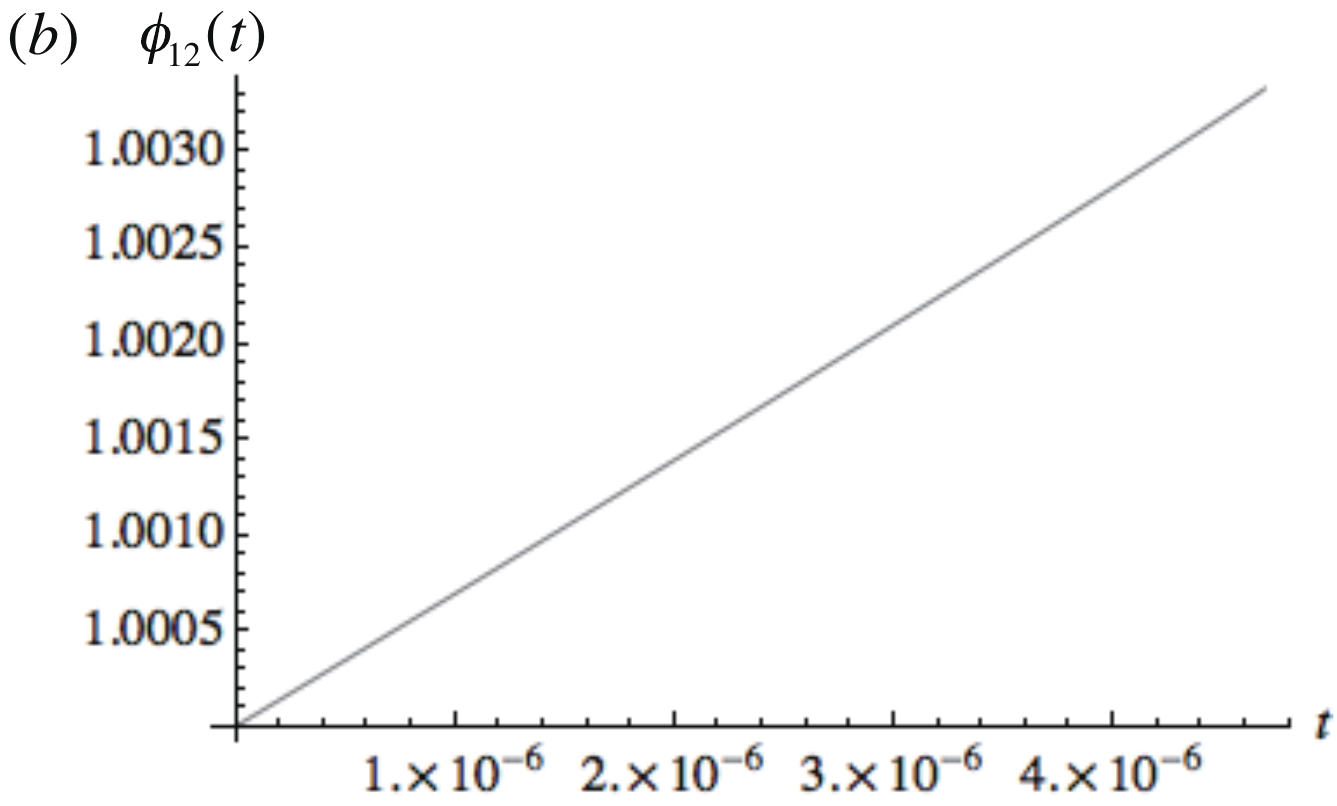}
}
 \caption{Plots of (a) $\phi_{11\pm}(t)$ and (b) $\phi_{12}(t)$.
These functions are approximated as $\phi_{11-}(t)\simeq 1-15.1t-40t^2$,
$\phi_{11+}(t)\simeq 1+13.6t-26t^2$,
  and $\phi_{12}(t)\simeq 1+700t$.}
  \label{phi15}
\end{figure}
Performing the derivatives of the GSE with respect to $V$ and $n$,
we obtain the pressure and sound velocity as follows: 
\beq
P=
\frac{n^2c_0^{(2)}}{2}+\frac{2q^2}{c_2^{(2)}}
+\frac{4\sqrt{M^3}}{15\pi^2\hbar^3}\Bigg[
(nc_0^{(2)})^{\frac{5}{2}}(3\phi_{8}(t_9)-4t_9\phi_{8}'(t_9))
+(2nc_1^{(2)})^{\frac{5}{2}}(6\phi_{7\pm}(t_8)-4t_8\phi_{7\pm}'(t_8))
\nonumber\\
+
\frac{3nc_1^{(2)}-2q+4c_1^{(2)}q/c_2^{(2)}}{nc_1^{(2)}
+q-2c_1^{(2)}q/c_2^{(2)}}
(2nc_1^{(2)}+2q-4c_1^{(2)}q/c_2^{(2)})^{\frac{5}{2}}
\Bigg],
\nonumber\\
\label{eq:spin-2-c-p}
\eeq
\beq
c&=&
\sqrt{\frac{nc_0^{(2)}}{M}}\Bigg[1+
\frac{\sqrt{M^3}}{15\pi^2\hbar^3}\sqrt{n(c_0^{(2)})^3}\phi_{10}(t_9)
+\frac{4\sqrt{M^3}}{15\pi^2\hbar^3}
\frac{c_1^{(2)}}{c_0^{(2)}}\sqrt{n(2nc_1^{(2)})^3}
\phi_{9\pm}(t_8)\nonumber\\
&&+\frac{4\sqrt{M^3}}{\pi^2\hbar^3}\frac{c_1^{(2)}}{c_0^{(2)}}
\frac{nc_1^{(2)}}{2nc_1^{(2)}
+2q-4c_1^{(2)}q/c_2^{(2)}}
\sqrt{n\left(2c_1^{(2)}+2q/n-4c_1^{(2)}q/(nc_2^{(2)})\right)^3}
\Bigg], 
\label{eq:spin-2-c-sv}
\eeq
where
\beq
\phi_{9\pm}(t)\equiv 15\phi_{7\pm}(t)
-12t\phi_{7\pm}'(t)+4t^2\phi_{7\pm}''(t),
\eeq
\beq
\phi_{10}(t)\equiv 15\phi_{8}(t)-16t\phi_{8}'(t)
+16t^2\phi_8''(t).
\eeq
The behaviors of $\phi_{9\pm}(t)$ and $\phi_{10}(t)$ plotted  in
Fig. \ref{phi13} show that for $q<0$ the sound velocities with respect to
the density and spin wave around the $x$ and $y$ axes decrease, 
while for $q>0$ those with respect to
the spin around the $x$ and $y$ axes increases,
as the quadratic Zeeman effect becomes stronger.
The quantum depletion is expressed as
\beq
\frac{N-N_0}{N}
=
\frac{\sqrt{M^3}}{3\pi^2\hbar^3}
\Bigg[\sqrt{n\left(2q/n+2c_1^{(2)}-4c_1^{(2)}q/(nc_2^{(2)})\right)^3}
+2\sqrt{n(2c_1^{(2)})^3}\phi_{11\pm}(t_8)\nonumber\\
+\sqrt{n(c_0^{(2)})^3}\phi_{12}(t_9)
\Bigg],
\label{eq:spin-2-c-dep}
\eeq
where the $+$ $(-)$ sign in $\phi_{11\pm}$ corresponds to the case
of $q>0$ $(q<0)$, and
\beq
\phi_{11\pm}(t)\equiv\frac{3}{\sqrt{2}}\int dxx^2\left(
\frac{x^2+1\mp t(1-2c_1^{(2)}/c_2^{(2)})}{
\sqrt{3t^2(4c_1^{(2)}/c_2^{(2)}-1)+x^2(x^2+2\mp 
2t(1-2c_1^{(2)}/c_2^{(2)}))}}
-1\right),
\eeq
\beq
\phi_{12}(t)\equiv\frac{6}{\sqrt{2}}\int_{0}^{\infty}dxx^2
(V_{11}^2+V_{12}^2+V_{21}^2+V_{22}^2).
\eeq
The behaviors of $\phi_{11\pm}(t)$ and $\phi_{12}(t)$ plotted
in Fig. \ref{phi15} show that
the quantum depletion from the density fluctuations 
increases, while that from the spin fluctuations 
around the $x$ and $y$ axes for $q<0$
(for $q>0$)
decreases (increases), 
as the external magnetic field increases.
The variations of the quantum corrections with respect to the density
component are small since the changes in $\phi_8(t)$,
$\phi_{10}(t)$, and $\phi_{12}(t)$ are of the order of $10^{-3}$.
On the other hand, the variations with respect to the spin components 
are of the order of 1. 
Considering the fact that $c_0^{(2)}\gg c_1^{(2)}$,
$c_2^{(2)}$, and $|q|/n$ for the alkali species, however,
the main contribution in the LHY corrections  stems from the
density fluctuations as in the cases of the other phases.

\subsubsection{Tetrahedral configuration}
\begin{figure}[!h]
 \begin{center}
  \includegraphics[width=0.4\linewidth]{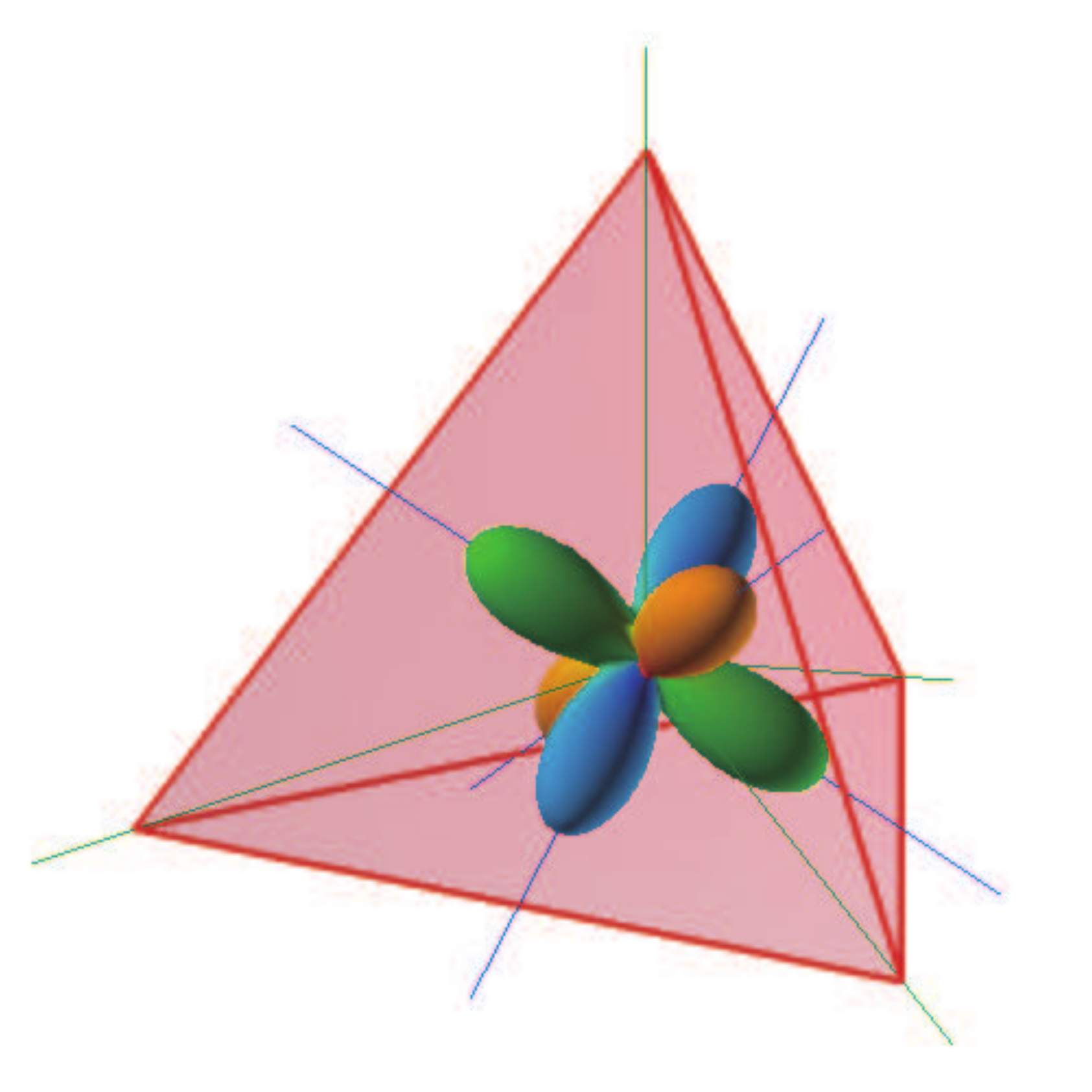}
  \caption{(Color online) Schematic illustration showing that the 
symmetry axes of the cyclic order parameter constitute a tetrahedron,
  where the color shows the phase of the order parameter as indicated in
the color gauge of Fig. 7.}
  \label{tetra}
 \end{center}
\end{figure}
Finally, we examine the mean-field configuration of
Eq. (\ref{cyclic}), which has the tetrahedral symmetry as shown
in Fig. \ref{tetra}.
We consider the following canonical transformations:
\beq
\ah_{\mk ,d}&=&\frac{1}{2}(\ah_{\mk ,2}-\ah_{\mk ,-2})+
\frac{1}{\sqrt{2}}\ah_{\mk ,0},\label{cdensity}\\
\ah_{\mk ,f_x}&=&\frac{1+\sqrt{3}}{2\sqrt{2}}\ah_{\mk ,1}
+\frac{-1+\sqrt{3}}{2\sqrt{2}}\ah_{\mk ,-1},\\
\ah_{\mk ,f_y}&=&i\frac{-1+\sqrt{3}}{2\sqrt{2}}\ah_{\mk ,1}
-i\frac{1+\sqrt{3}}{2\sqrt{2}}\ah_{\mk ,-1},\\
\ah_{\mk ,f_z}&=&\frac{1}{\sqrt{2}}(\ah_{\mk ,2}+\ah_{\mk ,-2}),\\
\ah_{\mk ,p}&=&\frac{1}{2}(-\ah_{\mk ,2}+\ah_{\mk ,-2})+\frac{1}{\sqrt{2}}\ah_{\mk ,0},\label{cpair}
\eeq
which represent the density $(d)$, spin $(f_x,f_y,f_z)$, 
and pair $(p)$ fluctuations, respectively. 
Then, the Hamiltonian is rewritten as follows:
\beq
\hat{H}_{\text{eff}}^{C}&=&2qN+
\frac{Vn^2\bar{c}_0^{(2)}}{2}
+{\sk}^{'}\Bigg\{\sum_{j=f_x,f_y}\Big[\left(\ek -q+2nc_1^{(2)}\right)
\ahd_{\mk ,j}\ah_{\mk ,j}
+nc_1^{(2)}(\ah_{\mk ,j}\ah_{\mmk ,j}
+\ahd_{\mk ,j}\ahd_{\mmk ,j})\Big]\nonumber\\
&&+\left(\ek+2q+2c_1^{(2)}\right)
\ahd_{\mk,f_z}\ah_{\mk,f_z}+
nc_1^{(2)}(\ahd_{\mk ,f_z}\ahd_{\mmk ,f_z}+\ah_{\mk ,f_z}\ah_{\mmk
,f_z})
\nonumber\\
&&+\left(\ek+nc_0^{(2)}\right)\ahd_{\mk ,d}\ah_{\mk ,d}
+\left(\ek+2nc_2^{(2)}\right)
\ahd_{\mk ,p}\ah_{\mk ,p}
-2q(\ahd_{\mk,d}\ah_{\mk,p}+\ahd_{\mk,p}\ah_{\mk,d})\nonumber\\
&&+\frac{nc_0^{(2)}}{2}(\ahd_{\mk ,d}\ahd_{\mmk ,d}
+\ah_{\mk ,d}\ah_{\mmk,d })
\Bigg\}.
\eeq
For nonzero $q$, the density fluctuation couples
with the pair fluctuation. 
For the spin modes, we consider the following transformations:
\begin{gather}
\bh_{\mk,j}=\sqrt{\frac{\ek-q+2nc_1^{(2)}
+E_{\mk,f_t}}{E_{\mk,f_t}}}\ah_{\mk,j}
+\sqrt{\frac{\ek-q+2nc_1^{(2)}
-E_{\mk,f_t}}{E_{\mk,f_t}}}\ahd_{\mmk,j},\\
\bh_{\mk,f_z}=\sqrt{\frac{\ek+2q+2nc_1^{(2)}
+E_{\mk,f_z}}{E_{\mk,f_z}}}\ah_{\mk,f_z}
+\sqrt{\frac{\ek+2q+2nc_1^{(2)}
-E_{\mk,f_z}}{E_{\mk,f_z}}}\ahd_{\mmk,f_z},
\end{gather}
where the Bogoliubov spectra are given by
\begin{gather}
E_{\mk ,f_t}=\sqrt{(\ek-q) \left(\ek-q
 +4nc_1^{(2)}\right)},\label{ct}\\
E_{\mk ,f_z}=\sqrt{(\ek+2q) \left(\ek+2q
 +4nc_1^{(2)}\right)}.\label{cz}
\end{gather}
On the other hand, for the density and pair modes, we
consider the following transformations:
\begin{gather}
\hat{\mathbf{B}}_{\mk}=U(k)\hat{\mathbf{A}}_{\mk,dp}
+V(k)\hat{\mathbf{A}^{\dagger}}_{\mmk,dp},
\end{gather}
where
$
\hat{\mathbf{B}}_{\mk}= ^{t}(\bh_{\mk,-},\bh_{\mk,+}), \ 
\hat{\mathbf{A}}_{\mk,dp}= ^{t}(\ah_{\mk,d},\ah_{\mk,p}),
$
\beq 
U(k)=\frac{1}{2}
\left( 
\begin{array}{cc}
\frac{1}{2C_{3-}(k)E_3(k)}
+C_{3-}(k)X_{3+}(k)
 &  4C_{3-}(k)q(\ek+nc_2^{(2)})
-\frac{X_{3-}(k)}{8C_{3-}(k)q(\ek+nc_2^{(2)})E_3(k)}\\
-\frac{1}{2C_{3+}(k)E_3(k)}
+C_{3+}(k)X_{3-}(k)
 & 4C_{3+}(k)q(\ek+nc_2^{(2)})
+\frac{X_{3+}(k)}{8C_{3+}(k)q(\ek+nc_2^{(2)})E_3(k)}
\end{array} 
\right) ,
\eeq
\beq 
V(k)=\frac{1}{2}
\left( 
\begin{array}{cc}
-\frac{1}{2C_{3-}(k)E_3(k)}
+C_{3-}(k)X_{3+}(k)
 &  4C_{3-}(k)q(\ek+nc_2^{(2)})
+\frac{X_{3-}(k)}{8C_{3-}(k)q(\ek+nc_2^{(2)})E_3(k)}\\
\frac{1}{2C_{3+}(k)E_3(k)}
+C_{3+}(k)X_{3-}(k)
 & 4C_{3+}(k)q(\ek+nc_2^{(2)})
-\frac{X_{3+}(k)}{8C_{3+}(k)q(\ek+nc_2^{(2)})E_3(k)}
\end{array} 
\right) ,
\eeq
\begin{gather}
X_{3\pm}(k)=-n(c_0^{(2)}-2c_2^{(2)})\ek
+2(nc_2^{(2)})^2\pm E_3(k),\\
E_3(k)=\sqrt{\left\{n(c_0^{(2)}-2c_2^{(2)})\ek
-2(nc_2^{(2)})^2\right\}^2
+16q^2(\ek+nc_2^{(2)})
\left\{\ek+n(c_0^{(2)}+c_2^{(2)})\right\}},\\
C_{3\pm}(k)=\sqrt{\frac{E_{\mk,\pm}}{X^2_{3\mp}(k)\ek+16q^2(\ek+nc_2^{(2)}
)\left\{(\ek+2nc_2^{(2)})
(\ek+nc_2^{(2)}
)-X_{3\mp}(k)
\right\}
}},
\end{gather}
where the Bogoliubov spectra are given by
\beq
E_{\mk,\pm}=\sqrt{\ek^2+n(c_0^{(2)}
+2c_2^{(2)})\ek+
2(nc_2^{(2)})^2+4q^2
\pm E_3(k)}.\label{cdp1}
\eeq
By using the above transformations, the effective Hamiltonian is 
diagonalized as follows:
\beq
\bar{H}_{\text{eff}}^{C}&=&E_{0}^C
+{\sk}^{'}\big[ 
E_{\mk ,f_t}(\bhd_{\mk ,f_x}\bh_{\mk ,f_x}+\bhd_{\mk ,f_y}\bh_{\mk ,f_y})
+ E_{\mk ,f_z} \bhd_{\mk ,f_z}\bh_{\mk ,f_z}
+ E_{\mk,+}\bhd_{\mk ,+}\bh_{\mk ,+}
+E_{\mk,-}\bhd_{\mk ,-}\bh_{\mk ,-}\big],
\nonumber\\
\eeq
where
\beq
E_{0}^C&=&2qN+\frac{Vn^2\bar{c}_{0}^{(2)}}{2}
-\frac{1}{2}{\sk}^{'}\Bigg[ 2\left(\ek-q 
+2nc_1^{(2)}- E_{\mk ,f_t}\right) 
+\left(\ek+2q +2nc_1^{(2)}- E_{\mk
,f_z}\right) 
\nonumber\\
&&+\left(2\ek+nc_0^{(2)}
+2nc_2^{(2)}-E_{\mk,+}-E_{\mk,-}\right)
\Bigg]
\eeq
is the GSE.
As can be seen from Eqs. (\ref{ct}) and (\ref{cz}),
the cyclic configuration of Eq. (\ref{cyclic}) suffers the dynamical 
instability unless $q=0$.
Moreover, the following inequality needs to be satisfied for
the Bogoliubov mode of Eq. (\ref{cdp1}) to be stable:
\beq
\ek^4&+&2n(c_0^{(2)}+2c_2^{(2)})\ek^3
+[8n^2c_0^{(2)}c_2^{(2)}
+4(nc_2^{(2)})^2-8q^2]\ek^2\nonumber\\
&+&[8n^3c_0^{(2)}(c_2^{(2)})^2
-16q^2n(c_0^{(2)}
+2c_2^{(2)})]\ek
+16q^2(q^2-n^2c_0^{(2)}c_2^{(2)})>0.\label{cyclicin}
\eeq 
Conversely, the dynamical instability occurs if
the above inequality is not satisfied.
Therefore, the tetrahedral configuration (\ref{cyclic}) is unstable for
an infinitesimal quadratic Zeeman effect in the thermodynamic limit.
For $q=0$, the spectra (\ref{ct}), (\ref{cz}), and (\ref{cdp1})
reduce to those of Ref. \cite{ueda}:
\begin{gather}
E_{\mk ,-}=\sqrt{\ek \left(\ek
 +2nc_0^{(2)}\right)},\\
E_{\mk ,f_i}=\sqrt{\ek \left(\ek
 +4nc_1^{(2)}\right)}, \ (i=x, \ y \ \text{or} \ z)\\
E_{\mk ,+}=\ek+ 
2nc_2^{(2)}. \label{modep}
\end{gather}
The above spectra are positive semidefinite if 
$c_0^{(2)}>0$, $c_1^{(2)}>0$, and $c_2^{(2)}>0$, 
consistent with the stability criteria of the mean-field cyclic state.

\section{Relation between the number of 
the spontaneously broken generators and that of the Nambu-Goldstone modes}

We discuss the number of the Nambu-Goldstone (NG) modes 
in light of the rule found by Nielsen and Chadha \cite{nielsen}
(see Ref. \cite{brauner2} for a review).
In general, if the Hamiltonian has a symmetry with respect to the
internal degrees of freedom, 
we can introduce the corresponding conserved charge operator 
defined as follows \cite{nagaosa}: 
\beq
\hat{G}=\int d\mathbf{x}\hat{J}^{0}(x),
\eeq
where
\beq
\hat{J}^0(x)=\frac{1}{\hbar}
\frac{\delta \hat{S}}{\delta \dot{\hat{\Psi}}_m(x)}\delta\hat{\Psi}_m(x),
\eeq
$\hat{S}$ is the action corresponding to the Hamiltonian,
$\delta \hat{\Psi}_m(x)$ is an infinitesimal transformation of the
field, and the overdot denotes the differentiation with respect to time.
The conserved charge operator commutes with the
Hamiltonian,
\beq
[\hat{G},\hat{H}]=0.
\eeq
When the spontaneous symmetry breaking occurs,
the ground state $|\text{GS}\rangle$ does
not have the full symmetry of the original Hamiltonian.
Mathematically, this implies that 
the following expectation value does not vanish:
\beq
\langle \text{GS}|[\hat{\Psi}_m(x),\hat{G}]|\text{GS}\rangle\ne 0.
\label{conditionng}
\eeq
Note that, in general, we can also substitute 
$[\hat{\Psi},\hat{G}]$
for the commutator in Eq. (\ref{conditionng}), where
$\hat{\Psi}$ is a composite field of 
$\hat{\Psi}_m^{\dagger}$ and $\hat{\Psi}_m$.
Then, the NG theorem predicts the appearance of the corresponding NG mode 
whose energy vanishes in the long-wavelength limit.

According to Ref. \cite{nielsen}, the energy of the NG mode 
obeys a power law of the wave number in the long-wavelength
limit, and the NG mode is classified as   
type-I or type-II according to whether 
this power is odd or even, respectively. 
Nielsen and Chadha formulated the following theorem: 
if the NG mode of type-I is counted once and that of type-II is
counted twice, then the total number of the NG modes is equal to or
greater than the number of the 
symmetry generators that correspond to the spontaneously broken symmetries:
\beq
N_{\text{NG}}\equiv N_{\text{I}}+2N_{\text{II}}\ge N_{\text{BG}},
\eeq
where $N_{\text{I}}$, $N_{\text{II}}$, $N_{\text{BG}}$
are 
the total number of the type-I NG mode, that of the type-II NG mode,
that of the symmetry generators whose symmetries 
are spontaneously broken, respectively.

Lorentz invariant theories can have only NG modes with linear
dispersion relations. 
In non-relativistic theories, however, 
there are examples that belong to type-II NG modes.
A well-known example of type-II NG modes is the Heisenberg ferromagnet, 
which has a NG mode satisfying
a quadratic dispersion relation. 
The authors of Ref. \cite{schafer} have proved that
if $\hat{G}_i$, $(i=1,2,...,n)$, 
constitute a full set of broken charges, and if
$\langle \text{GS}|[\hat{G}_i,\hat{G}_j]]|\text{GS}\rangle =0$ 
for any pair $(i,j)$,
the relation $N_{\text{I}}+N_{\text{II}}=N_{\text{BG}}$ is satisfied.
In Ref. \cite{brauner}, the author suggests that 
if the above commutators are not equal to zero, namely  
$\langle \text{GS}|[\hat{G}_i,\hat{G}_j]]|\text{GS}\rangle \ne 0$, 
a type-II NG mode appears.

We apply the above arguments to spinor BECs.
In the absence of an external
magnetic field, the low-energy Hamiltonian in a spinor BEC has the 
$U(1)\times SO(3)$ symmetry,
representing the global gauge and spin-rotation symmetries.
Specifically, 
$\delta \hat{\Psi}_m(x)$ in the $U(1)$ and $SO(3)$ transformations are 
given by 
\begin{gather}
\delta_{U(1)} \hat{\Psi}_m(x)=i\theta\hat{\Psi}_m(x),\\
\delta_{SO(3)} \hat{\Psi}_m(x)=i\theta' f^{j}_{mm'}\hat{\Psi}_{m'}(x), \ \ \
(j=x,\ y \ \text{or} \ z)
\end{gather}
where $\theta$ and $\theta'$ are infinitesimal parameters.
The conserved charges of the $U(1)$ and $SO(3)$
transformations are 
\begin{gather}
\hat{G}_{U(1)}\equiv \hat{N} = \int d\mathbf{x} 
\hat{\Psi}_m^{\dagger}(x)\hat{\Psi}_m(x),\\
\hat{G}_{SO(3)}\equiv \hat{F}^j = \int d\mathbf{x}
\hat{\Psi}_m^{\dagger}(x)f_{mm'}^{j}\hat{\Psi}_{m'}(x),
\end{gather}
where we drop the infinitesimal parameters.  
These conserved charge operators 
are nothing but the total number and spin operators.  
In the presence of an external magnetic field, however,
the $SO(3)$ symmetry reduces to $SO(2)$, representing
the spin-rotation symmetry about the direction of
the external magnetic field,
and the symmetry of 
the Hamiltonian therefore reduces to $U(1)\times SO(2)$.
Since 
some or all of these symmetries are spontaneously broken in
each phase, NG modes are expected to emerge.
To find the types of the relevant NG modes, 
it is important to specify the order parameter
manifold of each phase.
This is equivalent to finding the combination of gauge transformation 
and spin rotations  
to keep the order
parameter of each phase unchanged.
Such a programme has been carried out
in Refs. \cite{ho,song,zhou,makela,zhou2} for spin-1 and spin-2 BEC in
the absence of an external magnetic field, based on the mean-field theory.
The number of NG modes and the classification are 
investigated based on the Bogoliubov theory by analyzing
the long-wavelength limit of the massless modes.

We first discuss the cases of zero external
magnetic field.
TABLE I shows the summary in spin-1 and spin-2 BECs.
Interestingly, the number of symmetry generators that are
spontaneously broken $N_{\text{BG}}$ 
is equal to that of the NG modes $N_{\text{NG}}$ in the all phases
in spin-1 and spin-2 BECs, namely $N_{\text{NG}}=N_{\text{BG}}$, 
and to the best
knowledge of the present authors, 
the equality holds in all systems studied so far. 
For the spin-1 ferromagnetic BEC, 
it follows  
from Eqs. (\ref{ferroh}) and 
(\ref{eq:spin-1-f-b}) that
there exist
one type-I and one type-II NG modes, which
describe the density mode and transverse spin mode, respectively.
We note that the spin-1 ferromagnetic BEC is similar to
the Heisenberg ferromagnet in that
the type-II mode appears if the following condition
holds:
\beq
\langle \text{GS}|[\hat{F}^x,\hat{F}^y]|\text{GS}\rangle=
i\langle \text{GS}|\hat{F}^z|\text{GS}\rangle\ne 0.
\eeq
This rule applies also to a spin-2 ferromagnetic BEC. 
For the other spin-1 and spin-2 phases, however,
the expectation values of the commutators of 
the generators are equal to zero. Therefore,
$N_{\text{I}}+N_{\text{II}}=N_{\text{BG}}$.
In addition, all these phases only have
the Bogoliubov modes that are linear and massless.
Hence, $N_{\text{I}}=N_{\text{BG}}$ for all phases
except for the spin-1 and spin-2 ferromagnetic phases. 

The spin-2 nematic phase is special since
not all of the massless Bogoliubov modes 
can be interpreted as the NG modes.
Both of the uniaxial and biaxial nematic phases have the five Bogoliubov
modes that are  linear and massless,  
while the number of spontaneously broken generators is three for
the uniaxial nematic phase and four for the biaxial nematic phase.
However, since the expectation values of the commutators of
the generators are always zero, 
$N_{\text{I}}$ 
must be equal to $N_{\text{BG}}$ according to Ref. \cite{schafer}.
Therefore, there is one Bogoliubov mode that is not the NG mode for the
biaxial nematic phase and there are two such modes 
for the uniaxial nematic phases.
In each of the uniaxial and biaxial nematic phases,
there exists the Bogoliubov mode that originates from the fluctuations
with respect to $\eta$ and is not a NG mode 
because $\eta$ is not related to the symmetry of the Hamiltonian 
and we cannot introduce a conserved current on $\eta$.
Here, we note that 
this Bogoliubov mode can also be interpreted as a fluctuation mode
with respect to one of the $SO(5)$ directions
other than $SO(3)$ directions.
This is because we can define the fluctuation operator around the
one of $SO(5)$ directions as follows:
\beq
\frac{\cos\eta}{\sqrt{2}}(\ah_{\mk,2}+\ah_{\mk,-2})-\sin\eta\ah_{\mk,0},
\eeq
which is  the same as the nematic fluctuation operator defined in Eq.
(\ref{no}).
In addition, one of the spin fluctuation modes
is not the NG mode in the uniaxial nematic phase as mentioned in
Sec. V. B.
To show this,
we can choose the configuration $\eta =0$ without loss of generality.
Then the isotropy groups in the uniaxial spin nematic phase include
$SO(2)$, which represents the rotational symmetry around the $z$ axis.
Therefore, the spin mode around the $z$ axis is not the NG mode since
the symmetry is not broken spontaneously.  
This is also confirmed by the fact 
that the following expectation value 
is zero in the uniaxial nematic phase:
\beq
\langle \text{GS}|[\hat{\tilde{\Psi}}(\mathbf{x}),\hat{F}^z] |\text{GS}
\rangle =0,
\eeq
where $\hat{\tilde{\Psi}}(\mathbf{x})$ is an arbitrary polynomial 
in $\hat{\Psi}_m^{\dagger}(\mathbf{x})$ and $\hat{\Psi}_m(\mathbf{x})$.
This implies that the Bogoliubov mode under consideration
is not a NG mode.
However, this Bogoliubov mode defined in
Eq. (\ref{eq:spin-2-n-b4}) can be interpreted as the fluctuation mode around
the other direction of $SO(5)$, since we can define the 
fluctuation operator
\beq
\frac{1}{\sqrt{2}}(\ah_{\mk,2}-\ah_{\mk-2}),
\eeq
which is the same as the fluctuation operator defined in Eq. (\ref{fluc-ne-z}).
When taken together, 
these massless Bogoliubov modes that are not NG modes can be interpreted as
the modes related to the $SO(5)$ symmetry that the mean-field 
solution in the nematic phase has. 
This is worthy of special mention because 
it is rather exceptional that a massless mode is not a NG mode.
The other exceptions are a two-dimensional $SU(N)$ Thirring model 
\cite{witten}
and
a two-dimensional superfluid \cite{nagaosa}, both of which have massless
modes that are not a NG modes because of the Coleman-Hohenberg-Mermin-Wagner
theorem \cite{nagaosa}.
\begin{table}
\begin{center}
\begin{tabular}{cc} 
\begin{minipage}{0.5\hsize}
\begin{center}
\begin{tabular}{l|c|c|c}
\hline
Phase & $G/H$ & $N_{\text{BG}}$ & $N_{\text{NG}}$ \\ 
\hline
Spin-1 F & $SO(3)$ \cite{ho} & 3 & 3 \\ 
Spin-1 P & $(U(1)\times S^2)/Z_2$ \cite{zhou} & 3 & 3 \\
Spin-2 F & $SO(3)/Z_2$ \cite{makela} & 3 & 3 \\
Spin-2 UN  &$U(1)\times S^2/Z_2$ \cite{song} & 3 & 3 \\
Spin-2 BN & $(U(1)\times SO(3))/D_4$ \cite{song} & 4 & 4 \\
Spin-2 C  & $(U(1)\times SO(3))/T$ \cite{makela,zhou2} & 4 &4 \\ 
\hline
\end{tabular}
\caption{Order parameter manifold $G/H$, the number of spontaneously 
broken generators $N_{\text{BG}}$, and
that of NG modes $N_{\text{NG}}$ 
in each phase in the absence of an external
magnetic field, where
$D_4$ and $T$ represent the dihedral-four and tetrahedral groups,
respectively.}
\label{opm}
\end{center}
\end{minipage}
\begin{minipage}{0.5\hsize}
\begin{center}
\begin{tabular}{l|c|c|c}
\hline
Phase & $G/H$ & $N_{\text{BG}}$ & $N_{\text{NG}}$ \\ 
\hline
Spin-1 F & $U(1)$ & 1 & 1 \\ 
Spin-1 P & $U(1)$  & 1 & 1 \\
Spin-1 P'& $(U(1)\times SO(2))/Z_2$  & 2 & 2 \\
Spin-1 BA  &$U(1)\times SO(2)$  & 2 & 2 \\
Spin-2 F & $U(1)/Z_2$  & 1 & 1 \\
Spin-2 UN& $U(1)$ & 1 &1 \\
Spin-2 BN& $(U(1)\times SO(2))/Z_4$ &2 &2\\
Spin-2 C & $U(1)\times SO(2)/Z_2$ &2 &2\\ 
\hline
\end{tabular}
\caption{Order parameter manifold $G/H$, the number of spontaneously 
broken generators $N_{\text{BG}}$, and
that of NG modes $N_{\text{NG}}$ 
in each phase in the presence of an external
magnetic field.}
\label{opm2}
\end{center}
\end{minipage}
\end{tabular}
\end{center}
\end{table}

Finally, we discuss the cases in the presence of an external magnetic 
field, which are summarized in TABLE \ref{opm2}.
Since the symmetry of the Hamiltonian reduces to $U(1)\times SO(2)$
and the commutator of $U(1)$ and $SO(2)$
is zero, we must have $N_{\text{I}}+N_{\text{II}}=N_{\text{BG}}$.
The Bogoliubov theory shows that
for all phases that we have analyzed,
only type-I NG modes appear; therefore, $N_{\text{I}}=N_{\text{BG}}$.
To the best knowledge of the present authors,
this is always the case, that is,
only type-I NG modes appear,
when $\langle\text{GS}|[\hat{G}_i,\hat{G}_j]|\text{GS}\rangle=0$
for any pair $(i,j)$  of broken generators.
Hence, it is expected that for arbitrary phases in the presence of an external
magnetic field, only type-I NG modes emerge
whenever spontaneous symmetry breaking occurs.

\section{Summary and concluding remarks}
We have  
derived the Bogoliubov spectra and 
LHY corrections to the GSE, pressure, sound velocity, and quantum depletion
in the presence of a quadratic Zeeman effect. 
We have shown that 
in the absence of an external magnetic field, 
the Bogoliubov effective Hamiltonian reduces to the sum of
sub-Hamiltonians that can be diagonalized by
the standard Bogoliubov transformations. 
A nontrivial example is the nematic phase of a spin-2 BEC 
because we cannot determine the sufficient number of fluctuation
operators that
decompose the Hamiltonian from the symmetry of the Hamiltonian only.
However, because the nematic phase features an additional continuous parameter
$\eta$, the fluctuation operator with respect to $\eta$ can be constructed 
and allows the decomposition of 
the Hamiltonian into the sum of the sub-Hamiltonians.
Furthermore, because of the $\eta$ dependence in the Bogoliubov Hamiltonian,
the nematic phase is divided into the uniaxial and biaxial nematic phases
by quantum fluctuations even though these phases are degenerate at
the mean-field level. Finally, the $\eta$ dependence removes
the degeneracy and causes
the phase transition between the uniaxial and biaxial nematic phases. 

In the presence of the quadratic Zeeman effect,
the magnetic quantum number is no longer conserved and different modes
are coupled. Therefore, the Hamiltonian must be diagonalized by 
treating the multidimensional Hamiltonian explicitly. 
By explicitly constructing the Bogoliubov transformations,
we have obtained the LHY corrections to the GSE, 
sound velocity, and quantum depletion. 
In the case of a scalar BEC, the LHY corrections depend only on
the coupling constant and density. 
For spinor BECs, 
the LHY corrections also depend on the quadratic Zeeman effect except
for the ferromagnetic phase. 
Moreover, taking into consideration the fact that 
the spin-dependent coupling constants are small compared to
the spin-independent one for the alkali species,
the main contribution of the LHY corrections 
arises from the density fluctuation.
The LHY corrections in spinor BECs may be observed by making
the system strongly correlated by using an optical lattice
\cite{song2} or by controlling an external magnetic field. 
Since the enhancement of the quantum depletion in an optical lattice 
has already been demonstrated
by the MIT group in a scalar $^{23}$Na condensate 
\cite{xu}, it should also be possible to apply it to spinor condensates. 

Following the argument by Nielsen and Chadha,
we have pointed out the relation between the number of
symmetry generators that are spontaneously broken $N_{\text{BG}}$ 
and that of the
NG modes $N_{\text{NG}}$.
The NG modes are divided into  type-I and type-II
according to  the dispersion laws, and
the following inequality must be satisfied: 
$N_{\text{NG}}\equiv N_{\text{I}}+2N_{\text{II}}\ge N_{\text{BG}}$.
In contrast, for all the phases that we have analyzed,
it is shown that $N_{\text{NG}}= N_{\text{BG}}$.
The type-II NG modes only appear in the spin-1 and spin-2
ferromagnetic phases in the absence of an external magnetic field, 
as in the case of the Heisenberg ferromagnet.  
Although only type-I NG modes emerge for the other phases,  
nontrivial situations arise for the spin-2 nematic phase in
the absence of an external magnetic field: 
there exist the Bogoliubov modes that have linear dispersion relations
but do not belong to the NG modes.

We have shown that every configuration of the mean-field ground state
that
we have studied is
stable against quantum fluctuations. 
However, if the configuration is not a mean-field ground state, 
instabilities set in.
For the quadratic spectrum, it implies the Landau instability,
while for the linear spectrum, it implies the dynamical instability.
Furthermore, it is expected that the instabilities also occur
for the configurations that are not stationary solutions of the mean-field
theory.
We illustrated this for the tetrahedral configuration of the cyclic
phase, which is not a stationary solution of the mean-field theory for
nonzero $q$ and is unstable regardless of the sign of $q$. 
We hope that our analysis
helps the determination of the ground-state phase of the
spin-2 $^{87}$Rb condensate, which lies
in the vicinity of the phase boundary between the biaxial nematic and
cyclic phases for $q<0$ \cite{tojo}. 

States that are unstable in the thermodynamic limit may be stable
in a mesoscopic regime.
For example, for the case of the tetrahedral configuration of the cyclic phase,
if the lowest wave number is higher than a critical value 
$k_c=\sqrt{4M|q|/\hbar^2}$, the Bogoliubov spectra of 
the spin fluctuations are positive definite. For the
Bogoliubov spectrum of the density fluctuation, 
the condition of Eq. (\ref{cyclicin}) is needed to be positive. 
Hence, this configuration can be stabilized by 
the balance between the finite-size and quadratic Zeeman effects.
Similar arguments can be applied to other configurations.

Finally, we briefly comment on several applications related to
our analysis. Incorporating the trapping effect is  an important
extension of the present theory.
Meanwhile, changing of the sign of $q$ with the technique discussed in
Ref. \cite{leslie} also gives rise to nontrivial effect.
For example, the polar configuration 
in Eq. (\ref{polar}) 
becomes unstable and is expected to undergo the phase transition 
to the configuration in Eq. (\ref{polar'}) through
the dynamical instability
when the sign of $q$ changes from positive to negative.
In this process, the polar direction changes from the $z$ axis to
a transverse direction, triggering spontaneously symmetry breaking
of axisymmetry and dynamical creation of
half-quantum vortices around which both the polar axis and condensate
phase  rotate $\pi$ \cite{uchino2}.

\section*{Acknowledgements}
We thank T. Hatsuda, Y. Kawaguchi,
 N. Yamamoto, T. Kanazawa,
A. Rothkopf for fruitful
comments. This work was supported by KAKENHI (22340114),
Global COE Program ``the Physical Sciences Frontier'' and
the Photon Frontier Network Program, MEXT, Japan.

\appendix
\section{Derivation of the ground-state energies}
\subsection{Polar phase}
In the polar phase, the renormalized and
bare coupling constants are related up to the second Born approximation by
\beq
\bar{c}_0^{(1)}&=&\frac{\bar{g}_0+2\bar{g}_2}{3}
=\frac{g_0+2g_2}{3}+\frac{g_0^2+2g_2^2}{3V}{\sk}^{'}\frac{1}{2\ek}\nonumber\\
&=&c_0^{(1)}+[(c_0^{(1)})^2+2(c_1^{(1)})^2]\frac{M}{V\hbar^2}
{\sk}^{'}\frac{1}{k^2}.
\eeq
Therefore, the mean-field energy diverges as follows:
\beq
\frac{Vn^2c_0^{(1)}}{2}+\frac{\hbar^2}{8M}{\sk}^{'}\Bigg[
\left(\frac{2Mnc_0^{(1)}}{\hbar^2k}\right)^2+
2\left(\frac{2Mnc_1^{(1)}}{\hbar^2k}\right)^2
\Bigg].\label{renormp}
\eeq
On the other hand, the components of the GSE 
arising from quantum fluctuations involve the $k^{-2}$ terms
as follows: 
\beq
&&-\frac{1}{2}{\sk}^{'} 
\Bigg[ \left(\ek +nc_0^{(1)}- E_{\mk ,d}\right) 
+2\left(  \ek+q +nc_1^{(1)} -E_{\mk
,f_t}\right) \Bigg]\nonumber\\
&=&-\frac{\hbar^2}{4M}{\sk}^{'} 
\Bigg[\left( k^2+2Mnc_0^{(1)}-k\sqrt{k^2+\frac{4Mnc_0^{(1)}}{\hbar^2}}
\right) \nonumber\\
&&+2\left( k^2+\frac{2Mq}{\hbar^2}+\frac{2Mnc_1^{(1)}}{\hbar^2}
-\sqrt{\left(k^2+\frac{2Mq}{\hbar^2}
\right)\left(k^2+\frac{2Mq}{\hbar^2}+\frac{4Mnc_1^{(1)}}{\hbar^2}\right)}
 \right)
\Bigg]\nonumber\\
&&\xrightarrow[k\to \infty]{}
-\frac{\hbar^2}{8M}{\sk}^{'}\Bigg[
\left(\frac{2Mnc_0^{(1)}}{\hbar^2k}\right)^2+
2\left(\frac{2Mnc_1^{(1)}}{\hbar^2k}\right)^2
\Bigg].
\label{a3}
\eeq
Hence, the divergence in Eq. (\ref{renormp}) is canceled with
the last terms in Eq. \eqref{a3}
and the convergence of the GSE is ensured.
The GSE for $c_1^{(1)}>0$ in the polar phase $E_0^P$ is given by
\beq
\frac{E_0^P}{V}&=&\frac{n^2\bar{c}_0^{(1)}}{2}-\frac{1}{2V}{\sk}^{'} 
\Bigg[ \left(\ek +nc_0^{(1)}- E_{\mk ,d}\right) 
+2\left(  \ek+q +nc_1^{(1)} -E_{\mk
,f_t}\right) \Bigg]
\nonumber\\
&=&\frac{n^2c_0^{(1)}}{2}-\frac{\hbar^2}{4MV}{\sk}^{'} 
\Bigg\{ \left[ k^2+2Mnc_0^{(1)}-k\sqrt{k^2+\frac{4Mnc_0^{(1)}}{\hbar^2}}
-\frac{1}{2}\left(\frac{2Mnc_0^{(1)}}{\hbar^2k}\right)^2\right] \nonumber\\
&&+2\left[ k^2+\frac{2Mq}{\hbar^2}+\frac{2Mnc_1^{(1)}}{\hbar^2}
-\sqrt{\left(k^2+\frac{2Mq}{\hbar^2}
\right)\left(k^2+\frac{2Mq}{\hbar^2}+\frac{4Mnc_1^{(1)}}{\hbar^2}\right)}
-\frac{1}{2}\left(\frac{2Mnc_1^{(1)}}{\hbar^2k}\right)^2 \right]
\Bigg\}\nonumber\\
&=&\frac{n^2c_0^{(1)}}{2}-\frac{\hbar^2}{8\pi^2M}\left(
\frac{2Mnc_0^{(1)}}{\hbar^2}\right)^{\frac{5}{2}}
\int_{0}^{\infty}
dx x^2\left(x^2+1-x\sqrt{x^2+2}-\frac{1}{2x^2} \right)\nonumber\\
&&-\frac{\hbar^2}{4\pi^2M}\left(
\frac{2Mnc_1^{(1)}}{\hbar^2}\right)^{\frac{5}{2}}
\int_{0}^{\infty}
dx x^2\left(x^2+t_1+2-\sqrt{(x^2+t_1+1)(x^2+t_1+3)}-\frac{1}{2x^2}
\right)\nonumber\\
&=&\frac{n^2c_0^{(1)}}{2}\left(1+\frac{16\sqrt{M^3}}{15\pi^2\hbar^3}
\sqrt{n\left(c_0^{(1)}\right)^3}\right)
+\frac{16\sqrt{M^3}n^2c_1^{(1)}}{15\pi^2\hbar^3}
\sqrt{n\left(c_1^{(1)}\right)^3}
\phi_1(t_1+1).\label{cycgse}
\eeq
The GSE for $c_1^{(1)}<0$ can also be derived similarly.

\subsection{Broken-axisymmetry phase}
In the broken-axisymmetry phase, we note the following relations:
\beq
\frac{q^2}{\bar{c}_1^{(1)}}&=&\frac{q^2}{c_1^{(1)}}+
\frac{g_0^2-g_2^2}{3(c_1^{(1)})^2}
\frac{q^2M}{V\hbar^2}{\sk}^{'}\frac{1}{k^2}
=\frac{q^2}{c_1^{(1)}}+\frac{c_1^{(1)}-2c_0^{(1)}
}{c_1^{(1)}}\frac{q^2M}{V\hbar^2}{\sk}^{'}\frac{1}{k^2},
\eeq
\beq
\bar{c}_0^{(1)}+\bar{c}_1^{(1)}=c_0^{(1)}+c_1^{(1)}+(c_0^{(1)}+c_1^{(1)})^2
\frac{M}{V\hbar^2}{\sk}^{'}\frac{1}{k^2}.
\eeq
Therefore, the mean-field term diverges as
\beq
\frac{Vn^2(c_0^{(1)}+c_1^{(1)})}{2}
+\frac{Vq^2}{8c_1^{(1)}}+\frac{\hbar^2}{8M}{\sk}^{'}
\Bigg[\left(\frac{2Mn(c_0^{(1)}+c_1^{(1)})}{\hbar^2k}\right)^2+
\frac{c_1^{(1)}-2c_0^{(1)}}{c_1^{(1)}}
\left(\frac{Mq}{\hbar^2k}\right)^2
\Bigg].
\eeq
On the other hand, in the short-wavelength limit,
the components of the GSE arising from the
quantum fluctuations behave as
\beq
-\frac{1}{2}{\sk}^{'}\Bigg[
\left(\ek+\frac{q}{2}-E_{\mk,f_z}\right)
+\left(2\ek+nc_0^{(1)}-nc_1^{(1)}-E_{\mk,d}
-E_{\mk,\theta}\right)\Bigg]\nonumber\\
\xrightarrow[k\to \infty]{}
-\frac{\hbar^2}{8M}{\sk}^{'}\Bigg[
\left(\frac{2Mn(c_0^{(1)}+c_1^{(1)})}{\hbar^2k}\right)^2
+\frac{c_1^{(1)}-2c_0^{(1)}}{c_1^{(1)}}
\left(\frac{Mq}{\hbar^2k}\right)^2
\Bigg].
\eeq
Therefore, the above $k^{-2}$ terms are canceled  with those from the
mean-field terms. 
Thus, the convergence of the GSE is ensured and the GSE $E_0^{BA}$ is given by
\beq
\frac{E_0^{BA}}{V}&=&
\frac{nq}{2}+\frac{n^2(\bar{c}_0^{(1)}+\bar{c}_1^{(1)})}{2}
+\frac{q^2}{8\bar{c}_1^{(1)}}\nonumber\\
&&-\frac{1}{2V}{\sk}^{'}\Big[
\left(\ek+q/2-E_{\mk,f_z}\right)
+\left(2\ek+nc_0^{(1)}-nc_1^{(1)}-E_{\mk,d}
-E_{\mk,\theta}\right)\Big]\nonumber\\
&=&\frac{nq}{2}+\frac{n^2(c_0^{(1)}+c_1^{(1)})}{2}+
\frac{q^2}{8c_1^{(1)}}-\frac{\hbar^2}{4MV}{\sk}^{'}
\Bigg\{\left[k^2+\frac{Mq}{\hbar^2}-k\sqrt{k^2+\frac{2Mq}{\hbar^2}}
-\frac{1}{2}\left(\frac{Mq}{\hbar^2k}\right)^2\right]\nonumber\\
&&+\Bigg[2k^2+\frac{2Mn(c_0^{(1)}-c_1^{(1)})}{\hbar^2}
-\frac{2M}{\hbar^2}(E_{\mk,d}+
E_{\mk,\theta})\nonumber\\
&&-\left(\frac{2Mn}{\hbar^2k}\right)^2\left(
\frac{(c_0^{(1)}+c_1^{(1)})^2}{2}
-\frac{(c_0^{(1)}+c_1^{(1)})q^2}{4n^2c_1^{(1)}}
+\frac{3q^2}{8n^2}\right)
\Bigg]
\Bigg\}\nonumber\\
&=&\frac{nq}{2}+\frac{n^2(c_0^{(1)}+c_1^{(1)})}{2}
+\frac{q^2}{8c_1^{(1)}}+\frac{\sqrt{2M^3q^5}}{15\pi^2\hbar^3}
+\frac{8\sqrt{M^3}[n(c_0^{(1)}+c_1^{(1)})]^{\frac{5}{2}}}{15\pi^2\hbar^3}
\phi_4(t_3).
\eeq

\subsection{Nematic phase}
In the nematic phase, 
the following relation holds:
\beq
\bar{c}_0^{(2)}+\bar{c}_2^{(2)}&=&
\frac{7\bar{g}_0+10\bar{g}_2+18\bar{g}_4}{35}\nonumber\\
&&=\left( c_0^{(2)}+c_2^{(2)}\right) 
+ \Bigg[ \left( c_0^{(2)}+c_2^{(2)}\right)^2
+\left( c_2^{(2)}\right)^2
+\sum_{j=0}^2\left(c_3^{(2)}(\eta+\pi j/3)\right)^{2}
\Bigg] \frac{M}{V\hbar^2}
{\sk}^{'}\frac{1}{k^2},
\nonumber\\
\eeq
where we use 
\begin{gather}
\sin^2(\eta+\pi/3)+\sin^2(\eta-\pi/3)+\sin^2\left(\eta\right)=\frac{3}{2},\\
\sin^4(\eta+\pi/3)+\sin^4(\eta-\pi/3)+\sin^4\left(\eta\right)=\frac{9}{8}.
\end{gather}
Therefore, the mean-field energy is rewritten as follows: 
\beq
\frac{Vn^2(c_0^{(2)}+c_2^{(2)})}{2}+\frac{\hbar^2}{8M}{\sk}^{'}
\Bigg[\left(\frac{2Mn(c_0^{(2)}+c_2^{(2)})}{\hbar^2k}\right)^2
+\left(\frac{2Mn(-c_2^{(2)})}{\hbar^2k}\right)^2\nonumber\\
+\sum_{j=0}^2\left(\frac{2Mnc_3^{(2)}(\eta+\pi j/3
)}{\hbar^2k}\right)^2
\Bigg].
\label{renormn}
\eeq
On the other hand, the components of the GSE arising from the quantum
fluctuations diverge as follows:
\beq
&&-\frac{1}{2}{\sk}^{'}\Bigg[ \left(\ek+nc_0^{(2)}
+nc_2^{(2)}-E_{\mk ,d} \right)
+ \left(\ek+nc_3^{(2)}(\eta+\pi/3)
-E_{\mk ,f_x} \right) \nonumber\\
&&+\left(\ek+nc_3^{(2)}(\eta-\pi/3)
-E_{\mk ,f_y} \right)  
+\left(\ek+nc_3^{(2)}(\eta)-E_{\mk ,f_z} \right) 
+\left(\ek-nc_2^{(2)} -E_{\mk ,\eta} \right)
\Bigg]\nonumber\\
&&\xrightarrow[k\to \infty]{}
-\frac{\hbar^2}{8M}{\sk}^{'}
\Bigg[\left(\frac{2Mn(c_0^{(2)}+c_2^{(2)})}{\hbar^2k}\right)^2
+\left(\frac{2Mn(-c_2^{(2)})}{\hbar^2k}\right)^2
+\sum_{j=0}^2\left(\frac{2Mnc_3^{(2)}(\eta+\pi j/3
)}{\hbar^2k}\right)^2
\Bigg].
\nonumber\\
\eeq
As expected,
these divergences are canceled out with the last three terms of Eq. 
(\ref{renormn}), and
the GSE in the nematic phase $E_0^N$ is given by
\beq
\frac{E_0^N(\eta)}{V}&=&\frac{n^2
( \bar{c}_0^{(2)}+\bar{c}_2^{(2)})}{2}
-\frac{1}{2V}{\sk}^{'}\Bigg[ \left(\ek+nc_0^{(2)}
+nc_2^{(2)}-E_{\mk ,d} \right)
+ \left(\ek+nc_3^{(2)}(\eta+\pi/3)
-E_{\mk ,f_x} \right) \nonumber\\
&&+\left(\ek+nc_3^{(2)}(\eta-\pi/3)
-E_{\mk ,f_y} \right)  
+\left(\ek+nc_3^{(2)}(\eta)-E_{\mk ,f_z} \right) 
+\left(\ek-nc_2^{(2)} -E_{\mk ,\eta} \right)
\Bigg]\nonumber\\
&=&\frac{n^2( c_0^{(2)}+c_2^{(2)})}{2}-\frac{\hbar^2}{4MV}
{\sk}^{'}\Bigg\{\Bigg[(k^2+\frac{2Mn(c_0^{(2)}
+c_2^{(2)})}{\hbar^2}
-k\sqrt{k^2+\frac{4Mn(c_0^{(2)}+c_2^{(2)})}{\hbar^2}}\nonumber\\
&&-\frac{1}{2}\left(
\frac{2Mn(c_0^{(2)}+c_2^{(2)})}{\hbar^2k}
\right)^2 \Bigg]\nonumber\\
&&+ \Bigg[k^2+\frac{2Mn
(-c_2^{(2)})}{\hbar^2}-k\sqrt{k^2+\frac{4Mn
(-c_2^{(2)})}{\hbar^2}}
-\frac{1}{2}\left(\frac{2Mn(-c_2^{(2)})}{\hbar^2k}
\right)^2 \Bigg]\Bigg\}\nonumber\\
&&+ \sum_{j=0}^2\Bigg[k^2+\frac{2Mn
c_3^{(2)}(\eta+\pi j/3)}{\hbar^2}
-k\sqrt{k^2+\frac{4Mnc_3^{(2)}(
\eta+\pi j/3)}{\hbar^2}}\nonumber\\
&&-\frac{1}{2}\left(\frac{2Mnc_3^{(2)}(\eta+\pi j/3)
}{\hbar^2k}\right)^2 \Bigg]\nonumber\\
&=&\frac{n^2(c_{0}^{(2)}+c_2^{(2)})}{2}
\left(1+\frac{16\sqrt{M^3}}{15\pi^2\hbar^3}\sqrt{n
(c_{0}^{(2)}+c_2^{(2)})^{3}}
\right)
+\frac{8\sqrt{M^3}}{15\pi^2\hbar^3}\Big[
[n|c_2^{(2)}|]^{\frac{5}{2}}\nonumber\\
&&+[n(2c_1^{(2)}-c_2^{(2)})]^{\frac{5}{2}}
\sum_{j=0}^2(1+X\cos(2\eta+2\pi j/3))^{\frac{5}{2}}\Big]
.
\eeq
We note that the above renormalization procedure can be used 
in the cases of the uniaxial and biaxial nematic phases.
To see this, we focus on the uniaxial spin
nematic phase. As $\eta=0$ in the uniaxial nematic phase, 
the relation (\ref{renormn}) is rewritten as
\beq
\bar{c}_0^{(2)}+\bar{c}_2^{(2)}&=&
\frac{7\bar{g}_0+10\bar{g}_2+18\bar{g}_4}{35}\nonumber\\
&=&\left( c_0^{(2)}+c_2^{(2)}\right) 
+ \Bigg[ \left( c_0^{(2)}+c_2^{(2)}\right)^2
+2\left( 3c_1^{(2)}
-c_2^{(2)}\right)^2
+2\left( c_2^{(2)}\right)^2 \Bigg] \frac{M}{V\hbar^2}
{\sk}^{'}\frac{1}{k^2}.
\nonumber\\
\eeq
The mean-field term involves the $k^{-2}$ terms as follows:
\beq
\frac{Vn^2(c_0^{(2)}+c_2^{(2)})}{2}
+\frac{\hbar^2}{8M}{\sk}^{'}\Bigg[
\left(\frac{2Mn(c_0^{(2)}+c_2^{(2)})}{\hbar^2k}\right)^2
+2\left(\frac{2Mn(3c_1^{(2)}-c_2^{(2)})}{\hbar^2k}\right)^2
+2\left(\frac{2Mn(-c_2^{(2)})}{\hbar^2k}\right)^2
\Bigg].\nonumber\\
\label{renormn2}
\eeq
On the other hand, the components of the GSE arising from the quantum
fluctuations diverge as follows:
\beq
-\frac{1}{2}{\sk}^{'}\Bigg[(\ek+n(c_0^{(2)}+c_2^{(2)})-E_{\mk,d})
+2(\ek+q+n(3c_1^{(2)}-c_2^{(2)})-E_{\mk,f_t})
+2(\ek+4q-nc_2^{(2)}-E_{\mk,f_z})
\Bigg]\nonumber\\
\xrightarrow[k\to \infty]{}
-\frac{\hbar^2}{8M}{\sk}^{'}\Bigg[
\left(\frac{2Mn(c_0^{(2)}+c_2^{(2)})}{\hbar^2k}\right)^2
+2\left(\frac{2Mn(3c_1^{(2)}-c_2^{(2)})}{\hbar^2k}\right)^2
+2\left(\frac{2Mn(c_2^{(2)})}{\hbar^2k}\right)^2
\Bigg].
\nonumber\\
\eeq
As in the case of $q=0$,
these divergences are canceled out with the last three terms of 
Eq. (\ref{renormn2}) and the finite GSE is obtained.
The same holds in
the biaxial  nematic phase.
\subsection{Cyclic phase}
In the cyclic phase, we consider the following relations:
\beq
\frac{q^2}{\bar{c}_2^{(2)}}&=&\frac{q^2}{c_2^{(2)}}
-\Bigg[\frac{7g_0^2-10g_2^2+3g_4^2}{35(c_2^{(2)})^2}\Bigg]
\frac{q^2M}{V\hbar^2}{\sk}^{'}\frac{1}{k^2}
\nonumber\\
&=&\frac{q^2}{c_2^{(2)}}-\Bigg[\frac{25(c_2^{(2)})^2+30(c_1^{(2)})^2-60
c_1^{(2)}c_2^{(2)}+10c_0^{(2)}c_2^{(2)}}{5(c_2^{(2)})^2}\Bigg]
\frac{q^2M}{V\hbar^2}{\sk}^{'}\frac{1}{k^2},
\eeq
\beq
\bar{c}_0^{(2)}&=&\frac{4\bar{g}_2+3\bar{g}_4}{7}
=c_0^{(2)} + \Bigg[(c_0^{(2)})^2 +3(2c_1^{(2)})^2
\Bigg] \frac{M}{V\hbar^2}{\sk}^{'}\frac{1}{k^2}.\label{conreba}
\eeq
Hence, the mean-field terms behave as
\beq
2qN+\frac{Vn^2c_0^{(2)}}{2}-\frac{2Vq^2}{c_2^{(2)}}
+\frac{\hbar^2}{8M}{\sk}^{'}\Bigg[
\left(\frac{2Mnc_0^{(2)}}{\hbar^2k}\right)^2
+3\left(\frac{4Mnc_1^{(2)}}{\hbar^2k}\right)^2\nonumber\\
+4\left(\frac{2Mq}{\hbar^2k}\right)^2
\left(\frac{25(c_2^{(2)})^2+30(c_1^{(2)})^2-60c_1^{(2)}c_2^{(2)}
+10c_0^{(2)}c_2^{(2)}}{5(c_2^{(2)})^2}\right)
\Bigg].
\label{a21}
\eeq
At short-wavelength limit, the components of the GSE stemming from the quantum
fluctuations diverge as follows:
\beq
&&-\frac{1}{2}{\sk}^{'}\Bigg[2\left(\ek+2nc_1^{(2)}
+2c_1^{(2)}q/c_2^{(2)}
-q-E_{\mk,f_t}\right)
+\left(\ek+2nc_1^{(2)}+2q-4c_1^{(2)}q/c_2^{(2)}
-E_{\mk,f_z}
\right)\nonumber\\
&&+\left(2\ek+nc_0^{(2)}+2nc_2^{(2)}-E_{\mk,\theta}-E_{\mk,d}
\right)
\Bigg]\nonumber\\
&&\xrightarrow[k\to \infty]{}
-\frac{\hbar^2}{8M}{\sk}^{'}\Bigg[
\left(\frac{2Mnc_0^{(2)}}{\hbar^2k}\right)^2
+3\left(\frac{4Mnc_1^{(2)}}{\hbar^2k}\right)^2\nonumber\\
&&+4\left(\frac{2Mq}{\hbar^2k}\right)^2
\left(\frac{25(c_2^{(2)})^2+30(c_1^{(2)})^2-60c_1^{(2)}c_2^{(2)}
+10c_0^{(2)}c_2^{(2)}}{5(c_2^{(2)})^2}\right)
\Bigg].\nonumber\\
\eeq
These divergences are canceled out with $k$-dependent terms of Eq. \eqref{a21}.
The GSE $E_0^C$ is then given by
\beq
\frac{E_0^{C}}{V}
&=&2qn+\frac{n^2\bar{c}_0^{(2)}}{2}-\frac{2q^2}{\bar{c}_2^{(2)}}
-\frac{1}{2V}{\sk}^{'}\Bigg[2\left(\ek+2nc_1^{(2)}
+2c_1^{(2)}q/c_2^{(2)}
-q-E_{\mk,f_t}\right)\nonumber\\
&&+\left(\ek+2nc_1^{(2)}+2q-4c_1^{(2)}q/c_2^{(2)}
-E_{\mk,f_z}
\right)
+\left(2\ek+nc_0^{(2)}+2nc_2^{(2)}-E_{\mk,\theta}-E_{\mk,d}
\right)
\Bigg]\nonumber\\
&=&2qn+\frac{n^2c_0^{(2)}}{2}-\frac{2q^2}{c_2^{(2)}}
-\frac{\hbar^2}{4MV}{\sk}^{'}\Bigg\{
\Bigg[k^2
+\frac{2M(2nc_1^{(2)}+2q-4c_1^{(2)}q/c_2^{(2)})}{\hbar^2}\nonumber\\
&&-k\sqrt{k^2+\frac{4M(2nc_1^{(2)}+2q-4c_1^{(2)}q/c_2^{(2)})}{\hbar^2}}
-\frac{1}{2}\left(\frac{2M
(2nc_1^{(2)}+2q-4c_1^{(2)}q/c_2^{(2)})}{\hbar^2k}\right)^2\Bigg]\nonumber\\
&&+2\Bigg[k^2+\frac{2M(2nc_1^{(2)}-q+2c_1^{(2)}q/c_2^{(2)})}{\hbar^2}\nonumber\\
&&-\sqrt{\frac{2M(3n^2c_1^{(2)}c_q^{(2)}-3q^2)}{\hbar^2}+k^2\left(k^2+
\frac{4M(2nc_1^{(2)}+2c_1^{(2)}q/c_2^{(2)}-q)}{\hbar^2}\right)}\nonumber\\
&&-\frac{1}{2}\left(\frac{2M}{\hbar^2k}\right)^2
\left((2nc_1^{(2)}+2c_1^{(2)}q/c_2^{(2)}-q)^2
-(3n^2c_1^{(2)}c_q^{(2)}-3q^2)
\right)\Bigg]\nonumber\\
&&+\Bigg[2k^2+\frac{2Mn(c_0^{(2)}+2c_2^{(2)})}{\hbar^2}-
\frac{2M(E_{\mk,d}+E_{\mk,\theta})}{\hbar^2}
-\left(\frac{2Mn}{\hbar^2k}\right)^2
\left(\frac{(c_0^{(2)})^2}{2}
+\frac{4c_0^{(2)}c_2^{(2)}q^2}{(nc_2^{(2)})^2}
+\frac{4q^2}{n^2}
\right)\Bigg]
\Bigg\}\nonumber\\
&=&2qn+\frac{n^2c_0^{(2)}}{2}-\frac{2q^2}{c_2^{(2)}}\nonumber\\
&&+\frac{8\sqrt{M^3}}{15\pi^2\hbar^3}\Bigg[(
2nc_1^{(2)}+2q-4c_1^{(2)}q/c_2^{(2)})^{\frac{5}{2}}
+2(2nc_1^{(2)})^{\frac{5}{2}}\phi_{7\pm}(t_8)
+(nc_0^{(2)})^{\frac{5}{2}}\phi_{8}(t_9)
\Bigg].
\eeq
\clearpage
\section{Lists of equation numbers and symbols}
In this appendix, we list the properties in each phase of spin-1 and spin-2 BECs
in Table \ref{table:order-parameter},
equation numbers of
various physical quantities in Table \ref{table:number}, and
symbols in Table \ref{table:notation}.

\begin{table}[!h]
\caption{Possible ground-state phases, order parameters $\zeta_m$,
magnetization, and spin-singlet pair amplitude defined by
$\langle s_{-}\rangle=\frac{1}{2}\sum_m(-1)^m\zeta_m\zeta_{-m}$ 
of spin-1 and spin-2 BECs.
Here, F, P, BA, N, BN, UN, and C stand for ferromagnetic, polar,
broken-axisymmetry, nematic, biaxial nematic, uniaxial nematic, 
and cyclic phases, respectively.
In the spin-2 nematic phase, $\eta$ is an additional continuous parameter
that represents the degeneracy of the uniaxial and biaxial nematic
 phases.
In the spin-2 broken axisymmetry phase, $+$ $(-)$ sign corresponds to
the case of $c_1^{(2)}<0$ $(>0)$.  }
\label{table:order-parameter}
\begin{center}

\begin{tabular}{l|c|c|c}
\hline
Phase & order parameter $\zeta_m$ &$\langle\mathbf{f}\rangle$ & 
$2\times\langle s_{-} \rangle$  \\
\hline
Spin-1 F & $(1,0,0)$ &$\langle f^z\rangle=1$ & 0  \\ 
P &  $(0,1,0)$ &$\langle \mathbf{f}\rangle=\mathbf{0}$ & 1 \\ 
P' & $(1/\sqrt{2},0,1/\sqrt{2})$&$\langle\mathbf{f}\rangle=\mathbf{0}$ & 1\\
BA & {\footnotesize$\Big(\sqrt{1/4+q/(8nc_1^{(1)})},
\sqrt{1/2-q/(4nc_1^{(1)})},\sqrt{1/4+q/(8nc_1^{(1)})}\Big)$}
&{\footnotesize$\langle f^x\rangle=\sqrt{1-(q/2nc_{1}^{(1)})^2}$} 
&$-q/(2nc_1^{(1)})$ \\
 \hline
Spin-2 F & $(1,0,0,0,0)$ &$\langle f^z\rangle=2$ &  0 \\
N & $(\sin\eta/\sqrt{2},0,\cos\eta,0,\sin\eta/\sqrt{2})$ 
&$\langle\mathbf{f}\rangle=\mathbf{0}$ & 1 \\
BN & $(1/\sqrt{2},0,0,0,1/\sqrt{2})$ 
&$\langle\mathbf{f}\rangle=\mathbf{0}$ &1 \\
UN & $(0,0,1,0,0)$  &$\langle\mathbf{f}\rangle=\mathbf{0}$ &1 \\
C &{\footnotesize $\Big(\sqrt{1/4-q/(2nc_2^{(2)})},
0,\sqrt{1/2+q/(nc_2^{(2)})},0,\sqrt{1/4-q/(2nc_2^{(2)})}\Big)$} 
&$\langle\mathbf{f}\rangle=\mathbf{0}$ & $2q/(nc_2^{(2)})$\\
M &{\footnotesize $\Big(\sqrt{1/3-q/(3nc_1^{(2)})}
,0,0,\sqrt{2/3+q/(3nc_1^{(2)})},
0\Big)$}&$\langle f^z\rangle=-4q/(3nc_1^{(2)})$ & $0$\\
 &or {\footnotesize$\Big(0,\sqrt{2/3+q/(3nc_1^{(2)})},
0,0,\sqrt{1/3-q/(3nc_1^{(2)})}\Big)$ }
&$\langle f^z\rangle=4q/(3nc_1^{(2)})$ & $0$\\
BA & $(\pm a,b,c,,b,\pm a)$  
&{\footnotesize$\langle f^x\rangle=4b(\sqrt{3/2}c\pm a)$ }
&{\footnotesize $2(a^2-b^2)+c^2$}\\
 \hline
\end{tabular}

\end{center}
\end{table}
\clearpage
\begin{table}[!h]
\caption{Equation numbers of various physical quantities. }
\label{table:number}
\begin{center}
\begin{tabular}{l|r|r|r|r|r}
\hline
Phase & Bogoliubov spectra & GSE & Pressure & Sound velocity & Depletion \\
\hline
Spin-1 F &  (\ref{eq:spin-1-f-b}) &  (\ref{eq:spin-1-f-gse}) &
(\ref{eq:spin-1-f-p})
&  (\ref{eq:spin-1-f-sv}) &  (\ref{eq:spin-1-f-dep}) \\
 P & (\ref{eq:spin-1-p-b1}), (\ref{eq:spin-1-p-b2}) &
(\ref{eq:spin-1-p-gse}) &
(\ref{eq:spin-1-p-p}) & (\ref{eq:spin-1-p-sv}) &
(\ref{eq:spin-1-p-dep}) \\
 P' & 
(\ref{eq:spin-1-p'-b1})-(\ref{eq:spin-1-p'-b3}) 
& (\ref{eq:spin-1-p'-gse}) & (\ref{eq:spin-1-p'-p}) &
(\ref{eq:spin-1-p'-sv}) & (\ref{eq:spin-1-p'-dep})\\
BA & (\ref{eq:spin-1-ba-b1}), (\ref{eq:spin-1-ba-b2})
& (\ref{eq:spin-1-ba-gse}) & (\ref{eq:spin-1-ba-p}) &
(\ref{eq:spin-1-ba-sv}) &
(\ref{eq:spin-1-ba-dep})\\
\hline
Spin-2 F &  (\ref{eq:spin-2-f-b}) &  (\ref{eq:spin-2-f-gse}) &
(\ref{eq:spin-2-f-p}) &
 (\ref{eq:spin-2-f-sv}) &  (\ref{eq:spin-2-f-dep}) \\
N & (\ref{eq:spin-2-n-b1})-(\ref{eq:spin-2-n-b5}) &
(\ref{eq:spin-2-n-gse}) & (\ref{eq:spin-2-n-p}) &
(\ref{eq:spin-2-n-sv}) &
(\ref{eq:spin-2-n-dep})\\
BN & (\ref{eq:spin-2-bn-b1})-(\ref{eq:spin-2-bn-b4}) &
(\ref{eq:spin-2-bn-gse}) & (\ref{eq:spin-2-bn-p}) &
(\ref{eq:spin-2-bn-sv}) &
(\ref{eq:spin-2-bn-dep})\\
UN & (\ref{eq:spin-2-un-b1})-(\ref{eq:spin-2-un-b3}) &
(\ref{eq:spin-2-un-gse}) & (\ref{eq:spin-2-un-p}) &
(\ref{eq:spin-2-un-sv}) &
(\ref{eq:spin-2-un-dep})\\
C & (\ref{eq:spin-2-c-b1}), (\ref{eq:spin-2-c-b2}),
(\ref{eq:spin-2-c-b3}) &
(\ref{eq:spin-2-c-gse}) & (\ref{eq:spin-2-c-p}) &
(\ref{eq:spin-2-c-sv}) &
(\ref{eq:spin-2-c-dep})\\
\hline
\end{tabular}
\end{center}
\end{table}

\clearpage
\begin{table}[!h]
\caption{List of symbols. Note that $c_i^{(j)}$ has the dimension of the
coupling constant and $t_i$ is  dimensionless.}
\label{table:notation}
\begin{center}
\begin{tabular}{l|rr}
\hline
Common & $g_F$ & $4\pi\hbar^2a_F/M$ \\
 & $q$ & $(g\mu_BB)^2/E_{\text{hf}}$ \\
\hline
Spin-1 BEC & $c_0^{(1)}$ & $(g_0+2g_2)/3$  \\ 
 &  $c_1^{(1)}$ & $(g_2-g_0)/3$  \\ 
 & $c_q^{(1)}$ & $q^2/(4n^2c_1^{(1)})$  \\
 & $t_1$ & $q/(n|c_1^{(1)}|)-1$ \\
 & $t_2$ & $-q/(nc_1^{(1)})$\\
 & $t_3$ & $q^2/(nc_0^{(1)}+nc_1^{(1)})^2$ \\
 \hline
Spin-2 BEC & $c_0^{(2)}$ & $(4g_2+3g_4)/7$  \\
 & $c_1^{(2)}$  & $(g_4-g_2)/7$  \\
 & $c_2^{(2)}$ & $(7g_0-10g_2+3g_4)/35$  \\
 & $c_3^{(2)}(\eta)$ & $4c_1^{(2)}\sin^2\eta -c_2^{(2)}$ \\
 & $c_q^{(2)}$ & $4q^2/(n^2c_2^{(2)})$\\
 & $t_4$ & $-3q/(n|c_1^{(2)}-c_2^{(2)}|)-1$ \\
 & $t_5$ & $-4q/(n|c_2^{(2)}|)-1$ \\
 & $t_6$ & $q/(n|3c_1^{(2)}-c_2^{(2)}|)-1$ \\
 & $t_7$ & $q/(n|c_2^{(2)}|)-1$ \\
 & $t_8$ & $|q|/(2nc_1^{(2)})$ \\
 & $t_9$ & $(2q)^2/(nc_0^{(2)})^2$ \\
 \hline
\end{tabular}
\end{center}
\end{table}
%\clearpage


\begin{thebibliography}{99}
\bibitem{note:nematic}
Since, the spin fluctuation around the $z$ axis
vanishes when $\eta=0$,
we cannot identify Eq. (\ref{fluc-ne-z})  as the spin
fluctuation.
However, we can interpret Eq. (\ref{fluc-ne-z}) as the fluctuation
operator with respect to one of the $SO(5)$ directions as discussed
in Sec.  V\hspace{-.1em}I.
\end{thebibliography}
\end{document}